\documentclass[10pt]{article}
\usepackage[a4paper,margin=2.4cm]{geometry}
\usepackage[utf8]{inputenc}
\usepackage{lmodern}
\usepackage{url}
\usepackage[T1]{fontenc}
\usepackage{authblk}
\usepackage{graphicx}

\usepackage{amsmath,amssymb}
\usepackage[normalem]{ulem}
\usepackage[numbers,sort&compress]{natbib}

\usepackage{bm}
\usepackage{booktabs,tabularx,array,makecell,multirow,threeparttable}
\usepackage{xcolor,colortbl}
\definecolor{rxnblue}{RGB}{237,245,250}

\usepackage{caption}
\usepackage{longtable,pdflscape}
\usepackage[hidelinks]{hyperref}

\title{RxnScope: Accurate joint generation of product and transition-state pairs from reactants via flow matching}

\author[1,2]{Yanyang Li}
\author[1]{DaoJing Li}
\author[1,2]{Manyi Yang,*}
\author[1]{Shuhua Li,*}

\affil[1]{State Key Laboratory of Coordination Chemistry, Key Laboratory of Mesoscopic Chemistry of Ministry of Education, School of Chemistry, Nanjing University, Nanjing 210023, China}
\affil[2]{The Institute of Green Chemistry and Engineering, Nanjing University, Suzhou 215163, China}
\affil[*]{e-mail: manyi.yang@nju.edu.cn; shuhua@nju.edu.cn}

\date{}
\newcommand{\SIcaptionfont}{\fontsize{10}{12}\selectfont}

\DeclareCaptionFont{SIcaption}{\SIcaptionfont}

\newcolumntype{C}[1]{>{\centering\arraybackslash}p{#1}}
\newcolumntype{M}[1]{>{\centering\arraybackslash}m{#1}}
\newcolumntype{Y}{>{\centering\arraybackslash}X}

\newcommand{\KHPfigref}[1]{Supplementary Fig.~\ref{khp:#1}}

\newcommand{\KHPfigurecaption}[3]{\begingroup\captionsetup{list=false}\captionof{figure}[#2]{\textbf{#2.} #3}\label{khp:#1}\addcontentsline{sif}{figure}{\protect\numberline{\thefigure}#2}\endgroup}
\newcommand{\KHPtableheading}[2]{\refstepcounter{table}\label{khp:#1}\addcontentsline{sit}{table}{\protect\numberline{\thetable}#2}\noindent{\SIcaptionfont\bfseries Supplementary Table \thetable\ | #2}\par\vspace{2mm}}

\makeatletter
\newcommand{\SIlistfigures}{%
  \section*{Supplementary Figures}%
  \begingroup\hypersetup{linktoc=all}\fontsize{10}{12}\selectfont\setlength{\parskip}{0pt}%
  \renewcommand{\numberline}[1]{\hb@xt@\@tempdima{Fig.~##1\hfil}}%
  \def\l@figure{\@dottedtocline{1}{0em}{4em}}%
  \@starttoc{sif}%
  \endgroup}
\newcommand{\SIlisttables}{%
  \section*{Supplementary Tables}%
  \begingroup\hypersetup{linktoc=all}\fontsize{10}{12}\selectfont\setlength{\parskip}{0pt}%
  \renewcommand{\numberline}[1]{\hb@xt@\@tempdima{Table~##1\hfil}}%
  \def\l@table{\@dottedtocline{1}{0em}{4.5em}}%
  \@starttoc{sit}%
  \endgroup}
\makeatother

\begin{document}

\flushbottom
\maketitle
\thispagestyle{empty}

\begin{abstract}
Reaction networks are central to understanding chemical reactivity and guiding reaction design. Constructing them when products are unknown remains a major challenge in computational chemistry, requiring the identification of products and their corresponding transition states (TSs) across elementary steps.
Here, we introduce RxnScope, a flow-matching framework that jointly generates paired product-TS geometries from reactants, conditioned on a potentially reactive region.
Directed product-to-TS coupling allows evolving rather than fixed product geometries to guide TS generation, while paired outputs enable product-topology ranking to prioritize candidates for validation.
RxnScope recovered and validated 95.8\% of reference pathways on Transition1x, performed robustly on multicomponent reactions and enabled multistep pathway exploration.
In automatic reaction discovery, it achieved low-energy pathway coverage comparable to that of the single-ended growing string method while using only about 2\% of its computational cost. These results establish joint product-TS generation as a new strategy for exploring reaction networks from reactants. 
\end{abstract}

\newpage
Reaction networks are central to understanding and predicting chemical reactivity in fields ranging from catalysis and chemical synthesis to drug discovery and prebiotic chemistry
~\cite{Wen2023CRNML,Steiner2024HumanMachine,Ritson2012PrebioticSugars,Seumer2026Pharmaceutical,Zhao2022DeepReactionNetwork}. 
Yet short-lived intermediates and competing pathways make these networks difficult to resolve experimentally~\cite{Ying2025SingleMolecule}.
 Constructing them computationally requires identifying elementary steps that connect reactants, intermediates and products, and locating the associated transition states (TSs) to determine kinetic characteristics~\cite{Simm2019ReactionNetworks,Unsleber2020Exploration}.
 This task is particularly challenging when products are unknown~\cite{Zimmerman2015SEGSM,Lavigne2022IACTA}. The search space expands rapidly with molecular complexity, and each newly identified intermediate can open further reaction channels~\cite{Unsleber2023Templates}.
 Efficient network exploration therefore depends on proposing and prioritizing candidate pathways while minimizing costly electronic-structure calculations. 

Automatic reaction discovery methods~\cite{Joung2026OutcomeReview} broadly follow two strategies: reactant-side exploration~\cite{Unsleber2022Chemoton,Maeda2010AFIR,Maeda2023AFIRReview,Shang2013SSW,Zhang2015SSWRS,Yang2017MDCD,Yang2018MDCD,Li2023MDCD} and product enumeration~\cite{Zimmerman2013ZStruct,Dewyer2018ReactionSpace,Zhao2021YARP,Zhao2022YARP}. Reactant-side methods, such as artificial force-induced reaction (AFIR)~\cite{Maeda2010AFIR,Maeda2023AFIRReview}, stochastic surface walking (SSW)~\cite{Shang2013SSW,Zhang2015SSWRS} and molecular dynamics and coordinate driving (MD/CD)~\cite{Yang2017MDCD,Yang2018MDCD,Li2023MDCD}, identify intermediates, products and connecting TSs through guided or stochastic exploration of the potential-energy surface (PES). Product-enumeration approaches, including ZStruct~\cite{Zimmerman2013ZStruct,Dewyer2018ReactionSpace} and Yet Another Reaction Program (YARP)~\cite{Zhao2021YARP,Zhao2022YARP}, instead enumerate candidate products and use the resulting reactant--product pairs as inputs for TS searches. Despite their success in pathway exploration and reaction design, repeated electronic-structure energy and gradient evaluations remain a major bottleneck for complex systems~\cite{Joung2026OutcomeReview}.
Machine-learned interatomic potentials (MLIPs) can alleviate this bottleneck by providing lower-cost energy and gradient evaluations~\cite{Parrinello2007NNP,Yuan2024Analytical,Anstine2026AIMNet2rxn,Li2026MDCDNN}.

Generative models offer a complementary approach by directly proposing candidate products or TSs for individual elementary steps. FlowER~\cite{Joung2025FlowER} predicts products by modelling  electron redistribution in a bond--electron matrix representation, without generating TS geometries. 
OA-ReactDiff~\cite{Duan2023OAReactDiff} learns an object-aware SE(3)-equivariant~\cite{Batatia2025EquivariantPotentials} joint distribution of reactant, TS and product geometries and uses diffusion-based inpainting to generate TSs conditioned on specified reactant--product structures. 
 React-OT~\cite{Duan2025ReactOT}~reformulates this task as an optimal-transport problem and uses flow matching to generate TSs deterministically.
 These TS-generation workflows nevertheless require predefined products and conformationally compatible reactant--product structures~\cite{Zhao2022ConformationalSampling}, which can be difficult to obtain for unknown reactions.
Although MolGEN~\cite{Tuo2026MolGEN} uses a separate flow-matching model to generate candidate products, its subsequent TS generation is still conditioned on reactant--product pairs, with the product geometries held fixed. 
Other approaches instead generate TS candidates directly from reactant information and identify the products afterwards~\cite{Gugler2026ReactionAtlas,Han2026ReactNexus,Xu2026UniTSGen}.
UniTS-Gen~\cite{Xu2026UniTSGen},for example, generates TS candidates from two-dimensional (2D) reactant graphs and reactive-site indices, then uses TS optimization and intrinsic reaction coordinate (IRC) analysis to identify the products.
However, jointly generating unknown products and their corresponding TS geometries directly from reactants remains challenging.

Here we introduce RxnScope, an equivariant flow-matching framework that jointly generates paired product $P$ and TS geometries from reactants $R$, conditioned on a potentially reactive region (RR),  specified by atom indices $c$, without prescribing bond rearrangements. 
It models the  joint conditional distribution $p_\theta(P,TS\mid R,c)$, using a dual-head velocity model that combines structural transport and geometric refinement within a unified velocity field. 
Directed product-to-TS coupling allows evolving product geometries to guide TS generation, without requiring predefined products or conformationally aligned reactant--product pairs. 
By integrating paired product--TS generation with automatic RR enumeration, reactant conformer sampling, product-topology ranking and electronic-structure validation, RxnScope enables automatic reaction discovery directly from reactants.
On the Transition1x dataset~\cite{Schreiner2022Transition1x} and an external prebiotic aqueous-HCN reaction network~\cite{Zhao2023AqueousHCN}, RxnScope demonstrates robust performance in recovering reference product topologies and generating high-quality TS candidates, including for challenging multicomponent reactions involving up to four reactant components. It also enables multistep reaction-network exploration and identifies previously unreported low-barrier pathways.
In automatic reaction discovery starting from 3-hydroperoxypropanal (KHP)~\cite{Grambow2018KHP}, RxnScope achieves low-energy pathway coverage comparable to that of the single-ended growing string method (SE-GSM)~\cite{Lee2025ReactionSampling}, a well-established method for extensive reaction-pathway exploration, while using only about 2\% as many energy and gradient calculations in the first-round search. These results demonstrate that RxnScope provides an efficient workflow for reaction discovery from reactants.
\section*{Results}
\subsection*{Joint product--TS generation with RxnScope}
\begin{figure}[!t]
\centering
\includegraphics[width=0.95\linewidth]{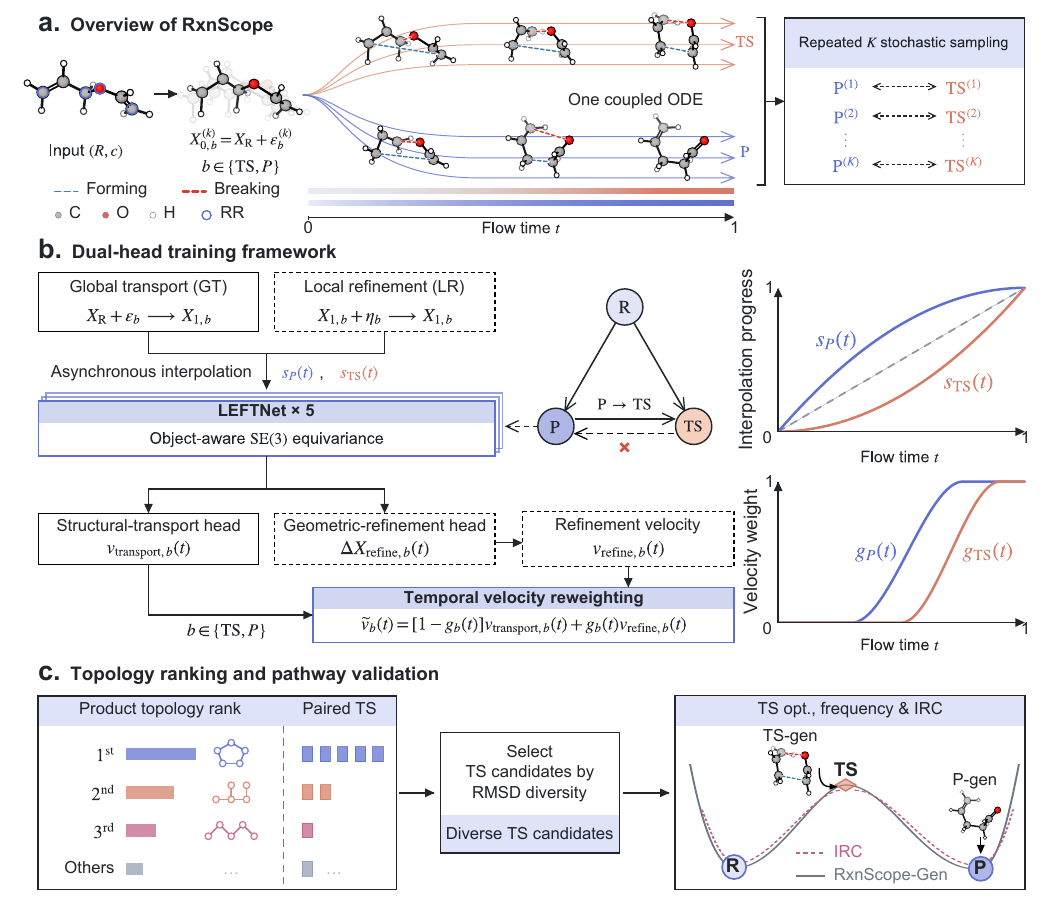}
\caption{
\textbf{Joint product and TS generation with RxnScope.}
\textbf{a}, Starting from reactant structures $R$, conditioned on a potentially reactive region $c$ specified by atom indices, RxnScope jointly generates paired product $P$ and TS geometries.
For each input ($R$, $c$), $K=16$ product-TS pairs are sampled. For each sample, the product and TS branches are initialized from perturbed reactant coordinates $\mathbf X_{0,b}^{(k)}=\mathbf X_R+\boldsymbol\epsilon_b^{(k)}$, where $b\in\{P,\mathrm{TS}\}$, 
and evolve jointly by integrating the learned ordinary differential
equation (ODE). 
\textbf{b}, Dual-head training framework. Global transport (GT) paths start from perturbed reactant coordinates, whereas local refinement (LR) paths start from perturbed reference $P$ and TS geometries. A structural-transport head and a geometric-refinement head share an object-aware SE(3)-equivariant LEFTNet backbone. At the same flow time, the product branch is further along its interpolation path  (upper; \(s_P(t)>s_{\mathrm{TS}}(t)\)) and shifts towards refinement earlier (lower; \(g_P(t)\geq g_{\mathrm{TS}}(t)\)) than the TS branch. Product-to-TS messages are retained, whereas direct TS-to-product messages are masked.
\textbf{c}, The generated pairs are grouped by product topology and ranked by sampling frequency. Geometrically diverse TS candidates are then selected from prioritized groups using a greedy procedure based on pairwise TS root-mean-square deviations (RMSDs),
followed by TS optimization, vibrational analysis and intrinsic reaction coordinate (IRC) validation.
In panels \textbf{a} and \textbf{b}, blue and orange denote the product and TS branches, respectively. 
In panel \textbf{c}, the grey solid and magenta dashed curves schematically illustrate the energy profiles of the RxnScope-generated and IRC-validated pathways, respectively.
}
\label{Figure1}
\end{figure}

Starting from reactant structures \(R\), RxnScope jointly generates paired product \(P\) and TS geometries  using a flow-matching framework, conditioned on a potentially reactive region \(c\) 
specified by atom indices without prescribing bond rearrangements (Fig.~\ref{Figure1}a). RxnScope models this joint distribution \(p_\theta(P,TS\mid R,c)\) using a dual-head architecture that integrates a structural-transport head and a geometric-refinement head within a unified velocity field. Both heads share an object-aware SE(3)-equivariant LEFTNet backbone~\cite{Du2023LEFTNet}, following OA-ReactDiff~\cite{Duan2023OAReactDiff} and React-OT~\cite{Duan2025ReactOT}.

RxnScope was trained on elementary-reaction data comprising corresponding reactant, product and TS geometries, using two complementary interpolation paths Fig.~\ref{Figure1}b).
Global transport (GT) starts from reactant coordinates perturbed by \(\epsilon_b\) (\(X_R+\epsilon_b\)), whereas local refinement (LR)~\cite{Darouich2025AEFM}~starts from reference product or TS coordinates perturbed by \(\eta_b\) (\(X_{1,b}+\eta_b\)). 
For each branch \(b\in\{P,\mathrm{TS}\}\), both paths share the reference target \(X_{1,b}\) and interpolation schedule \(s_b(t)\), with \(X_{1,P}=P^{\mathrm{ref}}\) and \(X_{1,\mathrm{TS}}=\mathrm{TS}^{\mathrm{ref}}\). For flow time \(t\in[0,1]\), the training interpolants are:

\begin{equation}
\begin{aligned}
\mathbf X_{t,b}^{(\mathrm{GT})}
&=
[1-s_b(t)]\bigl(\mathbf X_R+\boldsymbol\epsilon_b\bigr)
+s_b(t)\mathbf X_{1,b},\\[4pt]
\mathbf X_{t,b}^{(\mathrm{LR})}
&=
[1-s_b(t)]\bigl(\mathbf X_{1,b}+\boldsymbol\eta_b\bigr)
+s_b(t)\mathbf X_{1,b}.
\end{aligned}
\label{eq:interpolation_paths}
\end{equation} 

To control the relative contributions of  structural-transport and geometric-refinement velocities, \(v_{\mathrm{transport},b}(t)\) and \(v_{\mathrm{refine},b}(t)\), we introduce a branch-specific, time-dependent gating function \(g_b(t)\). The resulting branch velocity is

\begin{equation}
\tilde v_b(t)
=
[1-g_b(t)]v_{\mathrm{transport},b}(t)
+g_b(t)v_{\mathrm{refine},b}(t).
\label{eq:combined_velocity}
\end{equation}

The product and TS branch velocities then jointly define the unified velocity field \(v_\theta(t)\). 

To strengthen guidance from the evolving product geometry, we schedule the product branch to progress ahead of the TS branch and enter refinement earlier (Fig. 1b). Specifically, the interpolation schedules satisfy \(s_P(t)>s_{\mathrm{TS}}(t)\) for \(0<t<1\), giving the product branch greater interpolation progress at the same flow time. Meanwhile, the gating functions satisfy \(g_P(t)\geq g_{\mathrm{TS}}(t)\), favouring earlier refinement of the product branch.
Product-to-TS messages are retained, whereas direct TS-to-product messages are masked. The evolving product representation thus provides an internally generated condition for TS generation.

During training and benchmark evaluation, RRs are derived from connectivity changes between reactants and reference products. For reaction-network exploration, where products are unknown, candidate RRs are automatically enumerated as subsets of one to six reactant atoms, including explicit hydrogens. 

At inference, RxnScope generates \(K\) product--TS pairs for each input \((R,c)\), with sampling tests supporting \(K=16\) as the default (Supplementary Fig. 1). For each sample \(k\), both branches use the GT initialization, \(\mathbf X_{0,b}^{(k)}=\mathbf X_R+\boldsymbol\epsilon_b^{(k)}\), and evolve jointly by integrating the learned ordinary differential equation (ODE). Each integration produces one product--TS pair \(\bigl(P^{(k)},TS^{(k)}\bigr)\) in  approximately 0.3 s on an NVIDIA V100 GPU.

The generated product--TS pairs are grouped by product topology, and the groups are ranked by sampling frequency to prioritize electronic-structure validation  (Fig. 1c). Within each prioritized group, a greedy procedure based on pairwise TS root-mean-square deviations (RMSDs) selects geometrically diverse candidates (Methods).
Selected candidates undergo geometry optimization and vibrational analysis to assess their first-order saddle-point character, followed by IRC calculations to determine whether each TS connects the input reactants to its paired product.
Candidates satisfying both criteria are classified as intended, and their corresponding product--TS pairs are considered validated. 
A reference reaction is considered successfully recovered and validated if at least one generated product matches the reference product topology and its paired TS is validated as intended. Further computational details are provided in Methods.
\clearpage

\clearpage

\subsection*{Benchmarking paired product--TS generation on Transition1x}
 We first evaluated RxnScope on Transition1x ~\cite{Schreiner2022Transition1x}, a dataset of elementary reactions involving molecules with up to seven heavy atoms (C, N and O). The original training, validation and test splits were retained, with each reactant appearing in only one subset. Following optimization and IRC validation of the reference TSs at the $\omega$B97X/6-31G(d) level used for Transition1x, we retained 265 of the 287 test reactions with confirmed reactant--product connectivity (Supplementary Fig.~2). 
The evaluation used reference 3D reactant geometries (Ref.-3D) or, for comparison, conformers generated from 2D molecular graphs using reactive-region-guided distance geometry in RDKit (RR-DG)~~\cite{Riniker2015ETKDG}.
For each reaction, Ref.-3D sampled 16 product--TS pairs, whereas RR-DG pooled 80 pairs from five distinct reactant conformers that differed substantially from the reference geometry (16 pairs per conformer).

Under Ref.-3D, RxnScope recovered the reference product topology within the Top-3 groups for 254 of 265 reactions (95.8\%; Fig. 2a). Including all sampled groups yielded only two additional recoveries; the remaining nine reactions tended to have higher barriers and reaction energies  (Supplementary Figs. 3 and 4). 
In addition, the Top-3 groups accounted for approximately 99\% of generated product--TS pairs, with 95.0\% concentrated in Top-1 (Supplementary Fig. 5). These results support using sampling frequency to prioritize product-topology groups for electronic-structure validation. Subsequent TS and pathway analyses therefore focused on 3,837 product--TS  pairs from the reference-matching Top-3 groups across these 254 reactions.

Before optimization, the generated TS candidates closely matched the reference TSs, with a median structural RMSD of 0.089 \AA{} and a median absolute electronic-energy difference, $|\Delta E_{\mathrm{TS}}|$, of 2.6 kcal mol$^{-1}$ across all matched candidates (Fig.~2d,e and Table 1). Selecting the lowest-energy candidate per reaction gave corresponding medians of 0.070~\AA{} and 1.2~kcal mol$^{-1}$ (Fig.~2d,e and Table 1).
The latter values were comparable in magnitude to those reported for Confidence-based OA-ReactDiff~\cite{Duan2023OAReactDiff} (0.058 \AA{} and 1.6 kcal~mol$^{-1}$) and React-OT~\cite{Duan2025ReactOT} (0.053 \AA{} and 1.1 kcal~mol$^{-1}$), although the test splits and candidate-selection protocols differed (Table 1).
Notably, in those TS-generation evaluations, both reactant and product geometries were provided as inputs, whereas RxnScope jointly generates product--TS pairs using only reactant geometries and RR atom indices.

\begin{figure}[!htbp]
\centering
\includegraphics[width=\linewidth]{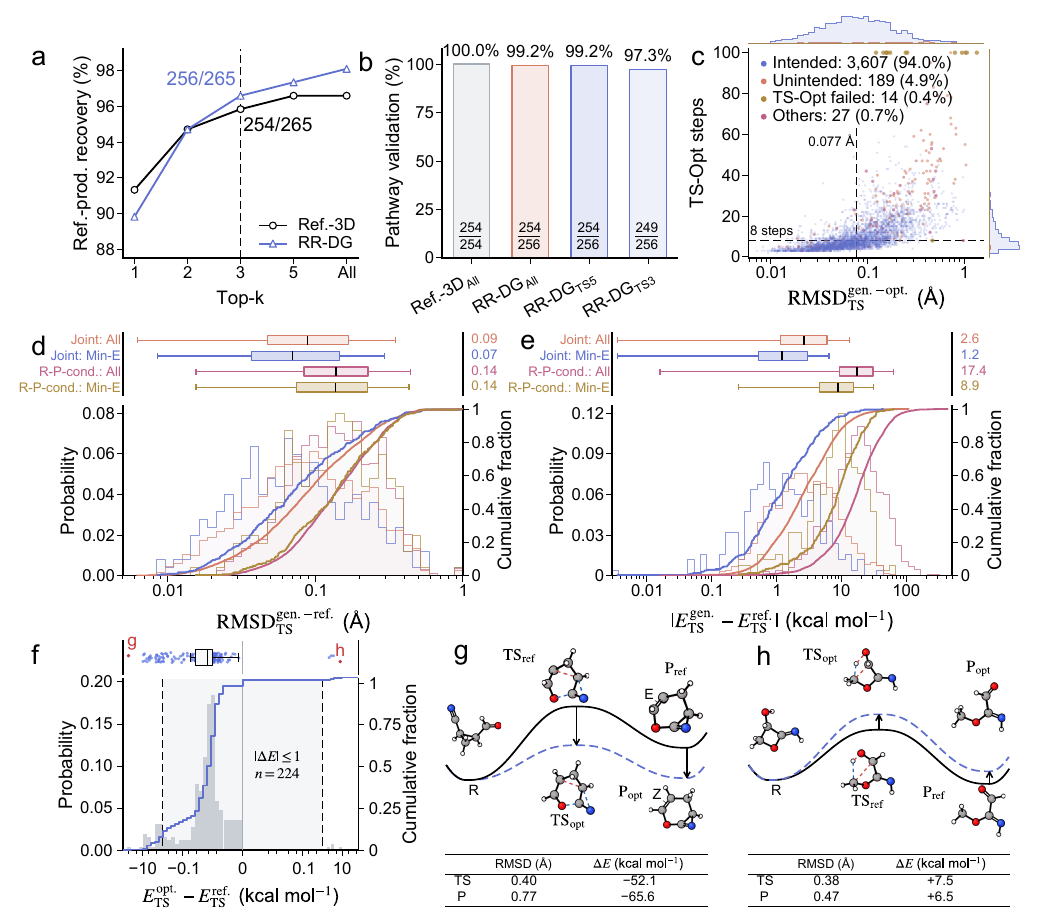}
\caption{\textbf{Overall performance of RxnScope on Transition1x.}
\textbf{a}, 
Top-$k$ recovery of reference product topologies across 265 reactions using reference 3D reactant geometries (Ref.-3D; 16 samples per reaction) or geometries generated by reactive-region-guided distance geometry (RR-DG; 5 conformers and 80 samples per reaction) in RDKit. Panels \textbf{c--f} show results for the reference-matching Top-3 groups across 254 Ref.-3D reactions (3,837 candidates). 
\textbf{b}, Reaction-level validation success among reactions with Top-3 reference-product recovery, defined as the fraction yielding at least one intended pathway after TS optimization and IRC validation. Numbers inside bars indicate validated/recovered reactions. ``All'' indicates that all matched candidates were evaluated, whereas TS5 and TS3 indicate selection of at most five and three geometrically diverse TS candidates per group, respectively.
\textbf{c}, Number of TS optimization steps versus the RMSD between generated and optimized TS geometries. Dashed lines indicate the median RMSD (vertical) and optimization step count (horizontal). Optimization was limited to 100 steps.
\textbf{d,e}, Histograms (left axis) and cumulative distributions (right axis) of structural RMSDs (\textbf{d}) and absolute electronic-energy differences (\textbf{e}) between generated and reference TSs. Joint generation (orange and blue) is compared with a reactant--product-conditioned variant (R-P-cond.; pink and olive), in which reactant and RxnScope-generated product geometries are held fixed during TS generation. Min-E (blue and olive) includes only the lowest-energy candidate per reaction ($n=254$), whereas the other distributions include all matched candidates ($n=3,837$). Boxes indicate medians (solid lines) and interquartile ranges (IQRs), with whiskers extending to the most extreme values within 1.5 times the IQR.
\textbf{f}, 
Histogram (left axis) and cumulative distribution (right axis) of the TS electronic-energy differences between the lowest-energy validated TS and the reference TS for each reaction. Dashed lines mark $\pm1$~kcal~mol$^{-1}$, and labels \textbf{g} and \textbf{h} identify the corresponding reactions  shown in panels \textbf{g} and \textbf{h}.
\textbf{g--h}, 
Electronic-energy profiles of the reference (solid black) and RxnScope-identified (dashed blue) pathways with the largest negative (\textbf{g}) and positive (\textbf{h}) TS energy differences, illustrating $E/Z$ product isomerism and alternative TS conformations, respectively.  
All electronic-structure calculations were performed using Gaussian 16~\cite{Frisch2016Gaussian16} at the $\omega$B97X/6-31G(d) level. The RxnScope-3000E model was used for this benchmark.
}
\label{Figure2}
\end{figure}

To assess the contribution of joint generation, we compared RxnScope with a reactant--product-conditioned variant (R-P-cond.) using the reactant and RxnScope-generated products as fixed inputs during TS generation.(Supplementary Fig.~6). In this comparison, RxnScope reduced the median $|\Delta E_{\mathrm{TS}}|$ from 17.4 to 2.6 kcal mol$^{-1}$ across all matched candidates and from 8.9 to 1.2 kcal mol$^{-1}$ for the lowest-energy candidates, corresponding to reductions of about 85\% and 87\%, respectively (Fig.~\ref{Figure2}e). Median RMSDs also decreased from 0.14 to 0.09 \AA{} and from 0.14 to 0.07 \AA{} (Fig.~\ref{Figure2}d). 
Further ablation studies evaluated asynchronous interpolation, temporal velocity reweighting and local-refinement training. With all three strategies enabled, the median \(|\Delta E_{\mathrm{TS}}|\) decreased by 19.7\% across all matched candidates and by 39.0\% in the lowest-energy subset relative to the model without them, while improving Top-3 reference-product recovery slightly (Supplementary Fig.~7). 
Even without these strategies, joint generation outperformed the R-P-cond. variant, supporting the role of evolving product geometries in TS generation.

The generated TS candidates also showed high success rates in optimization and IRC validation. Of the 3,837 candidates, 99.6\% converged to TSs with a single imaginary frequency (Fig. 2c), with an average of ten optimization steps across all candidates. IRC calculations confirmed 3,607 candidates (94.0\%) as intended. 
For these intended TSs, the median RMSD between generated and optimized geometries was 0.077 \AA{}.  At the reaction level, all 254 reactions with Top-3 reference-product recovery yielded at least one intended pathway, giving an overall validated reference-pathway recovery rate of 95.8\% (Fig. 2b).

For each reaction, we compared the lowest-energy validated TS with the reference.
For 224 reactions (88.2\%), the validated TS was lower in electronic energy than the reference TS by no more than 1 kcal~mol$^{-1}$. For another 26 (10.2\%), it was more than 1 kcal~mol$^{-1}$ lower, whereas for the remaining four (1.6\%), it was more than 1 kcal~mol$^{-1}$ higher (Fig. 2f). The largest negative difference ($-52.1$ kcal~mol$^{-1}$) corresponded to a pathway leading to the \textit{Z} rather than the reference \textit{E} stereoisomer (Fig. 2g). The largest positive difference (+7.5 kcal~mol$^{-1}$) was associated with a different relative orientation of the formyl and imino groups (Fig. 2h).

RxnScope also maintained high reference-product recovery and pathway-validation rates under RR-DG. Top-3 reference-product recovery reached 96.6\% (256/265), comparable to 95.8\% under Ref.-3D (Fig. 2a). Evaluating at most five geometrically diverse TS candidates per reference-matching group  validated 99.2\% (254/256) of the recovered reactions, matching the rate obtained using all matched candidates. 
Even with at most three candidates per group, validation success remained high at 97.3\% (249/256; Fig. 2b). These results support pathway recovery from graph-derived conformers and motivate subsequent use of RR-DG and diverse TS selection.

\clearpage

\begin{table*}[t]
\centering
\begingroup
\sffamily
\small
\definecolor{RxnTableHeader}{RGB}{216,216,195}
\definecolor{RxnTableStripe}{RGB}{239,238,229}
\captionsetup{font={small,sf},labelfont=bf,textfont=bf,
  labelsep=space,justification=raggedright,singlelinecheck=false}
\caption{\textbar\ Comparison of RMSDs and absolute electronic-energy differences for TS structures generated by various methods.}
\label{tab:ts-benchmarks}
\setlength{\tabcolsep}{1.4pt}
\setlength{\aboverulesep}{0pt}
\setlength{\belowrulesep}{0pt}
\renewcommand{\arraystretch}{1.28}
\arrayrulecolor{black}
\renewcommand{\tabularxcolumn}[1]{m{#1}}
\setlength{\arrayrulewidth}{0.3pt}
\newcommand{\RxnHeadCell}[1]{%
  \parbox[c][12mm][c]{\linewidth}{\centering\bfseries #1}%
}
\begin{tabularx}{\textwidth}{
  >{\raggedright\arraybackslash}m{36mm}
  >{\centering\arraybackslash}m{11.5mm}
  >{\centering\arraybackslash}m{13.5mm}
  >{\centering\arraybackslash}m{14.5mm}
  >{\centering\arraybackslash}m{11mm}
  >{\centering\arraybackslash}m{11mm}
  >{\centering\arraybackslash}m{15.6mm}
  >{\centering\arraybackslash}m{15.6mm}
  >{\raggedright\arraybackslash}X
}
\toprule[0.7pt]
\rowcolor{RxnTableHeader}
\RxnHeadCell{Method} &
\RxnHeadCell{Inputs} &
\RxnHeadCell{\shortstack[c]{Samples\\per input}} &
\RxnHeadCell{\shortstack[c]{Evaluated\\TSs}} &
\multicolumn{2}{>{\centering\arraybackslash}m{\dimexpr22mm+2\tabcolsep\relax}}{%
  \parbox[c][12mm][c]{\linewidth}{\centering
    \renewcommand{\arraystretch}{1.05}%
    \begin{tabular}{@{}>{\centering\arraybackslash}m{11mm}
      @{\hspace{\dimexpr2\tabcolsep\relax}}
      >{\centering\arraybackslash}m{11mm}@{}}
      \multicolumn{2}{c}{\textbf{RMSD (\AA)}}\\
      \cline{1-2}
      \textbf{Mean} & \textbf{Median}
    \end{tabular}%
  }%
} &
\multicolumn{2}{>{\centering\arraybackslash}m{\dimexpr31.2mm+2\tabcolsep\relax}}{%
  \parbox[c][12mm][c]{\linewidth}{\centering
    \renewcommand{\arraystretch}{1.05}%
    \begin{tabular}{@{}>{\centering\arraybackslash}m{15.6mm}
      @{\hspace{\dimexpr2\tabcolsep\relax}}
      >{\centering\arraybackslash}m{15.6mm}@{}}
      \multicolumn{2}{c}{{\boldmath$|\Delta E_{\mathrm{TS}}|$} \textbf{(kcal mol$^{-1}$)}}\\
      \cline{1-2}
      \textbf{Mean} & \textbf{Median}
    \end{tabular}%
  }%
} &
\RxnHeadCell{Selection} \\
\hline
TSDiff~~\cite{Kim2024TSDiff} & $R+P$ & 1 & 1,197 &
0.253 & 0.221 & 10.4 & 6.4 & Single prediction \\
\hline
\rowcolor{RxnTableStripe}
OA-ReactDiff~~\cite{Duan2023OAReactDiff} & $R+P$ & 1 & 1,073 &
0.180 & 0.075 & 6.3 & 1.7 & Single prediction \\
\hline
\mbox{OA-ReactDiff~~\cite{Duan2023OAReactDiff} + rec.\textsuperscript{a}} &
$R+P$ & 40 & 1,073 &
0.130 & 0.058 & 4.5 & 1.6 & Confidence-based \\
\hline
\rowcolor{RxnTableStripe}
React-OT~~\cite{Duan2025ReactOT} & $R+P$ & 1 & 1,073 &
0.103 & 0.053 & 3.3 & 1.1 & Deterministic \\
\hline
MolGEN~~\cite{Tuo2026MolGEN}\textsuperscript{b} & $R+P$ & 30 & \textemdash &
0.106 & \textemdash & 3.3 & \textemdash & Lowest energy \\
\hline
\rowcolor{RxnTableStripe}
RxnScope\textsuperscript{c} & $R+\mathrm{RR}$ & 16 & 254 &
0.112 & 0.070 & 2.5 & 1.2 &
Lowest energy \\
\bottomrule[0.7pt]
\end{tabularx}
\par\vspace{3pt}
\begin{minipage}{\textwidth}
\footnotesize
\setlength{\parindent}{0pt}
\setlength{\parskip}{0pt}
RxnScope jointly generates product--TS pairs from reactant geometries $R$ and reactive-region RR atom indices, whereas TS generation in the other methods is conditioned on reactant--product pairs. Inputs are 3D geometries except for TSDiff, which uses 2D molecular graphs. Test sets, input conditions and candidate-selection protocols differ across methods.
$|\Delta E_{\mathrm{TS}}|=|E_{\mathrm{TS}}^{\mathrm{gen}}-E_{\mathrm{TS}}^{\mathrm{ref}}|$, evaluated before TS optimization. Datasets: TSDiff, Grambow (train:validation:test = 8:1:1); OA-ReactDiff and React-OT, Transition1x (train:test = 9,000:1,073); MolGEN and RxnScope, original Transition1x split (train:validation:test = 9,561:225:287). For RxnScope, 265 test reactions passed reference-pathway validation.  \textsuperscript{a}\,rec. denotes recommender-based selection from 40 samples. \textsuperscript{b}\,Lowest-energy candidate selected from 30 generated samples. 
\textsuperscript{c}\, Ref.-3D inputs; the lowest-energy generated TS from 16 samples was selected from the reference-matching product-topology group within the Top-3. 
\end{minipage}
\endgroup
\end{table*}

\newpage
\subsection*{Generalization to multicomponent aqueous-HCN chemistry}
We next assessed RxnScope's generalization to multicomponent prebiotic chemistry using an external reaction set  derived from the aqueous-HCN network reported by Zhao \textit{et al.}~\cite{Zhao2023AqueousHCN}.
For this evaluation, we used RxnScope-500MIX, a model trained on the combined Transition1x and RGD1~\cite{Zhao2023RGD1} datasets.
After data screening and exclusion of training-data overlaps, TS optimization and IRC validation confirmed the assigned reactant--product connectivity for 253 reactions, which were retained for evaluation (Supplementary Figs. 8 and 9). 
Two- and three-component reactions accounted for 89.7\% of the set, compared with 0.14\% in Transition1x and 38.2\% in RGD1; 34.4\% involved three components, compared with approximately 4\% in RGD1 (Supplementary Fig. 10a).
The set also contains a higher proportion of reactions involving at least three bond formations or at least three bond cleavages (Supplementary Fig. 10b). The many possible molecular conformations and intermolecular arrangements make these reactions particularly challenging to explore ~\cite{Zhao2022ConformationalSampling}. 
To assess the influence of reactant conformer preparation, we compared three protocols: Ref.-3D, RR-DG and forming- and breaking-bond-guided distance geometry (FB-DG). Ref.-3D used the reference 3D reactant geometries, whereas RR-DG and FB-DG generated conformers from 2D molecular graphs. RR-DG sampled diverse conformers that differed substantially from the reference geometries, while FB-DG additionally used reference bond-formation and bond-cleavage information to favour arrangements consistent with the target bond changes.

RxnScope achieved high Top-3 reference-product recovery for one- and two-component reactions across all three protocols (96.2--100\% and 94.3--98.6\%, respectively; Fig.~3a). For the more challenging three-component reactions, recovery remained high under Ref.-3D and FB-DG but was somewhat lower under RR-DG.
Under FB-DG, Top-1 recovery nearly matched all-pool recovery, indicating that recovered reference products were generally ranked first. By contrast, for three-component reactions under RR-DG, recovery increased from 70.1\% (61/87) at Top-3 to 86.2\% (75/87) at Top-5, reaching 100\% when all generated candidates were considered. These differences highlight the influence of reactant-complex conformation on product-topology ranking (Supplementary Fig.~11).

Reference-product recovery was accompanied by high pathway-validation rates. Among reactions with reference-product recovery within the evaluated groups (Top-5 for three-component RR-DG and Top-3 otherwise), at least one pathway to the reference product was validated in 97.9\% (234/239) under Ref.-3D, 97.2\% (244/251) under FB-DG and 90.7\% (214/236) under RR-DG. 
Even for three-component reactions under RR-DG, the validation success rate remained high at 86.7\% (65/75; Fig. 3c).

Beyond reference-product recovery, RxnScope also identified reaction channels absent from the aqueous-HCN reference network. Under RR-DG, 52 of 87 three-component reactions (59.8\%) yielded at least two distinct validated reaction channels from conformers sharing the same 2D reactant graphs and RR indices within each reaction (Fig. 3b).
This proportion was higher than that obtained with the other protocols or for reactions with fewer components. Across these three-component reactions, RR-DG identified 113 additional product topologies not represented in the aqueous-HCN dataset, each supported by a validated reaction pathway. Among these additional channels, seven had TS electronic energies more than 1 kcal mol\(^{-1}\) below those of the reference TSs (Fig. 3d and Supplementary Fig. 12). Figure 3e shows three validated pathways from the same reaction. One recovered the reference product, while the other two led to alternative products with TS electronic energies 5.2 kcal mol\(^{-1}\) higher and 34.6 kcal mol\(^{-1}\) lower than the reference TS, respectively.

Guided multistep exploration identified a route to glycine, the simplest amino acid, absent from the reference aqueous-HCN network~\cite{Das2019HCNOrigin,Zhao2023AqueousHCN}. Starting from aminoacetonitrile, an intermediate in the aqueous-HCN network, the pathway proceeds through hydration to form an imidic-acid intermediate, tautomerization to glycinamide and subsequent hydrolysis to glycine.
This pathway involves up to three explicit water molecules, extending the exploration to four-component reactant complexes (Fig.~3f and Supplementary Fig.~13).

\subsection*{Automated exploration and expansion of the KHP reaction network}
Automatic reaction-pathway discovery from reactants alone was evaluated using 3-hydroperoxypropanal (KHP) decomposition~\cite{Jalan2013Korcek}. RxnScope-500MIX generated product--TS pairs for automatically enumerated RRs and ranked the product-topology groups by sampling frequency. Ref.-3D was used for the first-round method comparison, whereas RR-DG was used for subsequent network expansion.
Candidate pathways were validated at the g-xTB~\cite{Froitzheim2025gxTB} level, with selected low-energy pathways refined using density functional theory (DFT) at the B3LYP-D3/TZVP level. We compared first-round reaction-channel coverage and computational cost with SE-GSM~\cite{Zimmerman2015SEGSM} and YARP~\cite{Zhao2021YARP}.

In the first round, RxnScope identified 64 validated reaction channels, compared with 73 for SE-GSM and 30 for YARP (Supplementary Figs.~14--17). These totals include channels identified through unintended pathways (Supplementary Fig. 15).
Excluding IRC calculations, the computational cost was 0.108 million g-xTB energy and gradient evaluations for RxnScope, compared with 5.118 million for SE-GSM and 0.220 million for YARP (Fig. 4a).
 Among the 18 channels with electronic barriers below 70 kcal~mol$^{-1}$, RxnScope recovered 17, SE-GSM 18 and YARP 8 (Supplementary Fig. 18). Thus, RxnScope retained nearly all low-energy channels identified by SE-GSM while requiring only about 2\% as many energy and gradient evaluations. For comparison, the multi-method reaction set of Grambow \textit{et al.}~\cite{Grambow2018KHP} and MolGEN~\cite{Tuo2026MolGEN} contained product connectivities matching 14 and 7 of these 18 low-energy channels, respectively (Supplementary Fig.~18 and Supplementary Table~1).

A second round of searches was initiated from the products of the five DFT-validated first-round  pathways with the lowest free-energy barriers, yielding 473 additional reaction channels. Fig.~\ref{Figure4}c shows these five first-round pathways together with up to five DFT-validated pathways with the lowest free-energy barriers originating from each second-round starting product. The corresponding DFT-refined TS structures are shown in Supplementary Fig.~19. Supplementary Table~2 provides their detailed energetics, while Supplementary Fig.~20 and Supplementary Tables~3--8 list all identified reaction channels and their corresponding energies. 
The network reproduced the established Korcek sequence, in which KHP cyclizes to the five-membered cyclic peroxide IM1, followed by fragmentation to acetaldehyde and formic acid (IM2), with free-energy barriers of 36.5 and 27.1 kcal~mol$^{-1}$, respectively. RxnScope also recovered the competing fragmentation of IM1 to acetic acid and formaldehyde (IM1-1)~\cite{Jalan2013Korcek,Han2026ReactNexus}. For this pathway, we obtained an free-energy barrier of 29.2 kcal~mol$^{-1}$, comparable to that of the IM1\(\to\)IM2 pathway, and a reaction free-energy change of \ensuremath{-}71.7 kcal~mol$^{-1}$ relative to IM1. The expanded network also revealed previously unreported low-barrier pathways leading to IM3-3, IM4-4 and IM5-2~\cite{Jalan2013Korcek,Zhao2022YARP,Tuo2026MolGEN,Han2026ReactNexus}. RxnScope further identified different bond-rearrangement pathways between the same reactant and product topologies. For example, two independently validated IM2\(\to\)IM2-2 pathways involved methyl-hydrogen transfer to different oxygen atoms of formic acid. This illustrates how paired product--TS geometries can distinguish mechanistically distinct pathways that cannot be resolved by product topology alone.

\begin{figure}[!htbp]
\centering
\includegraphics[width=0.94\linewidth]{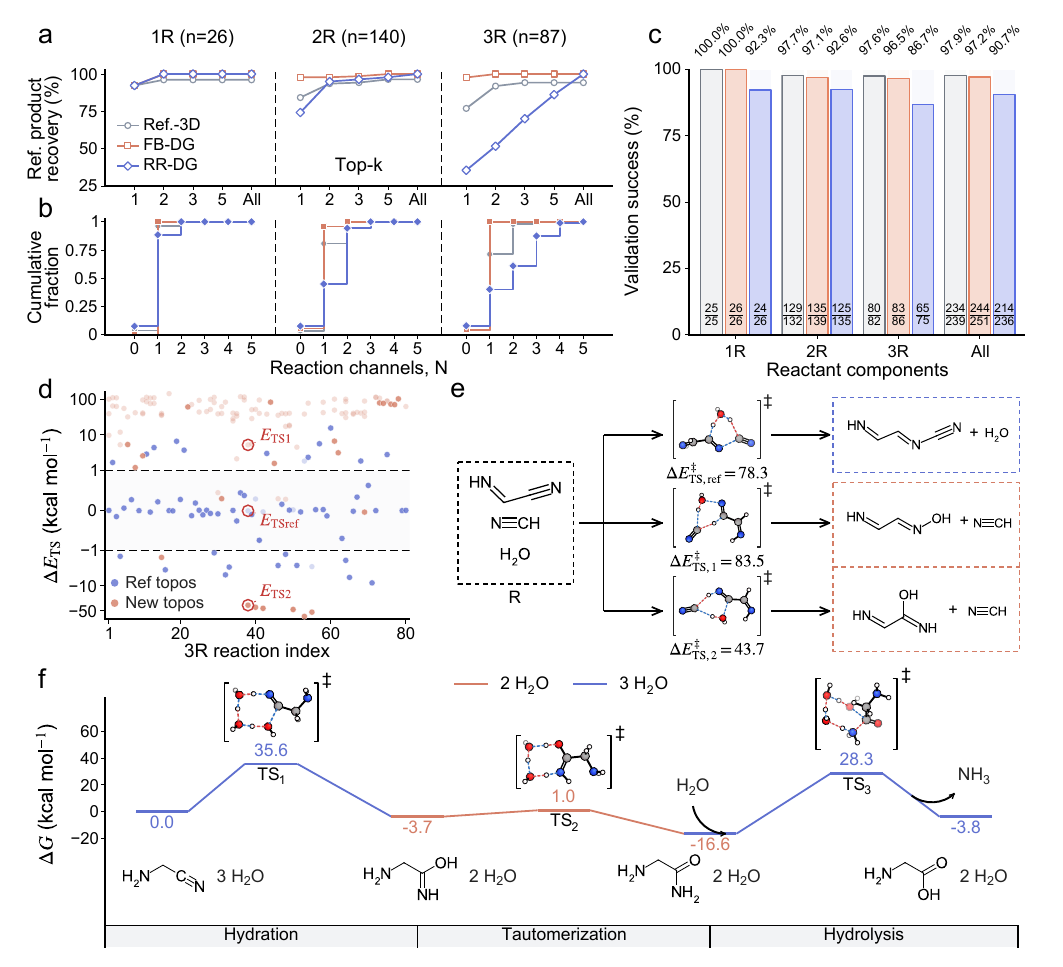}
\caption{
\textbf{Generalization of RxnScope to an external aqueous-HCN reaction network.}
\textbf{a}, Top-$k$ recovery of reference product topologies for reactions with one (1R, $n=26$), two (2R, $n=140$) or three (3R, $n=87$) reactant components, where $n$ denotes the number of reactions. Reactant geometries were obtained using three protocols: Ref.-3D (16 samples per reaction), RR-DG (10 conformers and 160 samples per reaction) and FB-DG (5 conformers and 80 samples per reaction).
\textbf{b}, Cumulative fraction of reactions with at most $N$ distinct intended reaction channels. Within each reaction, channels are generated using the same 2D reactant graphs and RR atom indices and are distinguished by product topology. Both reference and additional channels absent from the aqueous-HCN network are included.
\textbf{c}, Reaction-level validation success among reactions with Top-$k$ reference-product recovery. Numbers inside bars indicate validated/recovered reactions. 
For \textbf{b,c}, validation considered the Top-5 groups for 3R under RR-DG and the Top-3 groups otherwise, using at most five geometrically diverse TS candidates per group for both DG protocols and all candidates per group for Ref.-3D.
\textbf{d}, TS electronic-energy differences,  $\Delta E_{\mathrm{TS}}=E_{\mathrm{TS,opt}}-E_{\mathrm{TS,ref}}$, for validated reaction channels from 3R reactions under RR-DG.  The horizontal axis indexes 80 reactions, each represented by at least one point; points at the same index represent different channels from that reaction. Each point compares the lowest-energy validated TS for one product topology with that reaction's reference TS. Blue and salmon points denote reference-matching and additional channels, respectively. Shading indicates $\pm1$~kcal~mol$^{-1}$.
The circled points highlight the three channels illustrated in \textbf{e}.
\textbf{e}, Three representative validated pathways for one 3R reaction. One pathway recovers the reference product, while the other two lead to additional products. Electronic-energy barriers are defined as $\Delta E^{\ddagger}=E_{\mathrm{TS}}-E_{\mathrm{R}}$.
\textbf{f}, Validated multistep pathway from aminoacetonitrile to glycine, with reactant conformers prepared using FB-DG. Blue and salmon indicate steps involving three and two explicit water molecules, respectively.
Relative Gibbs free energies ($\Delta G$) were calculated using the harmonic approximation at 298.15~K and 1~atm, with the initial reactants as the reference. Electronic-structure calculations were performed using Gaussian~16~\cite{Frisch2016Gaussian16} at the B3LYP-D3/TZVP level, with implicit aqueous solvation described by the polarizable continuum model (PCM; $\varepsilon=80$)~\cite{Tomasi2005PCM}. 
}
\label{Figure3}
\end{figure}
\clearpage

\begin{figure}[!htbp]
\centering
\includegraphics[width=1\linewidth]{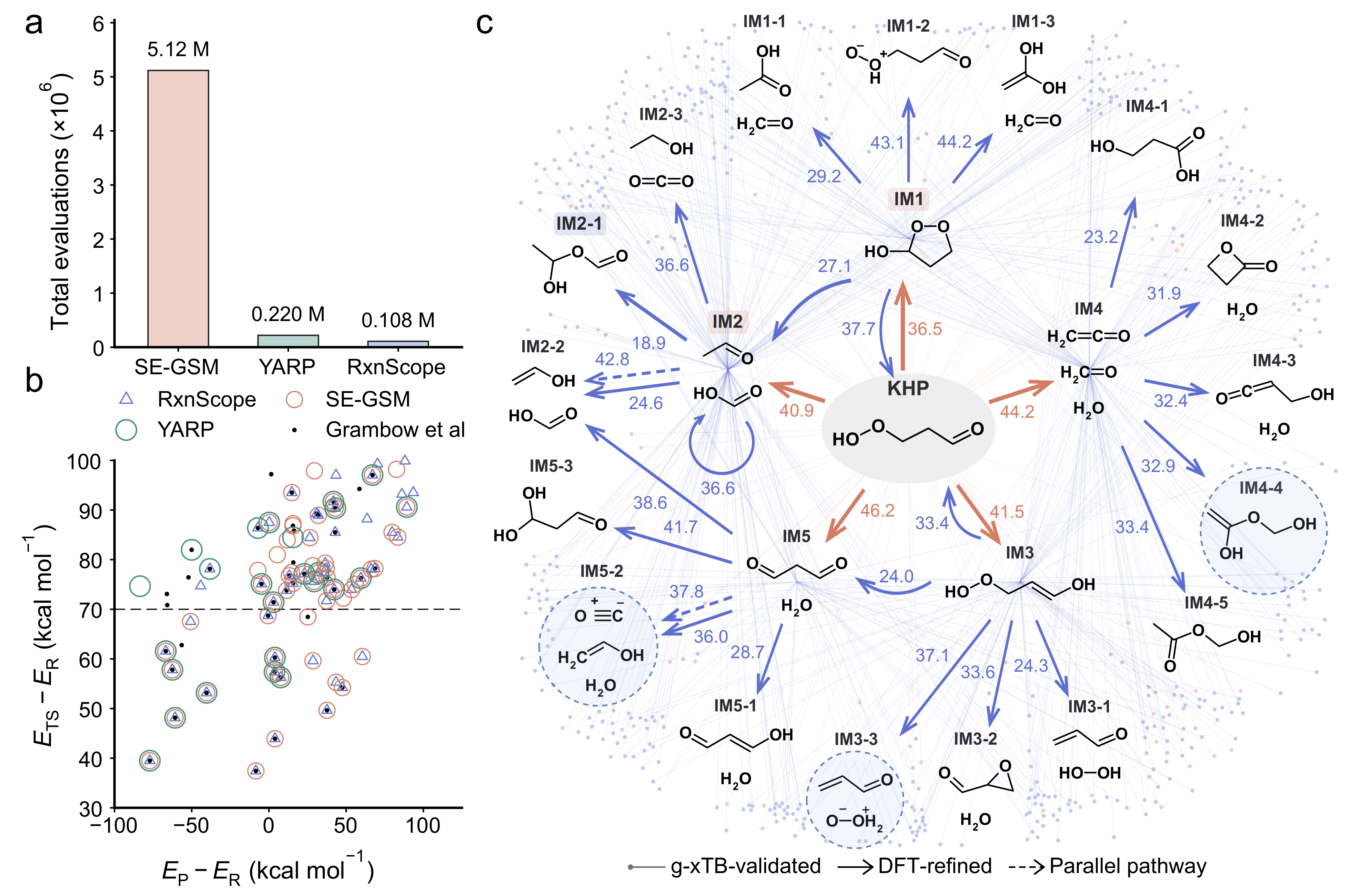}
\caption{
\textbf{Automated reaction-pathway discovery and iterative network expansion from KHP.}
\textbf{a}, Computational costs of first-round searches with SE-GSM, YARP and RxnScope, measured by the total number of g-xTB energy and gradient evaluations for candidate screening, pathway search and TS optimization, excluding IRC calculations.
\textbf{b}, Electronic-energy landscape of the first-round searches at the g-xTB level. Product and TS electronic energies ($E_{\mathrm{P}}-E_{\mathrm{R}}$ and $E_{\mathrm{TS}}-E_{\mathrm{R}}$, respectively) are reported relative to KHP. Black dots, blue triangles, orange circles and green circles denote reaction channels reported by Grambow \textit{et al.}~\cite{Grambow2018KHP} and identified by RxnScope, SE-GSM and YARP, respectively. RxnScope, SE-GSM and YARP recovered 17, 18 and 8, respectively, of the 18 validated channels with electronic barriers below 70~kcal~mol$^{-1}$ (dashed line). 
\textbf{c}, Reaction network obtained after two rounds of RxnScope exploration, comprising 64 first-round and 473 second-round reaction channels. Faint lines show the g-xTB-validated network, while highlighted arrows indicate 30 selected DFT-validated pathways connecting 23 structural nodes. Orange and blue arrows denote first- and second-round pathways, respectively; dashed arrows indicate alternative pathways connecting the same reactant and product nodes via distinct TSs. Labels indicate DFT free-energy barriers ($\Delta  G^{\ddagger}=G_{\mathrm{TS}}-G_{\mathrm{R}}$, kcal~mol$^{-1}$) relative to the corresponding reactant complexes. Gibbs free energies were calculated using the harmonic-oscillator approximation at 298.15~K and 1~atm.
Dashed blue circles highlight representative products of previously unreported low-barrier reaction pathways identified by RxnScope. DFT calculations were performed using Gaussian~16~\cite{Frisch2016Gaussian16} at the B3LYP-D3/TZVP level. 
}
\label{Figure4}
\end{figure}

\section*{Discussion}
We developed RxnScope, a flow-matching framework that jointly generates paired product and TS geometries from reactants, conditioned on a potentially reactive region (RR). At inference, it requires neither predefined products nor conformationally aligned reactant--product pairs.  Compared with a reactant--product-conditioned variant, RxnScope reduced the median absolute TS electronic-energy error from 8.9 to 1.2 kcal~mol$^{-1}$ for the lowest-energy candidates. On Transition1x~\cite{Schreiner2022Transition1x}, RxnScope achieved a 95.8\% overall validated reference-pathway recovery rate. 
It also demonstrated its applicability to multicomponent aqueous-HCN reactions involving up to four reactant components and enabled multistep reaction exploration.

Beyond joint product--TS generation, RxnScope provides an integrated workflow for automatic reaction discovery directly from reactants. 
In the first-round KHP decomposition search, this workflow achieved low-energy pathway coverage comparable to that of SE-GSM while requiring only about 2\% as many g-xTB energy and gradient evaluations, excluding IRC calculations. Subsequent network expansion also revealed previously unreported low-barrier pathways.

RR-DG and FB-DG offer complementary strategies for sampling 3D reactant conformers from 2D molecular graphs. FB-DG uses reference bond-change information to favour geometries associated with specific pathways, whereas RR-DG samples diverse conformers without prescribing bond rearrangements, enabling broader exploration of reaction channels. On Transition1x and one- or two-component aqueous-HCN reactions,  RxnScope maintained high reference-product recovery and pathway-validation rates with RR-DG conformers that differed substantially from the reference geometries, demonstrating robustness to these conformer ensembles.
For more complex multicomponent reactions, focused sampling can complement broader exploration when plausible bond changes are available. Although up to ten conformers per reaction were effective in the systems examined here, flexible molecules, stereoselective reactions and complex multicomponent systems may require more diverse sampling of conformations and intermolecular arrangements to capture competing pathways.

RR conditioning focuses sampling on potentially reactive atoms without prescribing products or specific bond rearrangements. For larger systems, however, the number of candidate regions can grow rapidly, complicating RR selection. 
Chemical heuristics~\cite{Joung2024ReactionMechanisms}, data-driven RR recommenders ~\cite{Lam2024ReactiveSites,Coley2019Reactivity} and reaction-search methods such as MD/CD~\cite{Yang2017MDCD}, could help identify and prioritize promising regions.

Reactive MLIPs could further accelerate downstream screening and pathway validation, as illustrated by MLIP-assisted reaction searches such as MDCD-NN~\cite{Li2026MDCDNN}.

Reliable generalization beyond the chemical and configurational space represented in training data remains challenging. Extending RxnScope beyond its current C, H, O and N training chemistry will require broader, validated datasets and more transferable learning strategies, potentially including self-supervised pretraining on equilibrium conformers~\cite{Darouich2026RobustTSGen}. In turn, RxnScope could help expand these datasets by proposing product--TS pairs for electronic-structure validation, supporting iterative improvements in data coverage and model generalization. Such advances could support automatic reaction discovery and mechanistic studies across chemistry, biology and materials science.

\section*{Methods}
\subsection*{Joint product--TS generation with RxnScope}
Starting from the reactant structures $R=(Z,\mathbf X_R)$, conditioned on a potential reactive region (RR) specified by atom indices $c$, RxnScope jointly generates paired product \(P\) and TS geometries. Here, $Z$ denotes the atomic species,
$\mathbf X_R\in\mathbb R^{N\times3}$ contains the Cartesian
coordinates, and $N$ is the number of atoms. 

During generation, the product and TS coordinates evolve as a
time-dependent joint state $\mathbf X_t$, with $t\in[0,1]$,
\begin{equation}
\mathbf X_t =
[\mathbf X_{t,\mathrm{TS}};\mathbf X_{t,P}]
\in\mathbb R^{2N\times3}.
\label{eq:rxnscope-methods-joint-state}
\end{equation}
Their joint evolution is governed by a learned velocity field
$\mathbf v_\theta$, parameterized by $\theta$:
\begin{equation}
\frac{\mathrm d\mathbf X_t}{\mathrm dt}
=\mathbf v_\theta(\mathbf X_t,t;R,c).
\label{eq:rxnscope-methods-joint-ode}
\end{equation}

\paragraph*{Training paths and asynchronous interpolation.}

RxnScope was trained on elementary-reaction data containing
corresponding reactant, product and TS geometries. We use two complementary training paths: global transport (GT)
and local refinement (LR). For each branch
$b\in\{\mathrm{TS},P\}$, GT starts from reactant coordinates
perturbed by Gaussian noise $\boldsymbol{\epsilon}_b$,
whereas LR starts from the corresponding reference geometry
perturbed by Gaussian noise $\boldsymbol{\eta}_b$.

Let $\mathbf X_{1,P}$ and $\mathbf X_{1,\mathrm{TS}}$
denote the reference product and TS geometries, respectively. Here, $\mathcal C$ denotes the centring operator that subtracts the mean position of the combined TS and product coordinates.

The initial joint states are then given by
\begin{align}
\mathbf X_0^{(\mathrm{GT})}
&=\mathcal C\bigl(
[\mathbf X_R+\boldsymbol{\epsilon}_{\mathrm{TS}};
 \mathbf X_R+\boldsymbol{\epsilon}_P]
\bigr),\\
\mathbf X_0^{(\mathrm{LR})}
&=\mathcal C\bigl(
[\mathbf X_{1,\mathrm{TS}}+\boldsymbol{\eta}_{\mathrm{TS}};
 \mathbf X_{1,P}+\boldsymbol{\eta}_P]
\bigr).
\end{align}

The Gaussian perturbations have standard deviations of 0.20 and 0.30~\AA{} for GT and LR, respectively.
During training, LR is selected with probability $p_{\mathrm{LR}}$ and GT otherwise. The LR probability increases linearly from 0 to 0.25 over the first 200 epochs and remains at 0.25 thereafter. For the ablation without LR training, $p_{\mathrm{LR}}=0$. The LR strategy is inspired by adaptive equilibrium flow matching (AEFM)~\cite{Darouich2025AEFM}.

To strengthen the guidance from the evolving product geometry, we schedule the product branch to progress ahead of the TS branch (Fig.~\ref{Figure1}b). For each training reaction, we sample a common flow time $t\sim\mathcal U(0,1)$ and use different interpolation schedules for the two branches:

\begin{equation}
\begin{aligned}
s_P(t) &= t+\alpha t(1-t),\\
s_{\mathrm{TS}}(t) &= t-\alpha t(1-t),
\end{aligned}
\label{eq:rxnscope-interpolation-schedules}
\end{equation}
where $0\leq\alpha<1$ controls the separation between
the two schedules. We set $\alpha=0.99$, such that
$s_P(t)>s_{\mathrm{TS}}(t)$ for $0<t<1$, giving the
product branch greater interpolation progress at the
same flow time. Setting $\alpha=0$ recovers synchronized
interpolation, which serves as an ablation control.

The jointly centred reference target is defined as $\mathbf X_1=\mathcal C([\mathbf X_{1,\mathrm{TS}};\mathbf X_{1,P}])$.
We introduce the branch-specific interpolation matrix $\mathbf S(t)=\operatorname{diag}(s_{\mathrm{TS}}(t)\mathbf I_N, s_P(t)\mathbf I_N)$, where $\mathbf I_N$ is the $N\times N$ identity matrix.
The interpolated state and its conditional target velocity~\cite{Lipman2023FlowMatching} are then defined as
\begin{equation}
\begin{aligned}
\mathbf X_t
&=\mathcal C\left(
[\mathbf I_{2N}-\mathbf S(t)]\mathbf X_0
+\mathbf S(t)\mathbf X_1
\right),\\
\mathbf u_t
&=\mathcal C\left(
\dot{\mathbf S}(t)(\mathbf X_1-\mathbf X_0)
\right),
\end{aligned}
\label{eq:rxnscope-methods-training-interpolants}
\end{equation}

where $\mathbf X_0$ is sampled from either the GT or LR
path, $\mathbf I_{2N}$ is the $2N\times2N$ identity matrix,
and $\dot{\mathbf S}(t)=\mathrm d\mathbf S(t)/\mathrm dt$. The same centring convention is applied to the coordinates and velocity field during integration.

\paragraph*{Network architecture and velocity-field construction.}

RxnScope uses an object-aware SE(3)-equivariant LEFTNet backbone~\cite{Duan2023OAReactDiff,Du2023LEFTNet} to jointly represent the fixed reactant and evolving product and TS geometries. The backbone consists of five message-passing layers with 128 hidden channels, 96 radial features and a 10~\AA{} cutoff. Two additional four-layer graph encoders extract 16-dimensional molecular and RR context features from the reactant. The RR context is obtained by averaging features over the RR atoms. These context features, together with flow-time, structural-role and RR-membership information, are provided to the generated branches. Message passing occurs both within each structure and between corresponding atoms across structures. The fixed reactant provides information to both generated branches, while product-to-TS messages are retained and TS-to-product messages are masked, allowing the evolving product geometry to guide TS generation. 

The shared backbone feeds two equivariant prediction heads. The structural-transport head predicts a velocity $\mathbf v_{\mathrm{transport}}$, whereas the geometric-refinement head predicts a displacement $\Delta\mathbf X_{\mathrm{refine}}$ towards the target geometry. The latter provides an estimate of the target geometry:
\begin{equation}
\widehat{\mathbf X}_1
=\mathcal C(\mathbf X_t+\Delta\mathbf X_{\mathrm{refine}}).
\label{eq:rxnscope-methods-endpoint-estimate}
\end{equation}

To convert the remaining displacement to the predicted endpoint into a refinement velocity, we introduce a branch-specific scaling factor:
\begin{equation}
D_b(t)=\frac{\dot s_b(t)}
{\max\{1-s_b(t),0.02\}},
\label{eq:rxnscope-methods-branch-conversion}
\end{equation}
where $b\in\{\mathrm{TS},P\}$ and the denominator is bounded below by 0.02 to limit numerical amplification near the endpoint. The refinement velocity is then defined as
\begin{equation}
\begin{aligned}
\mathbf v_{\mathrm{refine}}
=\mathcal C\bigl([&
D_{\mathrm{TS}}(t)
(\widehat{\mathbf X}_{1,\mathrm{TS}}-\mathbf X_{t,\mathrm{TS}});\\
&
D_P(t)
(\widehat{\mathbf X}_{1,P}-\mathbf X_{t,P})]
\bigr).
\end{aligned}
\label{eq:rxnscope-methods-endpoint-velocity}
\end{equation}

For each branch, a time-dependent gating function $g_b(t)$ combines the transport and refinement velocities:
\begin{equation}
\widetilde{\mathbf v}_b(t)
=[1-g_b(t)]\mathbf v_{\mathrm{transport},b}(t)
+g_b(t)\mathbf v_{\mathrm{refine},b}(t).
\label{eq:rxnscope-methods-velocity-blend}
\end{equation} 

The gating weights increase smoothly from zero to one, shifting each branch from structural transport towards geometric refinement. Specifically,
\begin{equation}
\begin{aligned}
g_b(t)&=3q_b(t)^2-2q_b(t)^3,\\
q_b(t)&=\operatorname{clip}\left(
\frac{t-t_{b,\mathrm{start}}}
{t_{b,\mathrm{end}}-t_{b,\mathrm{start}}},0,1
\right).
\end{aligned}
\label{eq:rxnscope-methods-gating}
\end{equation}

The gating windows $(t_{b,\mathrm{start}},t_{b,\mathrm{end}})$ are $(0.30,0.75)$ for the product branch and $(0.50,0.90)$ for the TS branch, allowing product refinement to begin earlier (Fig.~1b). Finally, the branch velocities are concatenated and jointly centred to obtain the unified velocity field:
\begin{equation}
\mathbf v_\theta(\mathbf X_t,t;R,c)
=\mathcal C\left(
[\widetilde{\mathbf v}_{\mathrm{TS}}(t);
 \widetilde{\mathbf v}_P(t)]
\right).
\label{eq:rxnscope-methods-unified-velocity}
\end{equation}

\paragraph*{Training losses and inference.}

The model is trained by supervising the structural-transport velocity, the unified velocity field and the predicted endpoint geometry. We use a role-weighted mean-squared error $E_w$, assigning
weights of 2 and 1 to the TS and product branches, respectively.
For predicted values $\mathbf Y_{\mathrm{pred}}$ and their
corresponding targets $\mathbf Y_{\mathrm{target}}$, we define
\begin{equation}
E_w(\mathbf Y_{\mathrm{pred}},\mathbf Y_{\mathrm{target}})
=
\frac{
\sum_i M_i w_i
\|\mathbf Y_{\mathrm{pred},i}
-\mathbf Y_{\mathrm{target},i}\|_2^2
}{
3\sum_i M_i w_i
},
\end{equation}

where $w_i$ is the branch weight and $M_i$ is 1 for an atom in the reaction and 0 for a padding entry. The dual-head loss is
\begin{equation}
\mathcal{L}_{\mathrm{dual}}
=
\mathbb{E}\left[
E_w(\mathbf v_{\mathrm{transport}},\mathbf u_t)
+E_w(\mathbf v_\theta,\mathbf u_t)
+0.5E_w(\widehat{\mathbf X}_1,\mathbf X_1)
\right].
\label{eq:rxnscope-methods-dual-loss}
\end{equation} 

The expectation is taken over training reactions, the sampled GT or LR initial states and the flow time.

To improve geometric accuracy around the reactive region, we introduce an additional RR distance loss that penalizes errors in interatomic distances involving at least one RR atom. We define the set of relevant atom pairs as
\begin{equation}
\mathcal Q_c=
\{(i,j):1\leq i<j\leq N,\;
i\in c\ \mathrm{or}\ j\in c\}.
\end{equation}
For each branch $b\in\{\mathrm{TS},P\}$, the RR distance loss is
\begin{equation}
\ell_b^{\mathrm{RR}}
=
\frac{1}{|\mathcal Q_c|}
\sum_{(i,j)\in\mathcal Q_c}
\left[
d_{ij}(\widehat{\mathbf X}_{1,b})
-d_{ij}(\mathbf X_{1,b})
\right]^2,
\label{eq:rxnscope-methods-rr-loss}
\end{equation}
where $d_{ij}$ denotes the distance between atoms $i$ and $j$. The total training loss is
\begin{equation}
\mathcal L
=
\mathcal L_{\mathrm{dual}}
+0.05\,\mathbb E\left[
\ell_{\mathrm{TS}}^{\mathrm{RR}}
+\ell_P^{\mathrm{RR}}
\right].
\label{eq:rxnscope-methods-total-loss}
\end{equation}

At inference, both branches use GT initial states derived from the reactant, without requiring reference product or TS geometries. The coupled ODE is integrated over 50 intervals using 49 Heun steps followed by one Euler step. We generate 16 product--TS pairs per reactant geometry and RR, with each pair generated in approximately 0.3~s on an NVIDIA V100 GPU. 

\subsection*{Training datasets and model settings}

\paragraph*{Datasets.}
Transition1x~\cite{Schreiner2022Transition1x} contains 10,073 elementary reactions involving H, C, N and O, with up to seven heavy atoms. Its reaction pathways were calculated using the climbing-image nudged elastic band (CI-NEB) method at the $\omega$B97X/6-31G(d) level. RGD1~\cite{Zhao2023RGD1} contains 176,992 elementary reaction records involving the same elements, with up to ten heavy atoms. Its TS geometries were optimized at the B3LYP-D3/TZVP level.

\paragraph*{Model training and applications.}
We trained two RxnScope models using different datasets and training durations. RxnScope-3000E was trained on the 9,561-reaction Transition1x training set for 3,000 epochs, whereas RxnScope-500MIX was trained on the same Transition1x training set combined with 176,898 RGD1 records for 500 epochs. RxnScope-3000E was used for the Transition1x benchmark, while RxnScope-500MIX was used for the external aqueous-HCN evaluation and KHP reaction-network exploration.

\paragraph*{Training settings.}
Both models were trained using the AdamW optimizer with AMSGrad, an initial learning rate of $5\times10^{-4}$ and a weight decay of $10^{-12}$. The learning rate was adjusted using the ReduceLROnPlateau scheduler with a reduction factor of 0.95, a patience of 20 epochs and a minimum learning rate of $10^{-6}$. Training was performed on eight NVIDIA V100 GPUs with a global batch size of 256. The inference used exponential-moving-average parameters with a decay of 0.999.

\subsection*{Automatic reactive-region assignment and enumeration}

For training and benchmarks with specified RRs, the RR comprises atoms involved in reactant--product connectivity changes; bond-order-only changes are excluded. For automatic reaction discovery, candidate RRs are enumerated from the reactant by selecting all subsets of one to six atoms, including explicit hydrogens. For a system containing $N$ atoms, this gives

\begin{equation}
N_{\mathrm{RR}}=\sum_{k=1}^{\min(6,N)}\binom{N}{k}.
\label{eq:rr-enumeration}
\end{equation}

\subsection*{Reactant conformer preparation}

Reactant geometries were prepared using three protocols. Ref.-3D used the reference reactant geometries from the datasets. The other two protocols, reactive-region-guided distance geometry (RR-DG) and forming- and breaking-bond-guided distance geometry (FB-DG), generated conformers from 2D reactant molecular graphs using RDKit-based distance geometry~\cite{Riniker2015ETKDG,Wang2020Conformer}. RR-DG used RR atom indices to guide conformer generation, whereas FB-DG additionally used reference bond-formation and bond-cleavage information as a reference-informed control. In all cases, only the resulting 3D reactant geometries and RR atom indices were provided to RxnScope.

\paragraph*{RR-DG protocol.}
RR-DG used the reactant connectivity graph, charge, spin multiplicity and RR atom indices $c$ to generate conformers. Distance constraints were constructed from combinations of atom pairs within the RR. RDKit was used to generate a candidate pool under these constraints, with 500 embeddings requested and a nonbonded radius factor of 1.50. Candidates were retained only if they preserved the original reactant connectivity. 
Geometrically diverse conformers were then selected using the RMSD-based procedure described in \textit{RMSD-based selection of diverse structures}, with the number of conformers determined by the evaluation protocol.

\paragraph*{FB-DG protocol.}
FB-DG used reference bond-forming and bond-breaking assignments to guide conformer generation. The corresponding atom pairs were identified by comparing reference reactant and product adjacency matrices constructed using a distance scale of 1.20. These pairs were incorporated as distance constraints during RDKit embedding. As in RR-DG, 500 embeddings were requested with a nonbonded radius factor of 1.50. 

\subsection*{Product-topology voting and ranking}

Generated product geometries were converted into all-atom binary connectivity graphs using a distance criterion adapted from Topology Automated Force-Field Interactions (TAFFI)~\cite{Seo2021TAFFI}. Two atoms $i$ and $j$ were considered connected when $0<d_{ij}<1.20(r_i+r_j)$, where $d_{ij}$ is their interatomic distance and $r_i$ and $r_j$ are modified UFF radii. The radii were 0.390, 0.757, 0.700 and 0.658~\AA{} for H, C, N and O, respectively. If an atom exceeded its allowed number of neighbours (H: 2; C: 4; N: 4; O: 2), the longest assigned connection was removed until the limit was satisfied.

For each reactant and RR, the product--TS pairs generated from all sampled reactant conformers were pooled. Products with identical adjacency matrices under the recorded atom correspondence were grouped by topology and ranked by their vote fractions, calculated as the number of pairs in each topology group divided by the total number of pairs generated for that reactant--RR combination.

\subsection*{RMSD-based selection of diverse structures}

We used the same RMSD-based procedure to select geometrically diverse reactant conformers in RR-DG and FB-DG and TS candidates from prioritized product-topology groups. Structural differences were measured using all-atom Kabsch RMSD with fixed atom correspondence, taking the smaller value from alignments with and without reflection. These calculations did not alter the original geometries.

The selection began with a representative structure, identified as the candidate with the smallest average RMSD to a subset of up to 64 candidates sampled at approximately uniform intervals in generation order. The second structure was chosen to maximize its RMSD from the first, while the third was selected with an RMSD from the first was closest to half the RMSD between the first two. Additional structures were then selected iteratively by maximizing their minimum RMSD to the already selected structures, until the required number was reached. If fewer candidates were available, all were retained.

\subsection*{Electronic-structure validation}

Selected TS candidates were subjected to geometry optimization, harmonic frequency analysis and intrinsic reaction coordinate (IRC) calculations. Calculations were performed using Gaussian~16 (Revision C.01)~\cite{Frisch2016Gaussian16}, with g-xTB calculations carried out through its external interface. All systems were treated as neutral singlets. Transition1x validation used $\omega$B97X/6-31G(d), whereas aqueous-HCN validation used B3LYP-D3/TZVP with the polarizable continuum model (PCM; $\varepsilon=80$). 
For KHP reaction-network exploration, initial validation was performed using g-xTB, followed by B3LYP-D3/TZVP calculations for selected pathways.

A TS candidate was accepted as a first-order saddle point if its optimization converged and frequency analysis confirmed exactly one imaginary frequency. TS optimization was limited to 100 steps. IRC calculations used the local quadratic approximation (LQA) with a step size of 0.05~\AA{} and a maximum of 100 points per direction. For validation of the Transition1x reference TSs, the maximum was 300 points per direction. Calculations were required to converge successfully in both directions.

The IRC endpoints were compared with the input reactant and the originally paired generated product based on their molecular connectivity. A pathway was classified as intended if the optimized TS connected the input reactant to its paired product. The corresponding product--TS pair was then considered validated. Pathways connecting the input reactant to a different product were classified as unintended and excluded from intended-pathway counts. For the Transition1x and aqueous-HCN evaluations, only intended pathways were included in the reported pathway-validation results. In KHP discovery, channels supported by unintended pathways were also included in the total validated-channel count, using the products identified by IRC. 

\subsection*{First-round KHP reaction searches}
\paragraph*{SE-GSM.}
The single-ended growing string method (SE-GSM) was implemented using the pyGSM version distributed with legacy YARP (\url{https://github.com/zhaoqy1996/YARP})~\cite{Zimmerman2015SEGSM}. Starting from the KHP reactant graph, candidate bond rearrangements were enumerated with up to two bond-forming and two bond-breaking (b2f2) operations, excluding cases without bond changes. The allowed numbers of neighbours for affected atoms were 0--1 for H, 1--4 for C and 0--3 for O. For each enumerated combination of bond-forming and bond-breaking operations, one SE-GSM search was performed using g-xTB energies and gradients. Each completed string provided a reactant geometry, an initial TS guess and a product-side endpoint.

\paragraph*{YARP.}
YARP 3.0.0 (\url{https://github.com/Savoie-Research-Group/yarp})~\cite{Zhao2021YARP,Zhao2022YARP} was used to enumerate candidate product graphs based on its Lewis-score and atom-pairing rules, allowing up to two bond-forming and two bond-breaking (b2f2) operations. Reactant and product conformers were sampled and optimized using CREST 2.12 (\url{https://github.com/crest-lab/crest}) with GFN2-xTB~\cite{Pracht2020CREST,Bannwarth2019GFN2}. For each product retained after screening, three distinct, conformationally aligned reactant--product pairs were prepared. Each pair was subjected to a separate double-ended growing string method (GSM) search using Pysisyphus (\url{https://github.com/eljost/pysisyphus})~\cite{Steinmetzer2021Pysisyphus}, with g-xTB energies and gradients. The highest-energy image from each string was retained as a TS guess, together with its corresponding reactant and product geometries.

\paragraph*{RxnScope.}
For a consistent comparison with SE-GSM and YARP, RxnScope used the reference 3D KHP reactant geometry (Ref.-3D). Candidate reactive regions (RRs) containing one to six atoms, including explicit hydrogens, were automatically enumerated, and 16 product--TS pairs were generated for each RR. Pairs whose generated products contained isolated hydrogen atoms were first excluded. For each RR, the remaining products were grouped by topology and ranked by sampling frequency. Only the highest-ranked (Top-1) topology group was retained if it accounted for at least 25\% of the remaining pairs. One product from each retained group was then optimized using g-xTB to check whether its connectivity was preserved. For groups passing this screening, up to three geometrically diverse product--TS pairs were selected, retaining their original product--TS associations.\\

For all three methods, the resulting TS candidates and their associated reactant--product structures were subjected to the same electronic-structure validation procedure, including TS optimization, frequency analysis and IRC calculations, as described in \textit{Electronic-structure validation}.

Computational cost was measured by the number of g-xTB energy and gradient evaluations, excluding IRC calculations. For SE-GSM and YARP, the counts included pathway searches and subsequent TS optimizations. For RxnScope, they included product preoptimization during screening and subsequent TS optimization of retained candidates. The GFN2-xTB calculations used for YARP conformer preparation were excluded from this comparison.

\subsection*{Second-round KHP exploration}

Second-round searches started from the products of the five first-round
pathways with the lowest DFT free-energy barriers. For each
enumerated RR, five RR-DG conformers were generated, with 16 product-TS pairs sampled per conformer. Samples from all five conformers were pooled for each RR, and up to three geometrically diverse product-TS pairs were selected from the Top-1 product-topology group for TS optimization and IRC validation. Selected low-energy pathways were then refined at the B3LYP-D3/TZVP level. The network 
visualization of Fig.~\ref{Figure4}c includes the five first-round pathways and up to five DFT-validated pathways with the lowest free-energy barriers from each second-round starting product.

\subsection*{Reactant--product-conditioned TS-generation control}

To compare joint product--TS generation with TS generation conditioned on fixed product geometries, we trained a reactant--product-conditioned model (R-P-cond.) based on React-OT~\cite{Duan2025ReactOT}. The model was trained from scratch using the same Transition1x training set as RxnScope. Its object-aware LEFTNet backbone consisted of six message-passing layers with 196 hidden channels, 96 radial features and a 10~\AA{} cutoff. Training used AdamW with AMSGrad, a learning rate of $10^{-4}$, zero weight decay, no learning-rate scheduler and an exponential moving average (EMA) decay of 0.999.

At inference, R-P-cond. used the reactant and RxnScope-generated product geometries as fixed inputs. The two geometries were centred independently, and the initial TS coordinates were set to the midpoint of corresponding atomic positions. TS generation used 50 midpoint integration steps without additional noise. R-P-cond. models trained for 500 and 1,000 epochs showed similar TS accuracy (Supplementary Fig.~6), and the 1,000-epoch model was used for the main-text comparison.

\subsection*{Evaluation metrics}
\paragraph{Reference-product recovery rate.}
The fraction of evaluated reactions for which the reference product topology was recovered within the selected Top-$k$ product-topology groups.
\paragraph{Reaction-level pathway-validation rate.}
Among reactions with reference-product recovery within the selected Top-$k$ groups, the fraction for which at least one intended pathway to the reference product was confirmed by TS optimization and IRC calculations.
\paragraph{Overall validated reference-pathway recovery rate.}
The fraction of all evaluated reactions for which the reference product topology was recovered and at least one intended pathway to that topology was validated.
\paragraph{Structural and energetic accuracy.}
All-atom RMSDs between generated and reference geometries were calculated using symmetry-aware alignment, allowing same-element atom permutations and reflection. The RMSD-based selection of geometrically diverse candidates used fixed atom correspondence. Absolute TS electronic-energy differences were calculated as $\lvert E_{\mathrm{TS,gen}}-E_{\mathrm{TS,ref}}\rvert$, with both energies evaluated at the same level of theory. All TS geometry and energy statistics refer to generated structures before optimization.
\paragraph{Candidate selection for accuracy evaluation.}
For \textit{Min-E}, the generated TS with the lowest electronic energy was selected from the reference-matching Top-3 pool for each recovered reaction. The same selected TS was used to evaluate both RMSD and energy error. For \textit{All matched candidates}, every TS in that pool was evaluated.

\section*{Data availability}

The Transition1x and RGD1 datasets are publicly available at \url{https://doi.org/10.6084/m9.figshare.19614657.v4} and \url{https://doi.org/10.6084/m9.figshare.21066901.v6}, respectively. The aqueous-HCN reference reaction data were obtained from Zhao~\textit{et al.}~\cite{Zhao2023AqueousHCN} and their supporting data.

The processed benchmark datasets and split definitions, generated product--TS pairs, TS optimization and IRC validation results, KHP reaction-network data and associated energies, and source data underlying the figures and tables are maintained in a private repository. These data are available to editors and reviewers during peer review and will be publicly released upon acceptance of the manuscript.

\section*{Code availability}

The RxnScope source code, model configurations, trained weights, and scripts for data preparation, training, inference, reaction-network exploration and benchmark reproduction are maintained in a private GitHub repository. The code and associated documentation are available to editors and reviewers during peer review and will be publicly released upon acceptance of the manuscript.

\section*{Author contributions}
S.L., M.Y. and Y.L. conceived the project. Y.L. developed the model, designed the computational experiments and performed the simulations. D.L. implemented the conformer-generation workflow. S.L. and M.Y. supervised the project. Y.L. wrote the initial draft of the manuscript. All authors discussed the results, reviewed the manuscript and approved its final version.

 \subsection*{Funding}

S.L. and M.Y. acknowledge support from the Fundamental and Interdisciplinary Disciplines Breakthrough Plan of the Ministry of Education of China (No. JYB2025XDXM309). M.Y. also acknowledges support from the National Natural Science Foundation of China (No. 22403042), the Suzhou Innovative Leading Talent Project (No. ZXL2025318), the Natural Science Foundation of Jiangsu Province (No. BK20241192) and the National Key Research and Development Program of China (No. 2025YFA1511500). S.L. also acknowledges support from the National Natural Science Foundation of China (No. 22433001) and the robotic AI-Scientist platform of the Chinese Academy of Sciences for dataset construction using the Automated Design of Chemical Reactions (ADCR) program.

\section*{Competing interests}
The authors declare no competing interests.

\section*{Supplementary information}
Supplementary Information accompanies this manuscript.

\bibliographystyle{unsrtnat}
\bibliography{references}

\clearpage
\newgeometry{left=18mm,right=18mm,top=18mm,bottom=18mm}
\setcounter{figure}{0}
\setcounter{table}{0}
\renewcommand{\theHfigure}{S\arabic{figure}}
\renewcommand{\theHtable}{S\arabic{table}}
\renewcommand{\figurename}{Supplementary Figure}
\renewcommand{\tablename}{Supplementary Table}
\captionsetup{font=SIcaption,labelfont=bf,textfont=normalfont,labelsep=period,justification=raggedright,singlelinecheck=false,skip=4pt,hypcap=false}
\setlength{\parindent}{0pt}
\setlength{\parskip}{5pt}
\setlength{\emergencystretch}{3em}
\raggedbottom

{\Large\bfseries Supplementary information\par}
{\large\bfseries RxnScope: Accurate joint generation of product and transition-state pairs from reactants via flow matching\par}
\vspace{2mm}
{\normalsize Yanyang Li\textsuperscript{1,2}, DaoJing Li\textsuperscript{1}, Manyi Yang\textsuperscript{1,2,*} and Shuhua Li\textsuperscript{1,*}\par}
{\small \textsuperscript{1}State Key Laboratory of Coordination Chemistry, Key Laboratory of Mesoscopic Chemistry of Ministry of Education, School of Chemistry, Nanjing University, Nanjing 210023, China\par
\textsuperscript{2}The Institute of Green Chemistry and Engineering, Nanjing University, Suzhou 215163, China\par
\textsuperscript{*}Correspondence to \href{mailto:manyi.yang@nju.edu.cn}{manyi.yang@nju.edu.cn} and \href{mailto:shuhua@nju.edu.cn}{shuhua@nju.edu.cn}.\par}
\vspace{2mm}
\SIlistfigures
\SIlisttables
\clearpage

\section*{Supplementary figures}

\addcontentsline{toc}{section}{Supplementary figures 1--19}

\noindent\begin{minipage}{\linewidth}
\refstepcounter{figure}\label{fig:s04_sampling_budget_product_geometry}
\addcontentsline{sif}{figure}{\protect\numberline{\thefigure}Sampling budget and accuracy of generated product geometries}

{\centering\includegraphics[width=\linewidth,height=0.76\textheight,keepaspectratio]{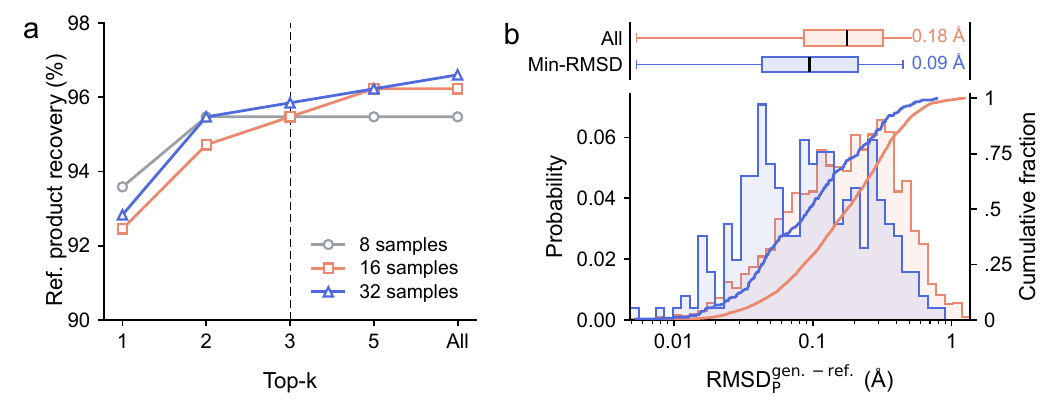}\label{si:asset_04}\par}
\vspace{2mm}
{\SIcaptionfont
\textbf{Supplementary Figure~\ref{fig:s04_sampling_budget_product_geometry}. Sampling budget and accuracy of generated product geometries.} (a) Top-$k$ recovery of reference product topologies using 8, 16 or 32 product--TS pairs per reaction under Ref.-3D ($n=265$). Topology groups were ranked by sampling frequency; All includes all generated groups.
(b) RMSDs between generated and reference product geometries under Ref.-3D with 16 samples per reaction. All includes 3,837 candidates from the reference-matching Top-3 product-topology groups across 254 reactions, whereas Min-RMSD selects the candidate closest to the reference product for each reaction. Boxes show medians and interquartile ranges (IQRs), with whiskers extending to the most extreme values within 1.5 times the IQR. Histograms and cumulative curves show the corresponding RMSD distributions.\par}
\end{minipage}

\clearpage

\newgeometry{left=10mm,right=10mm,top=10mm,bottom=12mm}
\noindent\begin{minipage}{\linewidth}
\refstepcounter{figure}\label{fig:s01_reference_ts_audit}\label{fig:transition_audit_second}
\addcontentsline{sif}{figure}{\protect\numberline{\thefigure}Excluded Transition1x test reactions}
{\centering\includegraphics[width=\linewidth]{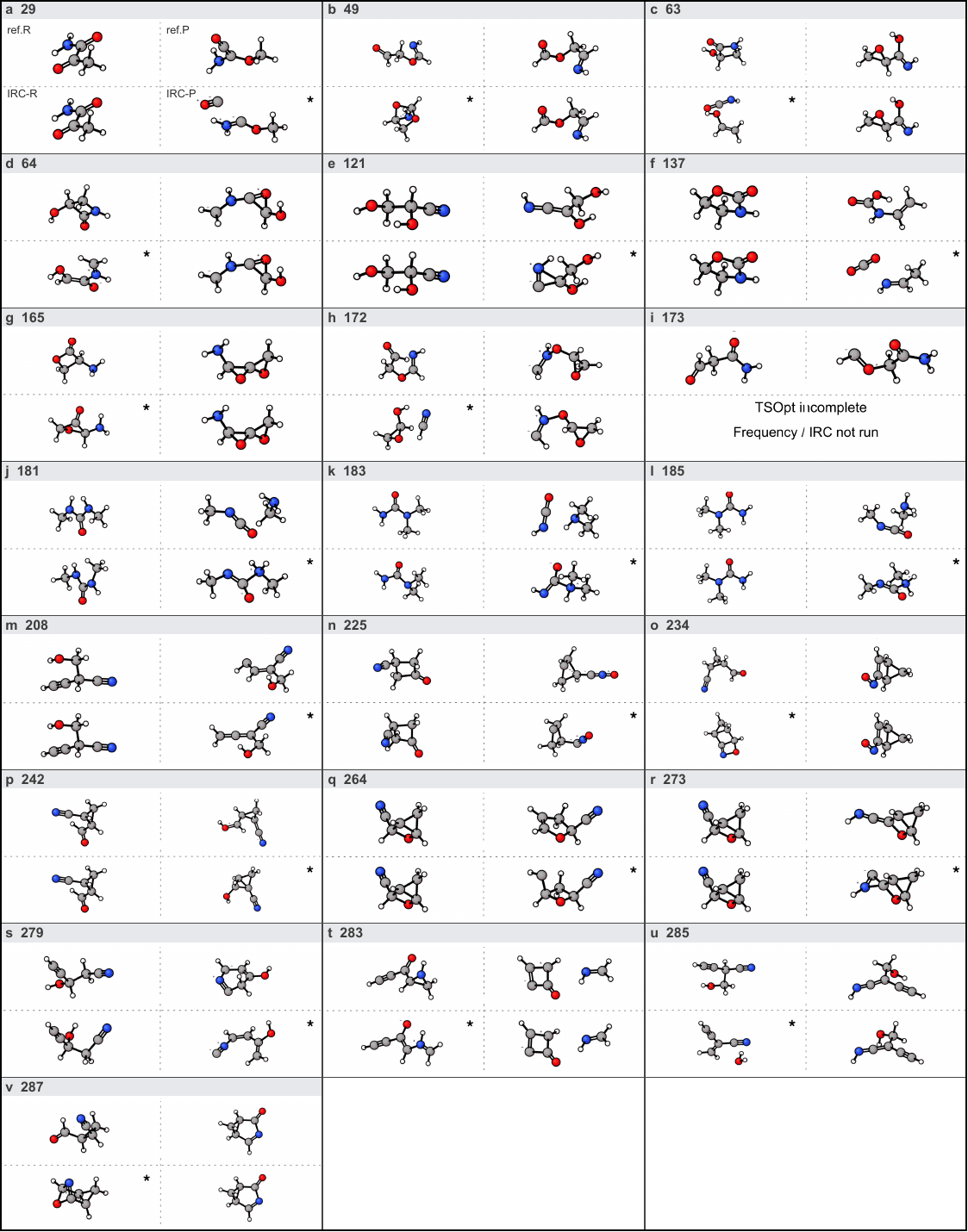}\par}
\vspace{1mm}
{\SIcaptionfont
\textbf{Supplementary Figure~\ref{fig:s01_reference_ts_audit}. Excluded Transition1x test reactions.}
The reference TSs of 287 Transition1x~\cite{Schreiner2022Transition1x} test reactions were optimized and validated by IRC calculations using Gaussian~16~\cite{Frisch2016Gaussian16} at the $\omega$B97X/6-31G(d) level. Of these, 265 reactions were retained with confirmed reactant--product connectivity. The 22 excluded reactions are shown, with reference reactant and product structures above their corresponding IRC endpoints, where available. Among the 21 reactions with completed IRC calculations, 12 matched only the reference reactant and nine matched only the reference product. TS optimization was incomplete for reaction 173. Asterisks mark unmatched IRC endpoints, and $\ddagger$ denotes structures with uncertain electronic states (four in total).\par}
\end{minipage}\clearpage\newgeometry{left=18mm,right=18mm,top=18mm,bottom=18mm}

\clearpage

\noindent\begin{minipage}{\linewidth}
\refstepcounter{figure}\label{fig:s05_reference_energy_landscape}
\addcontentsline{sif}{figure}{\protect\numberline{\thefigure}Distributions of electronic energies for Transition1x test reactions}

{\centering\includegraphics[width=170mm,height=0.76\textheight,keepaspectratio]{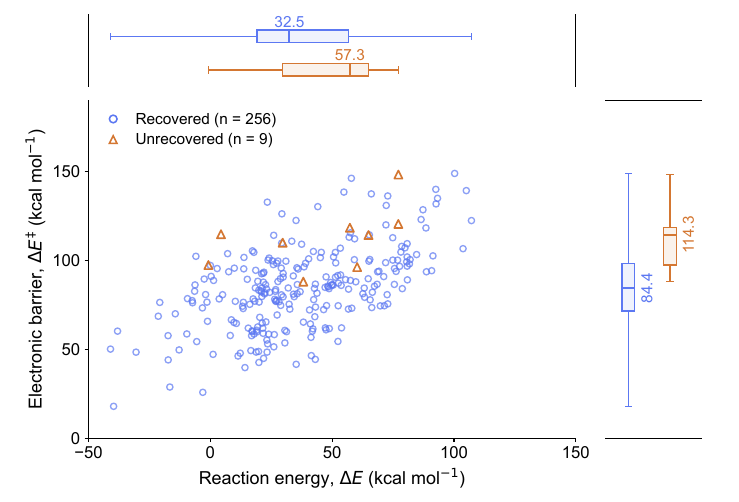}\label{si:asset_05}\par}
\vspace{2mm}
{\SIcaptionfont
\textbf{Supplementary Figure~\ref{fig:s05_reference_energy_landscape}. Distributions of electronic reaction energies and barriers for Transition1x test reactions.}
Each point represents one of the 265 test reactions retained after
reference-pathway validation. Reactions are classified according to
whether their reference product topologies were recovered among the
16 product--TS pairs generated under Ref.-3D: recovered ($n=256$)
or unrecovered ($n=9$). The energies are taken from the Transition1x
dataset, with $\Delta E=E(P)-E(R)$ and
$\Delta E^{\ddagger}=E(\mathrm{TS})-E(R)$.
Marginal box plots show medians and interquartile ranges (IQRs),
with whiskers spanning the minimum to maximum values.\par}
\end{minipage}

\newpage

\noindent\begin{minipage}{\linewidth}
\refstepcounter{figure}\label{fig:s06_unrecovered_reactions}
\addcontentsline{sif}{figure}{\protect\numberline{\thefigure}Unrecovered reference products in Transition1x test reaction.}
{\centering \includegraphics[width=160mm,height=0.76\textheight,keepaspectratio]
{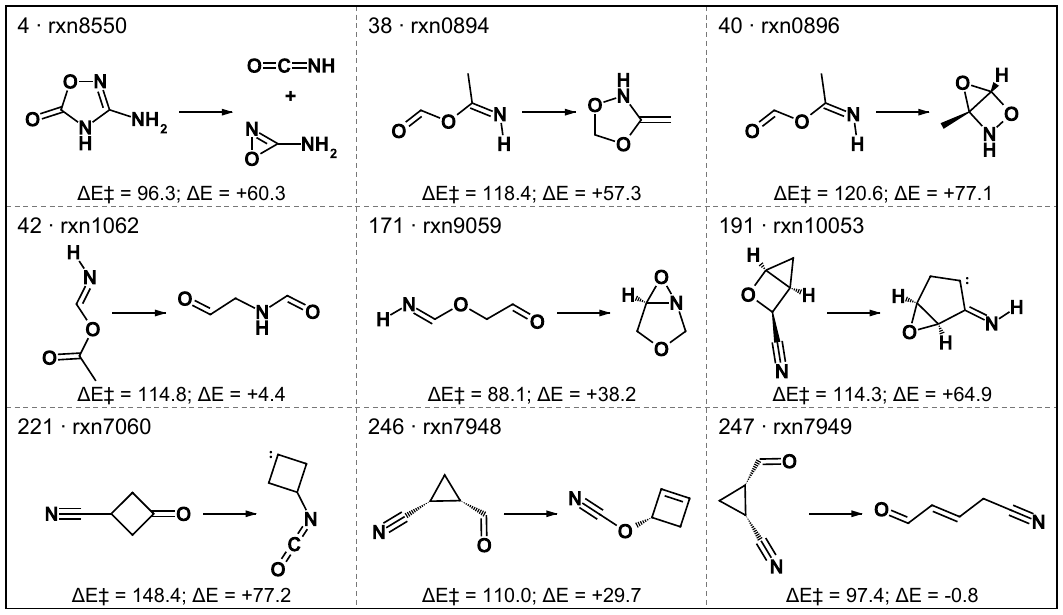}
\label{si:asset_06}\par}
\vspace{1mm}
{\SIcaptionfont
\textbf{Supplementary Figure~\ref{fig:s06_unrecovered_reactions}.
Nine Transition1x test reactions whose reference products were not recovered by RxnScope.} 
Reference reactant and product structures are shown for the nine reactions whose reference product topologies were not recovered among the 16 product--TS pairs generated under Ref.-3D. Each panel shows the test-set index, source reaction ID, electronic barrier ($\Delta E^{\ddagger}$) and electronic reaction energy ($\Delta E$), with energies in kcal~mol$^{-1}$.
The values are taken from Transition1x, where $\Delta E^{\ddagger}=E(\mathrm{TS})-E(R)$ and $\Delta E=E(P)-E(R)$.\par}
\end{minipage}
\clearpage

\noindent\begin{minipage}{\linewidth}
\refstepcounter{figure}\label{fig:s07_topology_votes}
\addcontentsline{sif}{figure}{\protect\numberline{\thefigure}Product-topology vote distribution}

{\centering\includegraphics[width=112mm]{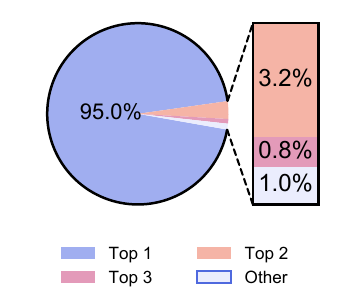}\label{si:asset_07}\par}
\vspace{1mm}
{\SIcaptionfont
\textbf{Supplementary Figure~\ref{fig:s07_topology_votes}. Product-topology vote distribution.}
Fractions of the 4,240 product-TS pairs generated across 265 Transition1x test reactions under Ref.-3D (16 samples per reaction), 
assigned to each product-topology rank. Products were grouped by topology and ranked by sampling frequency within each reaction. Sample counts at each rank were then summed across reactions and divided by 4,240. Top-1 groups accounted for 95.0\% of all generated pairs, and groups ranked within the Top 3 collectively accounted for 99.0\%. Other includes ranks above 3.\par}
\end{minipage}
\clearpage

\noindent\begin{minipage}{\linewidth}
\refstepcounter{figure}\label{fig:s03_two_stage_control}
\addcontentsline{sif}{figure}{\protect\numberline{\thefigure}TS geometric accuracy of joint and R-P-cond. generation on Transition1x}
{\centering
\includegraphics[width=\linewidth]
{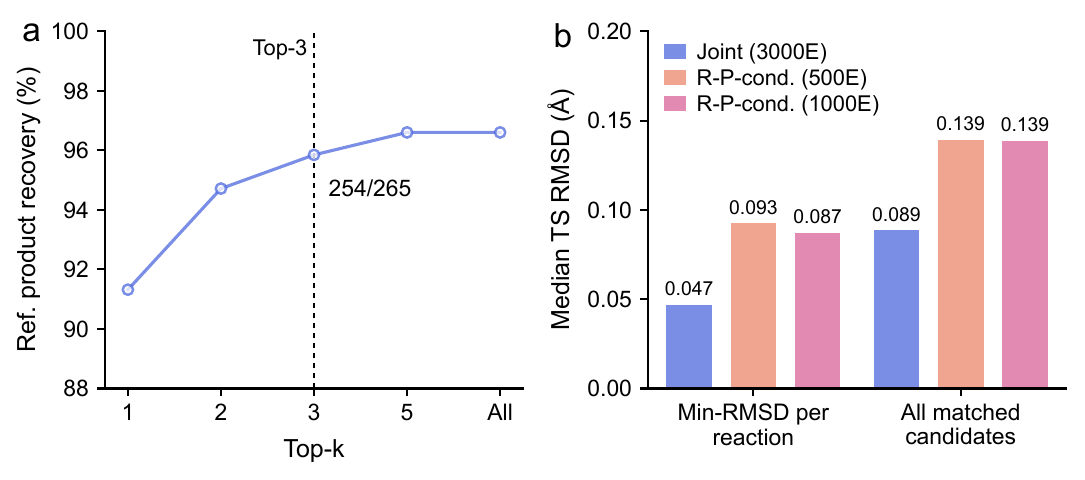}
\label{si:asset_02}\par}

\vspace{2mm}
{\SIcaptionfont
\textbf{Supplementary Figure~\ref{fig:s03_two_stage_control}.
TS geometric accuracy of joint and reactant--product-conditioned
generation on Transition1x.}
(a) Top-$k$ recovery of reference product topologies by RxnScope across 265 Transition1x test reactions under Ref.-3D (16 samples per reaction).
The reactant geometries and generated products from the reference-matching Top-3 groups were used as fixed inputs for the reactant--product-conditioned variant (R-P-cond.) in (b). (b) Median RMSDs between generated and reference TS geometries for joint generation and R-P-cond., using 3,837 candidates per model from the 254 reactions with Top-3 reference-product recovery. Min-RMSD selects the candidate closest to the reference TS for each reaction, whereas All matched candidates includes all 3,837  candidates. Joint denotes RxnScope trained for 3,000 epochs. Results are shown for R-P-cond. models trained for 500 and 1,000 epochs; the 1,000-epoch model was used for the main-text comparison.\par}
\end{minipage} 
\clearpage

\noindent\begin{minipage}{\linewidth}
\refstepcounter{figure}\label{fig:s02_training_strategy_ablation}
\addcontentsline{sif}{figure}{\protect\numberline{\thefigure}Comparison of RxnScope variants}

{\centering
\includegraphics[width=\linewidth,height=0.76\textheight,keepaspectratio]
{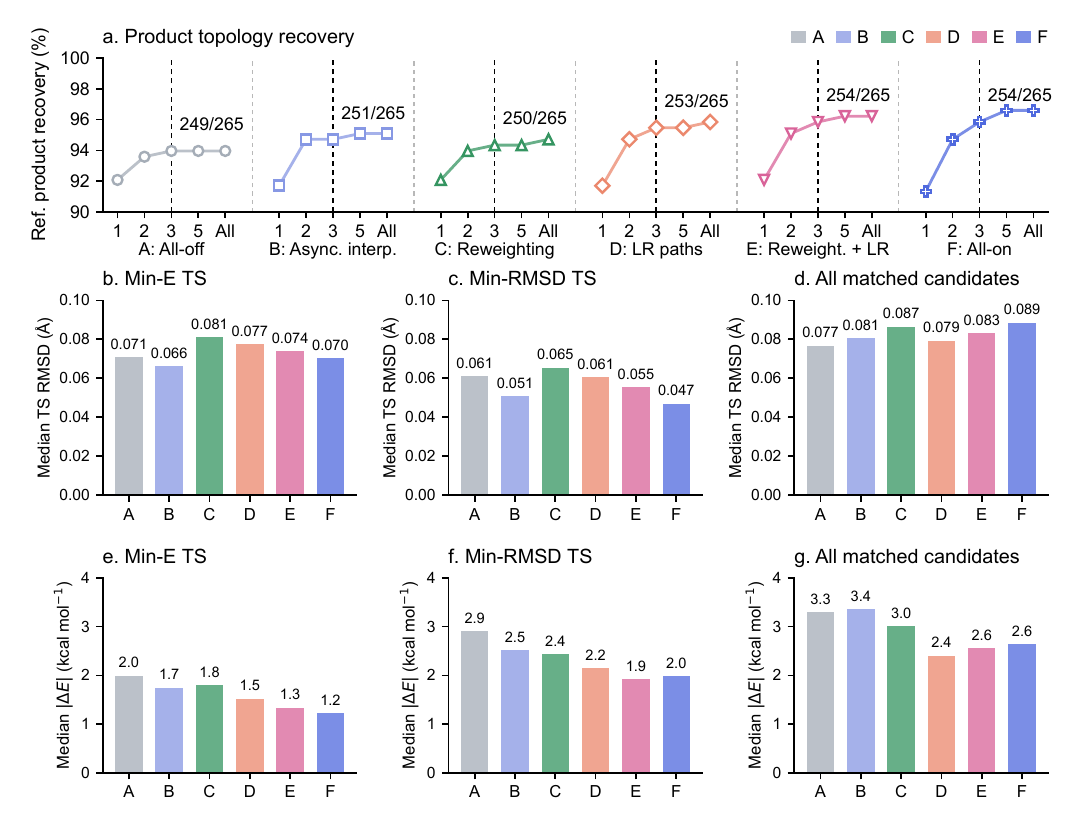}
\label{si:asset_01}\par}

\vspace{2mm}
{\SIcaptionfont
\textbf{Supplementary Figure~\ref{fig:s02_training_strategy_ablation}.
Product recovery and TS accuracy of RxnScope variants on Transition1x.}
Comparison of six model variants (A--F) with different combinations of asynchronous interpolation, temporal velocity reweighting and local-refinement (LR) training. A, all three strategies disabled; B, asynchronous interpolation only; C, temporal velocity reweighting only; D, LR training only; E, temporal velocity reweighting and LR training; F, all three  strategies enabled.
All variants used the same Transition1x training split, 3,000 training epochs and random seed (42), and were evaluated on 265 test reactions under Ref.-3D (16 samples per reaction).
(a) Top-$k$ recovery of reference product topologies. For (b--g), candidates were drawn from reference-matching Top-3 product-topology groups for each variant. (b,e) Median RMSDs (b) and absolute TS electronic-energy differences (e) for Min-E TS candidates. Within each group, the candidate with the smallest absolute electronic-energy difference from the reference TS was selected, and both metrics were calculated for that candidate. (c,f) Corresponding metrics for Min-RMSD TS candidates, selected within each group by the lowest RMSD to the reference TS. (d,g) Corresponding metrics for all matched candidates. All RMSDs and energy differences were evaluated for generated TS geometries before optimization.\par}
\end{minipage}
\clearpage

\noindent\begin{minipage}{\linewidth}
\refstepcounter{figure}\label{fig:s11_reference_ts_audit_251}
\addcontentsline{sif}{figure}{\protect\numberline{\thefigure}Excluded aqueous-HCN reference reactions (part 1)}

{\centering\includegraphics[width=160mm]{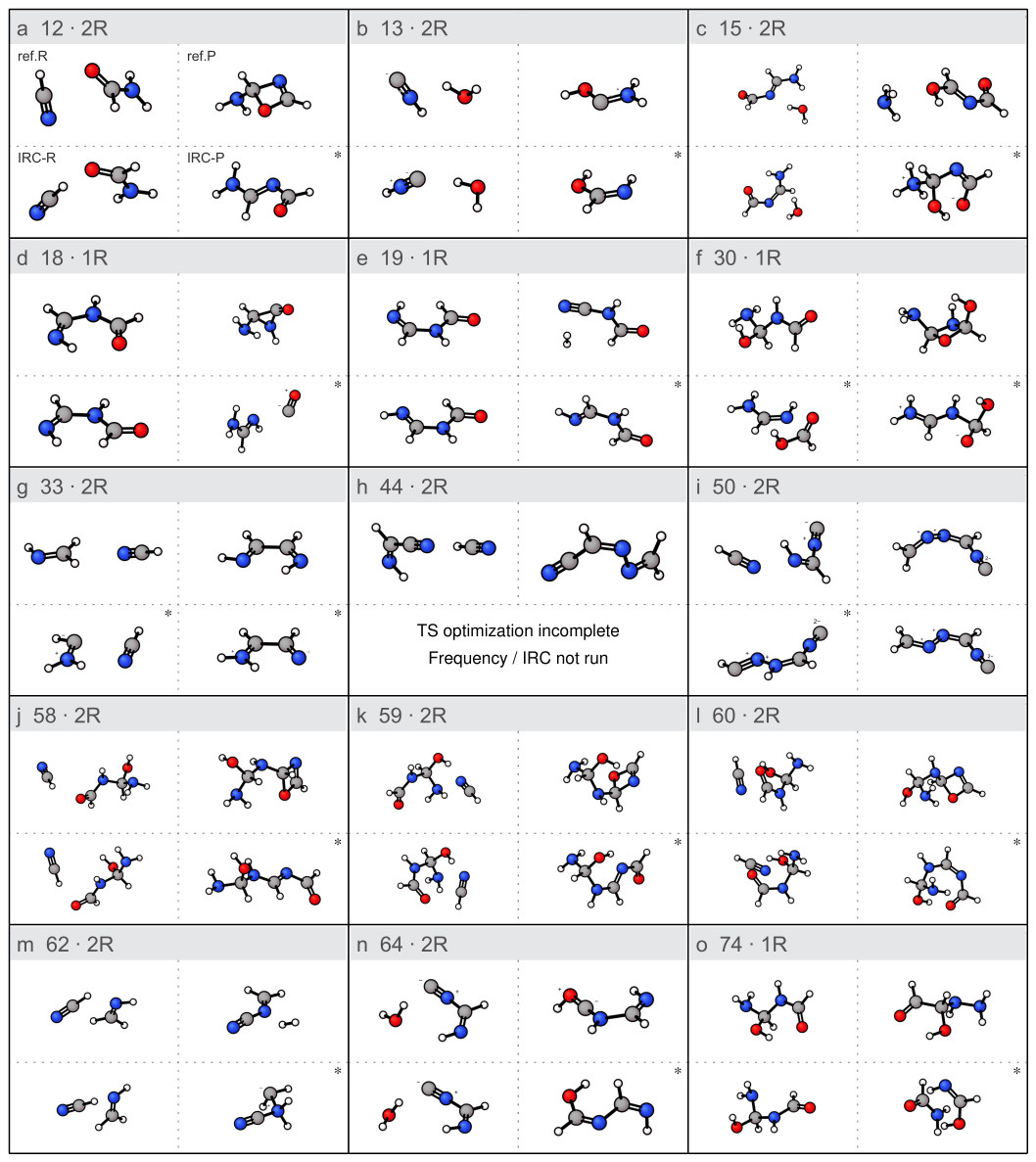}\label{si:asset_08}\par}
\vspace{2mm}
{\SIcaptionfont
\textbf{Supplementary Figure~\ref{fig:s11_reference_ts_audit_251}. Excluded aqueous-HCN reference reactions (part 1).}
Starting from 331 reactions reported by {Zhao et al. (2023)}~\cite{Zhao2023AqueousHCN}, we excluded three with inconsistent atom ordering, 19 overlapping the training datasets (16 with RGD1~\cite{Zhao2023RGD1} and three with Transition1x), and 22 without sufficient bond-change information for RR assignment. Of the remaining 287 reactions, 253 had reference pathways confirmed by TS optimization and IRC validation using Gaussian~16 at the B3LYP-D3/TZVP level with PCM~\cite{Tomasi2005PCM} ($\varepsilon=80$). The 34 excluded reactions are shown here and in Supplementary Fig.~\ref{fig:hcn_audit_second}, with reference reactant and product structures above their corresponding IRC endpoints, where available. Among the 32 reactions with completed IRC calculations, 25 matched only the reference reactant, one matched only the reference product and six matched neither. One reaction had incomplete TS optimization, and another had no imaginary frequency. Asterisks mark unmatched IRC endpoints; 1R, 2R and 3R indicate the number of reactant components.\par}
\end{minipage}

\clearpage
\noindent\begin{minipage}{\linewidth}
\refstepcounter{figure}\label{fig:hcn_audit_second}
\addcontentsline{sif}{figure}{\protect\numberline{\thefigure}Excluded aqueous-HCN reference reactions (part 2)}

{\centering\includegraphics[width=160mm,height=0.88\textheight,keepaspectratio]{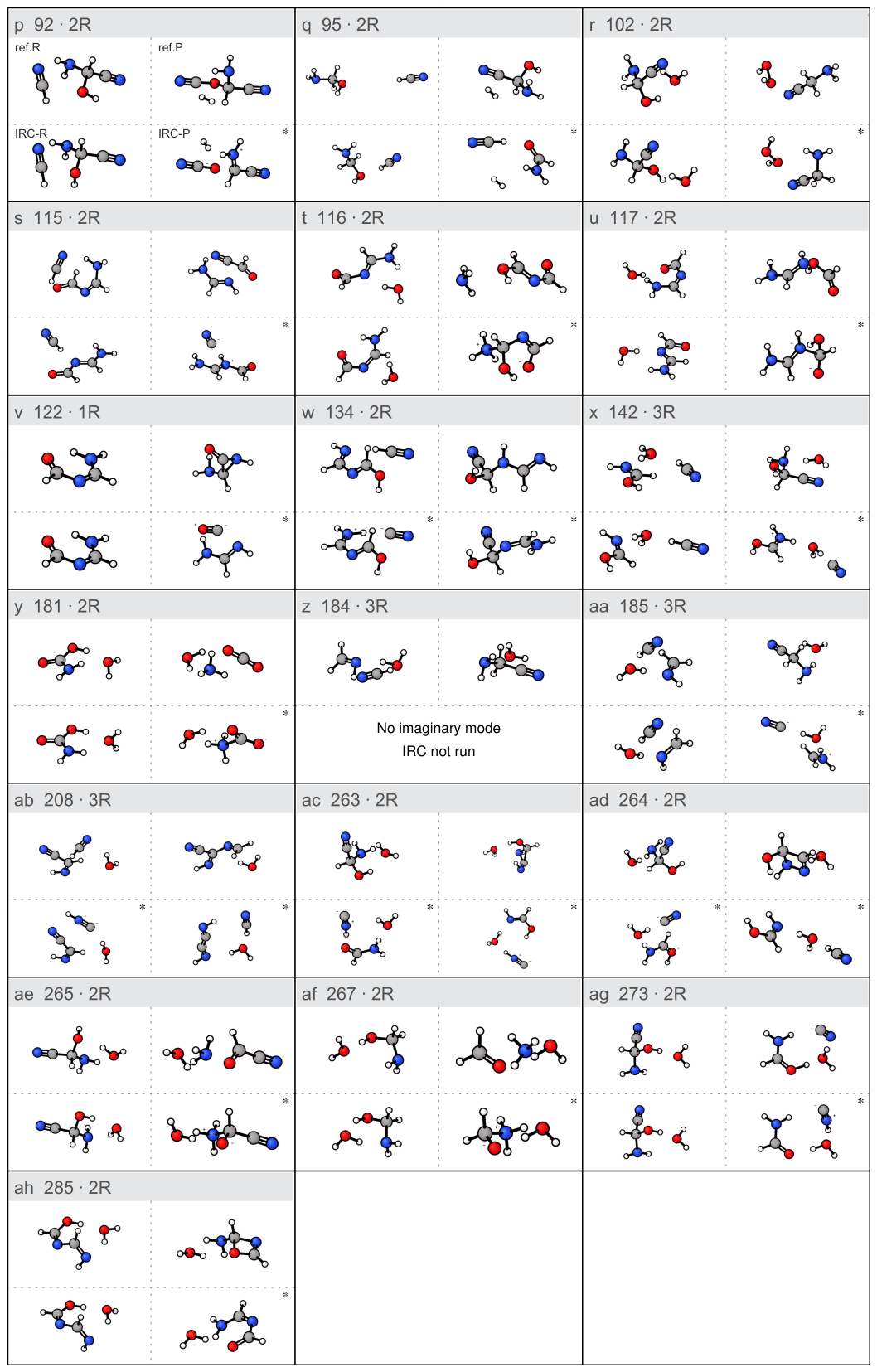}\label{si:asset_09}\par}
\vspace{2mm}
{\SIcaptionfont
\textbf{Supplementary Figure~\ref{fig:hcn_audit_second}. Excluded aqueous-HCN reference reactions (part 2).}
Reference reactant and product structures and their corresponding IRC endpoints are shown as described in Supplementary Fig.~\ref{fig:s11_reference_ts_audit_251}.\par}
\end{minipage}

\clearpage
\newgeometry{left=18mm,right=18mm,top=12mm,bottom=14mm}
\noindent\begin{minipage}{\linewidth}
\refstepcounter{figure}\label{fig:composition_datasets_251}
\addcontentsline{sif}{figure}{\protect\numberline{\thefigure}Reaction composition and bond changes across datasets}

{\centering\includegraphics[width=164mm]{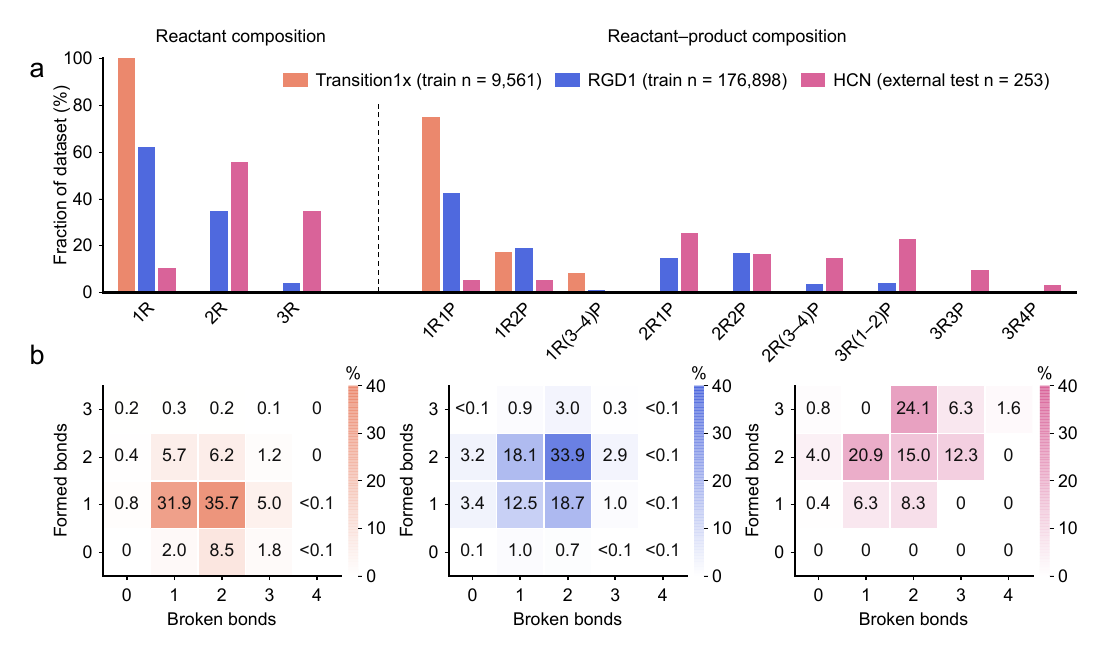}\label{si:asset_10}\par}
\vspace{2mm}
{\SIcaptionfont
\textbf{Supplementary Figure~\ref{fig:composition_datasets_251}. Reaction composition and bond changes in Transition1x, RGD1 and aqueous-HCN datasets.}
Comparison of the Transition1x~\cite{Schreiner2022Transition1x} training set ($n=9,561$), RGD1~\cite{Zhao2023RGD1} training set ($n=176,898$) and external aqueous-HCN test set ($n=253$). (a) Distributions of reactant component counts (left) and reactant--product component combinations (right). The notation $x$R$y$P indicates $x$ reactant and $y$ product components. (b) Distributions of bond formations and cleavages, determined from changes in reactant--product connectivity; bond-order-only changes were excluded. Percentages were calculated separately for each dataset.\par}
\end{minipage}

\newpage
\par\vspace{3mm}

\noindent\begin{minipage}{\linewidth}
\refstepcounter{figure}\label{fig:s13_topology_vote_concentration_251}
\addcontentsline{sif}{figure}{\protect\numberline{\thefigure}Product-topology voting in aqueous-HCN reactions}

{\centering
\includegraphics[width=164mm]
{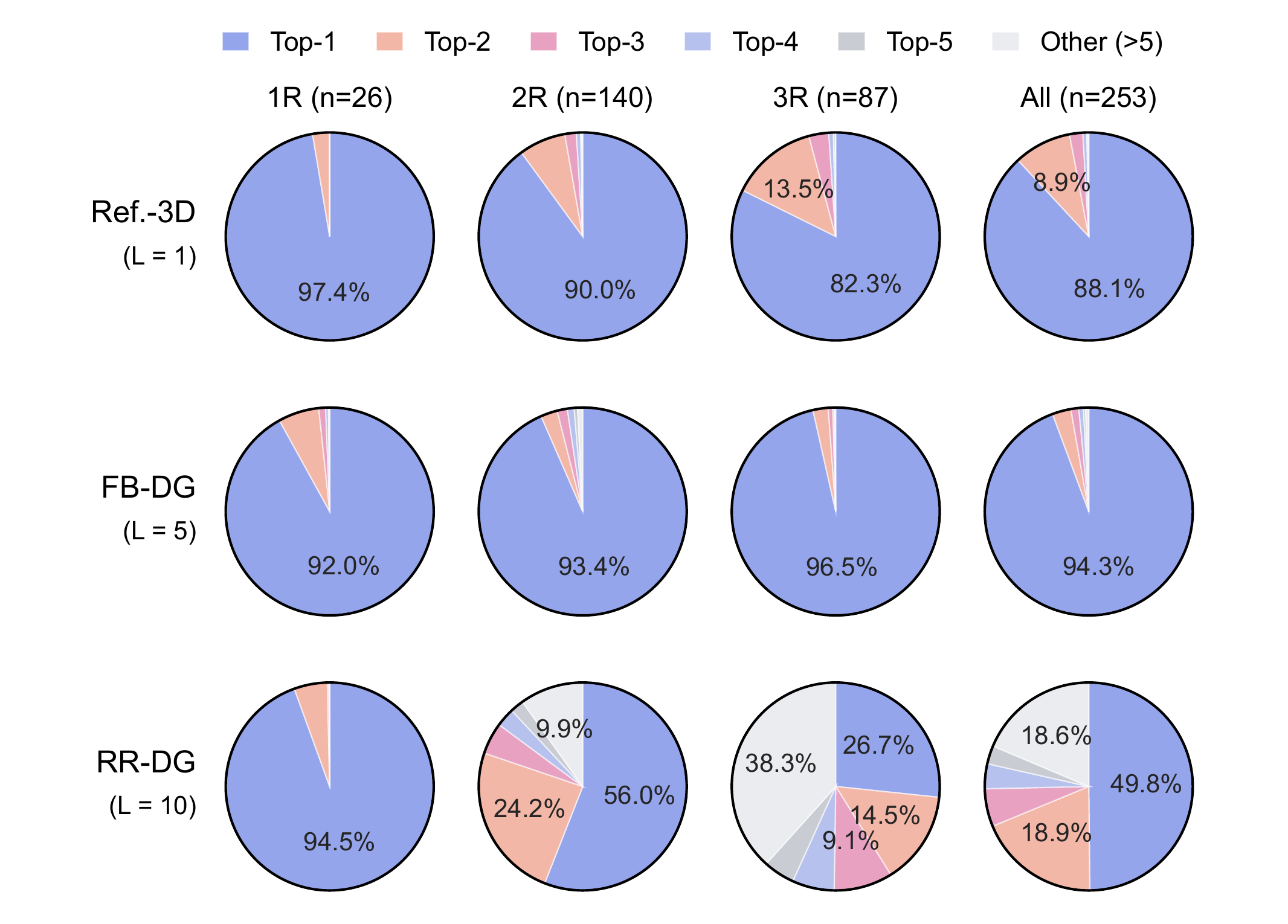}
\label{si:asset_11}\par}

\vspace{2mm}
{\SIcaptionfont
\textbf{Supplementary Figure~\ref{fig:s13_topology_vote_concentration_251}.
Product-topology vote distributions for aqueous-HCN reactions
under three input protocols.}
Fractions of generated product--TS pairs assigned to each product-topology rank under Ref.-3D, FB-DG and RR-DG. Results are shown for reactions with one (1R, $n=26$), two (2R, $n=140$) or three (3R, $n=87$) reactant components, and for all 253 aqueous-HCN reactions.
Within each reaction, generated products were grouped by topology and the groups were ranked by sampling frequency. For each protocol and reaction set, sample counts at each rank were then summed across reactions and divided by the total number of generated product--TS pairs in that set. Other includes groups ranked beyond the Top 5.
$L$ denotes the number of reactant conformers per reaction, with 16 product--TS pairs generated per conformer.\par}
\end{minipage}

\clearpage\newgeometry{left=18mm,right=18mm,top=18mm,bottom=18mm}

\clearpage
\noindent\begin{minipage}{\linewidth}
\refstepcounter{figure}\label{fig:s15_additional_product_energy_profiles}
\addcontentsline{sif}{figure}{\protect\numberline{\thefigure}Additional aqueous-HCN pathways}

{\centering
\includegraphics[width=\linewidth]
{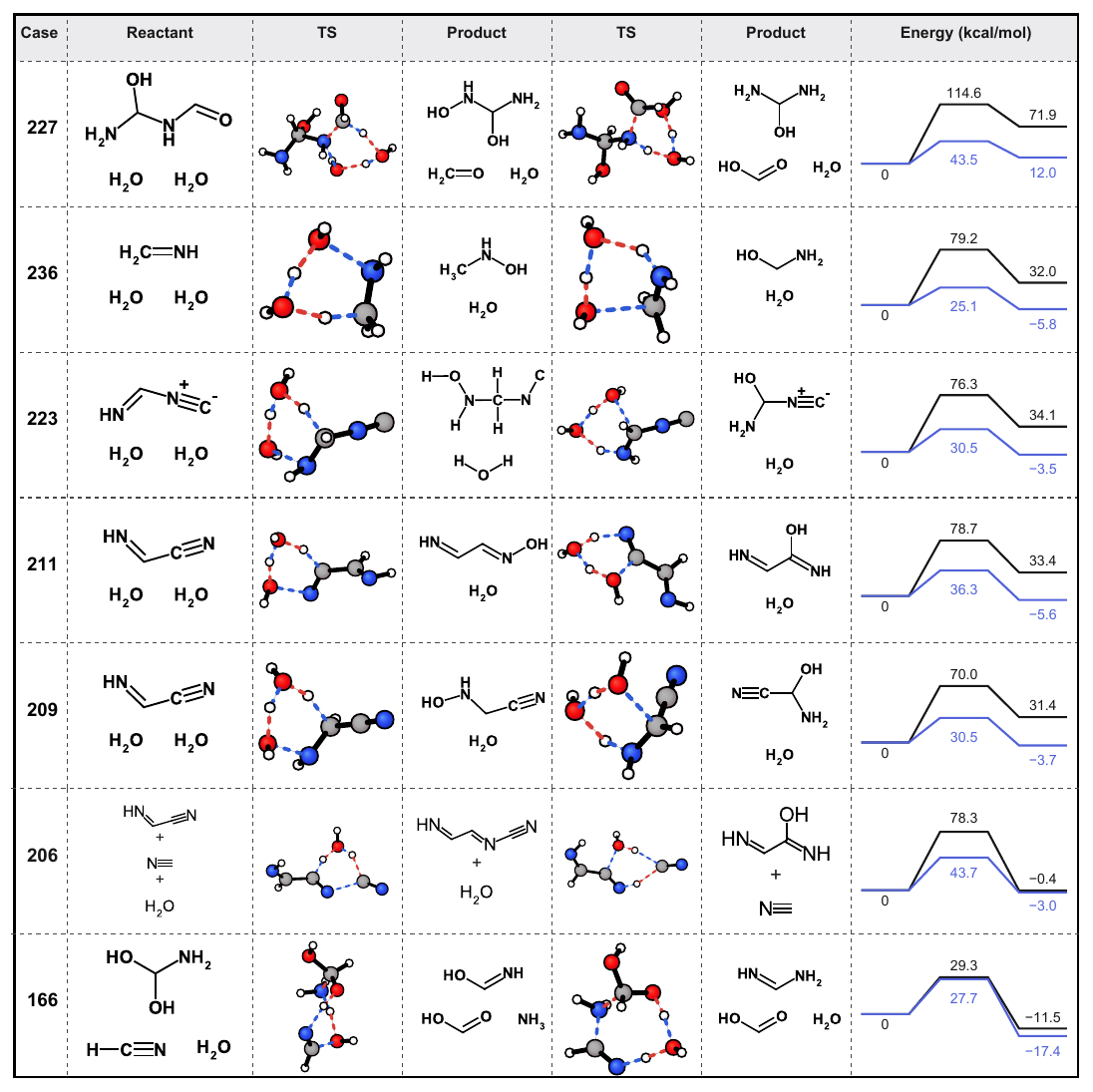}
\label{si:asset_12}\par}

\vspace{2mm}
{\SIcaptionfont\textbf{Supplementary Figure~\ref{fig:s15_additional_product_energy_profiles}.Additional aqueous-HCN pathways with lower TS electronic energies.}
For each of the seven aqueous-HCN reactions shown, the lowest-energy validated TS identified leads to an alternative product and lies more than 1~kcal~mol$^{-1}$ below the reference TS in electronic energy (main-text Fig.~3d). Alternative and reference pathways are shown in blue--purple and black, respectively. TS and product electronic energies are referenced to the optimized reactant for each reaction (kcal~mol$^{-1}$). Generated TSs were optimized and validated by frequency analysis and IRC calculations.For reaction 223, the reference product is shown by connectivity only. Red and blue dashed lines on the TS structures indicate broken and formed bonds, respectively. Calculations used B3LYP-D3/TZVP with the polarizable continuum model (PCM; $\varepsilon = 80$). 
In the rightmost column, blue curves represent the lower-energy generated pathways, whereas black curves represent reference pathways.\par}
\end{minipage} 
\clearpage

\noindent\begin{minipage}{\linewidth}
\refstepcounter{figure}\label{fig:s17_glycine_water_controls}
\addcontentsline{sif}{figure}{\protect\numberline{\thefigure}Water-assisted glycine formation}

{\centering
\includegraphics[width=\linewidth,height=0.76\textheight,keepaspectratio]
{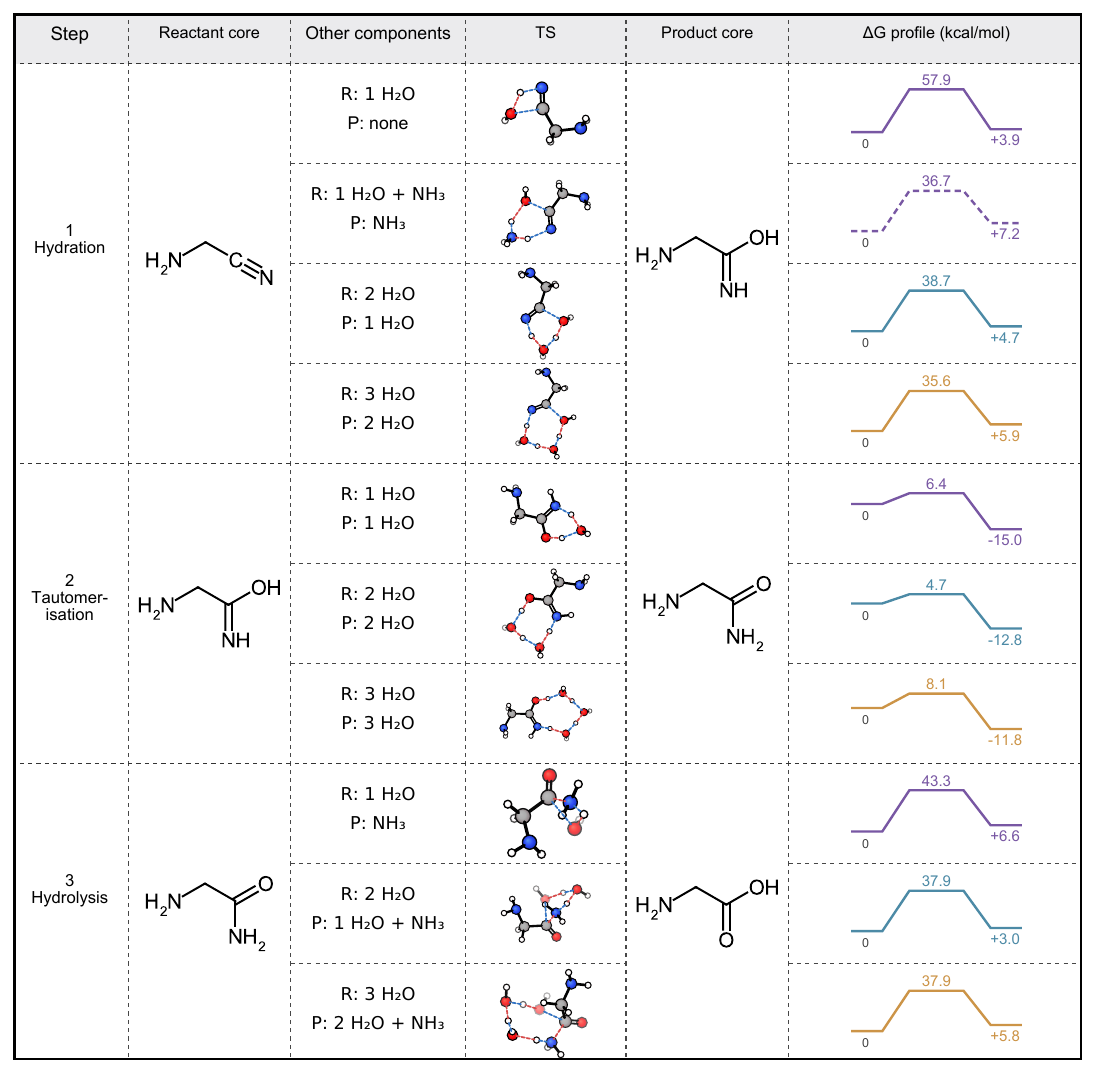}
\label{si:asset_13}\par}

\vspace{2mm}
{\SIcaptionfont
\textbf{Supplementary Figure~\ref{fig:s17_glycine_water_controls}.
Free-energy profiles for glycine formation with different numbers
of explicit water molecules.}
The profiles show hydration, tautomerization and hydrolysis
steps in the conversion of aminoacetonitrile to glycine. For each step and reactant composition, the validated pathway with the lowest TS Gibbs free energy among the calculated candidates was selected.
Labels for R and P specify the numbers of water and ammonia molecules in the reactant and product complexes, respectively. Gibbs free energies include all molecules in the corresponding complexes.
Each free-energy profile is referenced to its optimized reactant
complex, with $\Delta G^{\ddagger}=G_{\mathrm{TS}}-G_{\mathrm{R}}$
and $\Delta G_{\mathrm{rxn}}=G_{\mathrm{P}}-G_{\mathrm{R}}$
(kcal~mol$^{-1}$). Colours indicate the number of explicit water molecules, and the dashed hydration profile includes ammonia. Red and blue dashed lines on TS structures indicate broken and formed bonds, respectively.\par}
\end{minipage}

\newgeometry{margin=14mm}
\noindent\begin{minipage}{\linewidth}
\vspace{1mm}
{\centering
\includegraphics[width=.97\linewidth]
{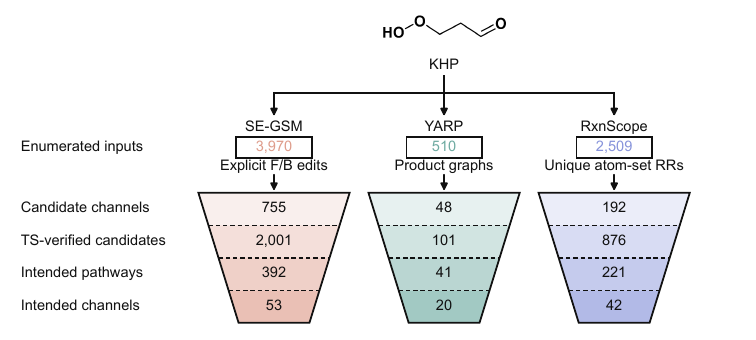}
\label{si:asset_14}\par}

\vspace{1mm}
\KHPfigurecaption{F1}
{First-round KHP candidate generation and validation}
{Comparison of candidate generation and pathway validation for SE-GSM~\cite{Zimmerman2015SEGSM}, YARP~\cite{Zhao2021YARP,Zhao2022YARP} and RxnScope starting from KHP. The three methods enumerate bond-forming and bond-breaking operations, candidate product graphs and reactive regions (RRs), respectively. Candidate channels are defined by reactant--product connectivity, with each channel potentially containing multiple TS candidates. TS-verified candidates passed TS optimization and frequency analysis with exactly one imaginary frequency. Intended pathways were confirmed by IRC calculations to connect the input reactant to the originally paired product. The final row counts intended channels only, with each distinct channel supported by at least one intended pathway. Total reaction-channel counts for each method are shown in \KHPfigref{F2}. Funnel widths are schematic.}
\end{minipage}

\par\medskip

\newpage
\noindent\begin{minipage}{\linewidth}
\vspace{1mm}
{\centering
\includegraphics[width=.62\linewidth]
{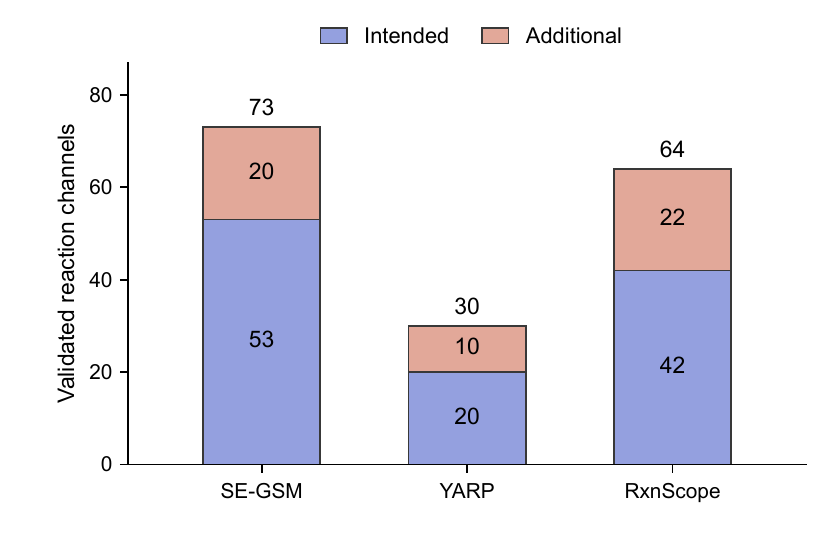}
\label{si:asset_15}\par}

\vspace{1mm}
\KHPfigurecaption{F2}
{Reaction channels identified in first-round KHP searches}
{Numbers of validated reaction channels identified by SE-GSM, YARP and RxnScope. Intended channels contain at least one validated pathway connecting the input reactant to its originally paired product. Unintended pathways connect the input reactant to a product
different from the originally paired product. For each method, Additional comprises channels identified through unintended pathways after excluding those already counted as Intended. Each channel is counted once per method, and total counts are shown above the bars.}
\end{minipage}
\clearpage 

\newgeometry{left=10mm,right=10mm,top=10mm,bottom=11.0mm}
\noindent\begin{minipage}{\linewidth}
{\centering\includegraphics[width=\linewidth]{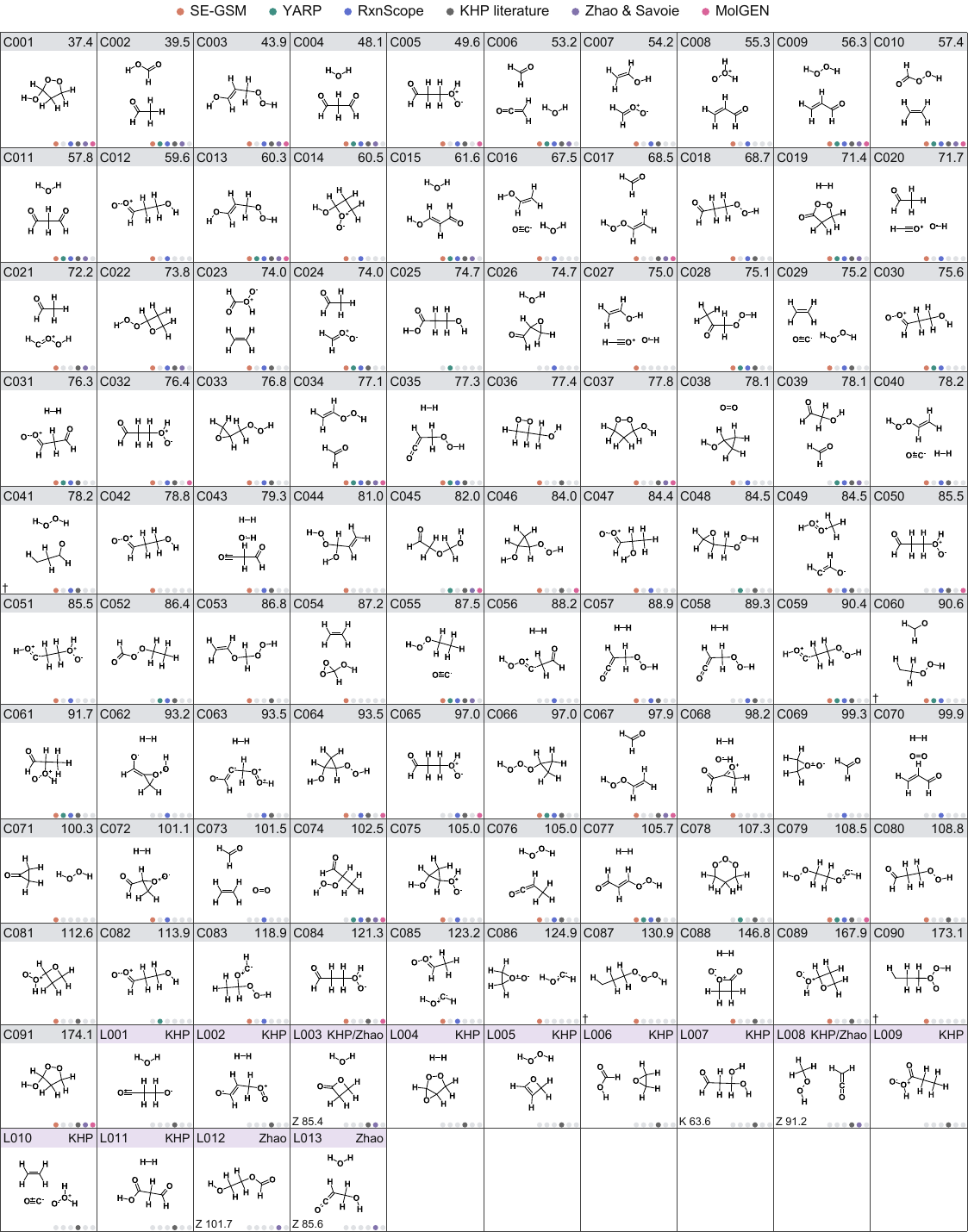}\par}
\vspace{0mm}
\KHPfigurecaption{F5}{First-round KHP reaction channels and comparison with literature}{Cards C001--C091 show distinct reaction channels validated in the first-round KHP searches with SE-GSM, YARP and RxnScope. Channels are ranked by the lowest optimized g-xTB~\cite{Froitzheim2025gxTB} TS electronic energy obtained for each channel among validated TSs from the three methods. Header values are relative to the original KHP single-point energy (kcal~mol$^{-1}$). Coloured dots indicate reaction-channel recovery by the three search methods and product-topology matches to reactions reported by Grambow \textit{et al}, Zhao--Savoie and MolGEN~\cite{Tuo2026MolGEN}, as shown in the legend. Cards L001--L013 show literature-reported product topologies not recovered in these searches. K and Z give the electronic barriers reported by Grambow \textit{et al.}~\cite{Grambow2018KHP} and Zhao and Savoie~\cite{Zhao2022YARP}, respectively; MolGEN energies are not shown. Daggers ($\dagger$) mark structures shown by connectivity only.}
\end{minipage}
\clearpage\newgeometry{left=18mm,right=18mm,top=18mm,bottom=18mm}

\clearpage
\noindent\begin{minipage}{\linewidth}
\vspace{1mm}
\begin{center}
\includegraphics[width=\linewidth,height=195mm,keepaspectratio]{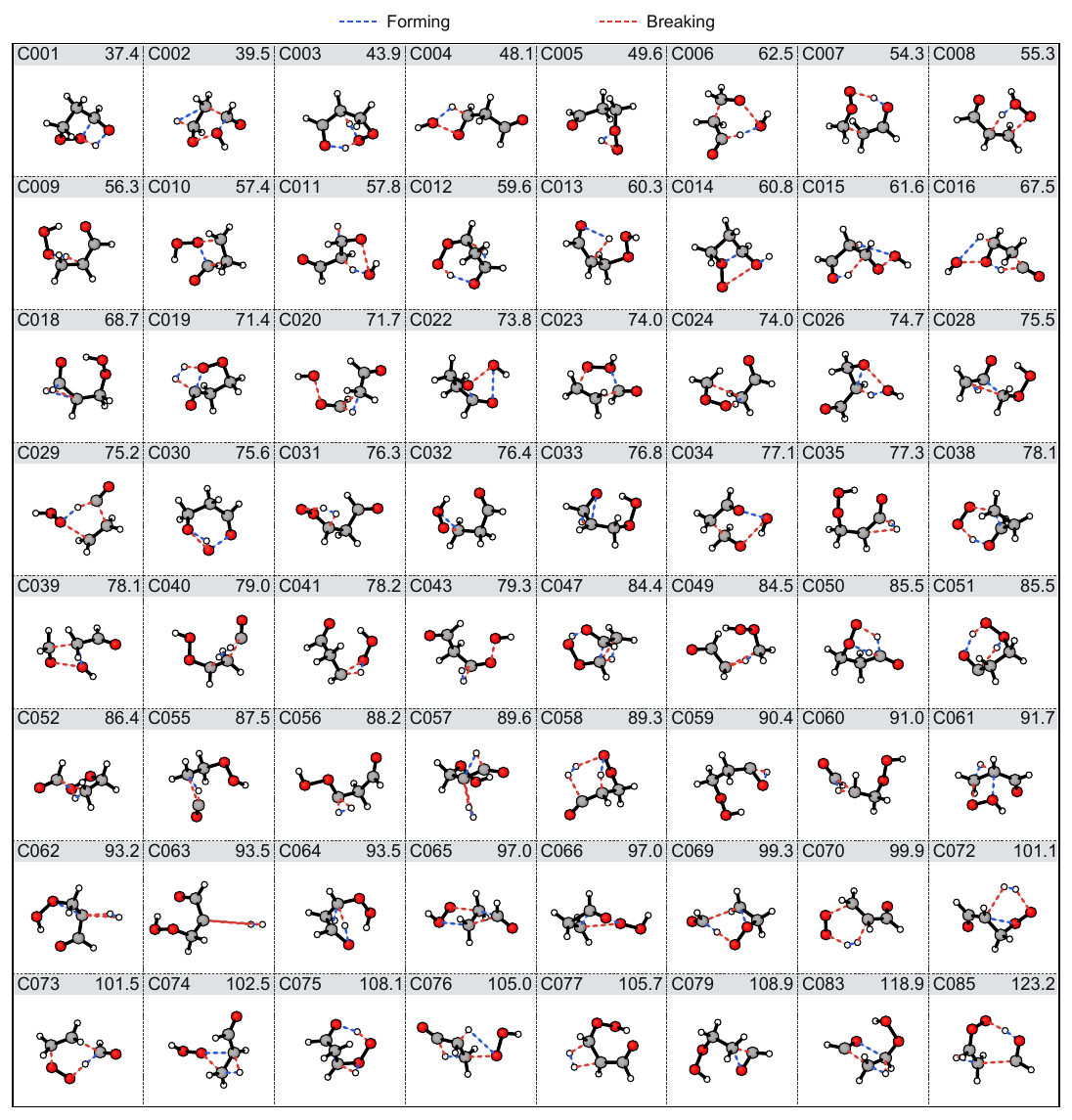}\label{si:asset_17}
\end{center}
\vspace{1mm}
\KHPfigurecaption{F7}{g-xTB-validated TS structures from RxnScope searches of KHP}{Each card shows the lowest-energy validated TS among the RxnScope candidates for one reaction channel. Channel identifiers and ordering follow \KHPfigref{F5}. Header values give the g-xTB electronic energy of the displayed TS relative to the original KHP single-point energy (kcal~mol$^{-1}$). All 12 atoms are shown, and structures are displayed at the same scale. Carbon, oxygen and hydrogen are shown in gray, red and white, respectively. Blue and red dashed lines indicate bonds formed and broken between the IRC endpoints, respectively.}
\end{minipage}

\clearpage
\noindent\begin{minipage}{\linewidth}
\vspace{2mm}
\begin{center}
\includegraphics[width=\linewidth,height=163mm,keepaspectratio]{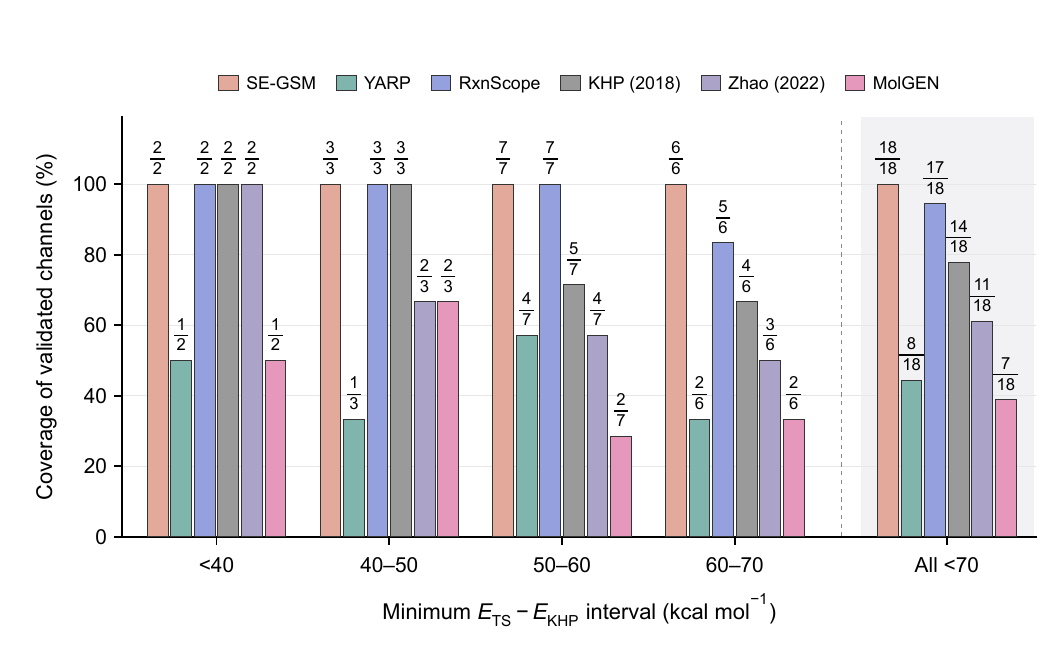}\label{si:asset_18}
\end{center}
\vspace{0mm}
\KHPfigurecaption{F6}
{Coverage of lower-energy KHP reaction channels}
{Fractions of distinct validated first-round KHP reaction channels identified by the combined SE-GSM, YARP and RxnScope searches that are matched by each search method or the literature reaction sets reported by Grambow et al., Zhao and Savoie, and MolGEN.
Channels are grouped by their lowest optimized g-xTB TS electronic energy across the three search methods, relative to the original KHP single-point energy (kcal~mol$^{-1}$).
For SE-GSM, YARP and RxnScope, a match requires a validated reactant--product channel, whereas for the literature datasets, matching is based on product-topology correspondence.
Numbers above the bars indicate the number of matched channels divided by the total number of distinct validated channels in each energy interval.
The four individual intervals contain 2, 3, 7 and 6 channels, respectively, totaling 18 channels below 70~kcal~mol$^{-1}$; ``All $<70$'' includes all 18 channels.}
\end{minipage}

\clearpage
\newgeometry{left=10mm,right=10mm,top=10mm,bottom=12mm}
\noindent\begin{minipage}{\linewidth}
{\centering\includegraphics[width=\linewidth]{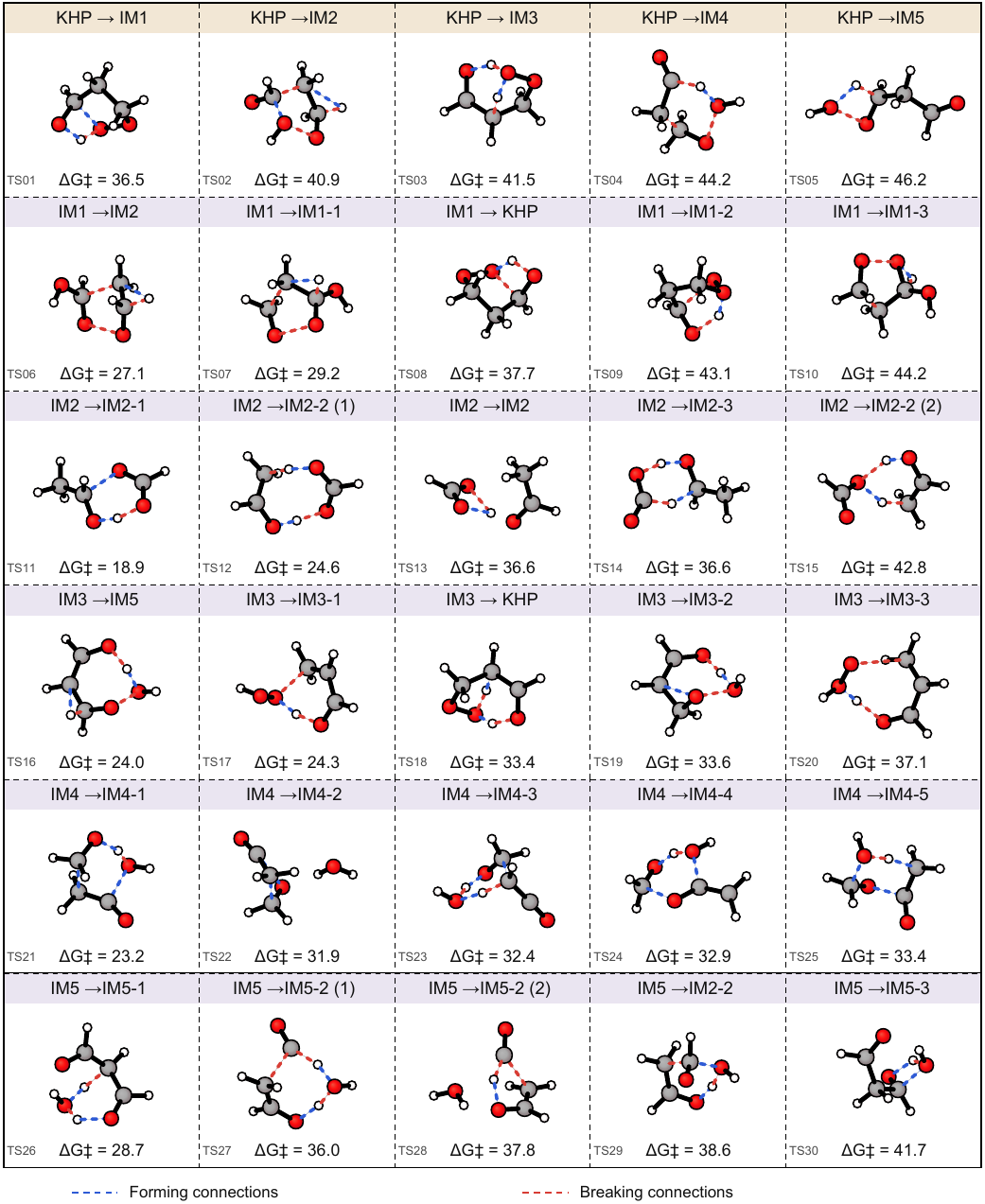}\par}
\vspace{1mm}
\KHPfigurecaption{F8}{DFT-validated TS structures in the two-round KHP network}{DFT-optimized TS structures for the 30 selected reaction pathways shown in main-text Fig.~4c. Structures are grouped by their starting reactant and ordered within each group by free-energy barrier. Orange and purple headers indicate first- and second-round pathways, respectively. TS01--TS30 correspond to the pathways listed in Supplementary Table~2. Free-energy barriers ($\Delta G^{\ddagger}$, kcal~mol$^{-1}$) were calculated at the B3LYP-D3/TZVP level, using a common optimized reactant reference for pathways from the same starting state. Blue and red dashed lines indicate bonds formed and broken between the IRC endpoints, respectively. The IM2 self-link represents a proton transfer within formic acid that changes connectivity at fixed atom indices while retaining the IM2 product label.}
\end{minipage}
\clearpage\newgeometry{left=18mm,right=18mm,top=18mm,bottom=18mm} 

\clearpage
\noindent\begin{minipage}{\linewidth}
\begin{center}
\includegraphics[width=\linewidth,height=220mm,keepaspectratio]{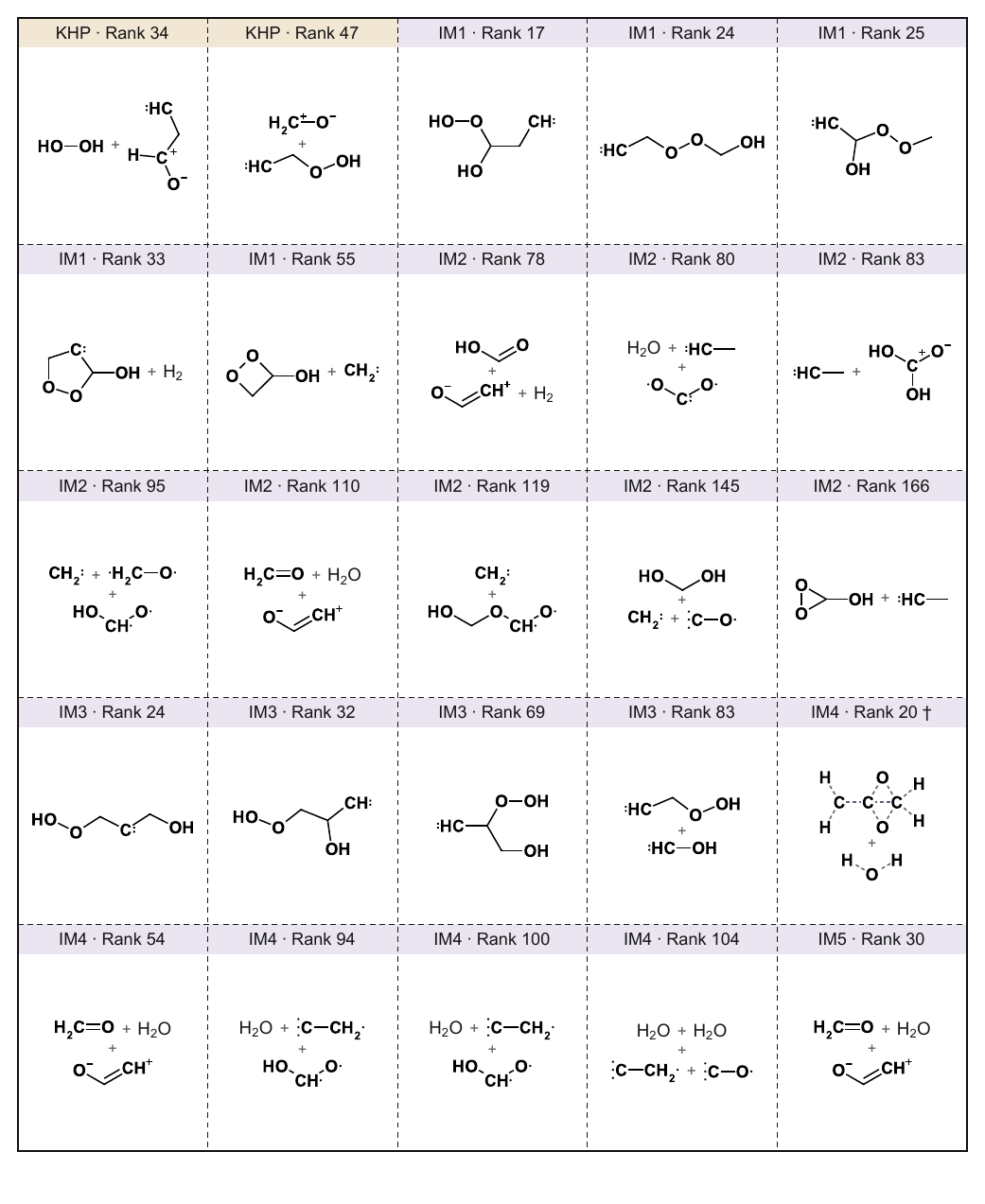}
\end{center}
\vspace{0mm}
\KHPfigurecaption{F9}{Structures of selected KHP reaction products}{The 25 product species shown here were identified during the first- and second-round KHP searches and are marked with a dagger ($\dagger$) in Supplementary Tables~3--8. Their candidate bond-order representations were assigned from atom-connectivity graphs using valence-based rules and may include formal charge separation and radical centres. Candidate SMILES representations were obtained for 24 species. The product from IM4 (Rank 20) is shown only by its connectivity graph because no bond-order-resolved SMILES could be assigned. Other products are represented by SMILES in Supplementary Tables~3--8 and are not shown here.}

\end{minipage}
\clearpage

\section*{Supplementary tables}

\addcontentsline{toc}{section}{Supplementary tables 1--8}
\KHPtableheading{T1}{KHP reference reaction channels from Grambow \textit{et al.}}
{\SIcaptionfont Reference reactions from Grambow \textit{et al.}~\cite{Grambow2018KHP} were retained when the optimized products matched the reported products and passed frequency checks. Records corresponding to equivalent reactant--product reaction channels were grouped according to the matching criteria described in Methods. R and G denote the original reaction records and fixed-index connectivity graphs, respectively. TS electronic energies were calculated using g-xTB single-point calculations on the reference DFT geometries and are reported relative to KHP (kcal~mol$^{-1}$). Training-set overlap indicates whether the corresponding reaction was included in the RxnScope-500MIX training set.\par}
\vspace{0mm}
\begingroup
\fontsize{8.5}{10.03}\selectfont
\renewcommand{\arraystretch}{1.03}
\setlength{\tabcolsep}{3.5pt}
\begin{center}
\begin{tabular*}{\linewidth}{@{\extracolsep{\fill}}lllll@{}}
\toprule
Literature records & Fixed-index graphs & Representative & $E_{\mathrm{TS,ref}}-E_{\mathrm{KHP}}$ & Training match \\
\midrule
R001; R003; R004; R005 & G001; G003 & R001 & 49.191 & No \\
R002 & G002 & R002 & 60.636 & No \\
R006 & G004 & R006 & 93.633 & No \\
R007 & G005 & R007 & 84.601 & No \\
R008 & G006 & R008 & 108.848 & No \\
R009 & G007 & R009 & 117.336 & No \\
R011 & G008 & R011 & 104.260 & No \\
R015 & G009 & R015 & 78.500 & No \\
R016 & G010 & R016 & 75.292 & No \\
R017 & G011 & R017 & 54.636 & No \\
R018 & G012 & R018 & 77.521 & No \\
R019 & G013 & R019 & 51.277 & No \\
R021 & G014 & R021 & 93.017 & No \\
R022 & G015 & R022 & 108.189 & No \\
R023 & G016 & R023 & 94.251 & No \\
R024 & G017 & R024 & 101.189 & No \\
R025 & G018 & R025 & 103.224 & No \\
R026 & G019 & R026 & 87.509 & No \\
R027 & G020 & R027 & 90.596 & No \\
R028 & G021 & R028 & 55.312 & No \\
R030 & G022 & R030 & 77.299 & No \\
R033 & G023 & R033 & 84.511 & No \\
R035; R036 & G024 & R036 & 73.086 & No \\
R037; R038 & G025 & R037 & 39.399 & No \\
R039 & G026 & R039 & 75.045 & No \\
R040 & G027 & R040 & 94.566 & No \\
R041 & G028 & R041 & 112.850 & No \\
R042 & G029 & R042 & 101.682 & No \\
R043 & G030 & R043 & 76.435 & No \\
R044 & G031 & R044 & 75.564 & No \\
R045 & G032 & R045 & 101.770 & No \\
R046 & G033 & R046 & 99.125 & No \\
R047 & G034 & R047 & 74.432 & No \\
R048 & G035 & R048 & 69.874 & No \\
R049 & G036 & R049 & 104.844 & No \\
R050 & G037 & R050 & 81.768 & No \\
R051 & G038 & R051 & 89.073 & No \\
R052 & G039 & R052 & 83.330 & No \\
R053 & G040 & R053 & 45.242 & No \\
R054 & G041 & R054 & 62.782 & No \\
R055 & G042 & R055 & 101.945 & No \\
R056 & G043 & R056 & 75.654 & No \\
R057 & G044 & R057 & 70.842 & No \\
R058 & G045 & R058 & 105.172 & No \\
R060 & G046 & R060 & 92.514 & No \\
R061 & G047 & R061 & 92.466 & No \\
R063 & G048 & R063 & 63.793 & No \\
R064 & G049 & R064 & 81.502 & No \\
R065 & G050 & R065 & 79.428 & No \\
R066 & G051 & R066 & 85.837 & No \\
R067; R068 & G052 & R068 & 81.849 & No \\
R069 & G053 & R069 & 41.033 & No \\
R070 & G054 & R070 & 97.217 & No \\
R071; R072 & G055 & R071 & 60.388 & Yes \\
R073 & G056 & R073 & 70.512 & No \\
R074 & G057 & R074 & 61.010 & No \\
R075 & G058 & R075 & 109.467 & No \\
\bottomrule
\end{tabular*}
\end{center}
\endgroup

\clearpage

\vspace{3mm}
\noindent\begin{minipage}{\linewidth}
{\centering\KHPtableheading{T8}{DFT energetics of validated pathways in the KHP reaction network}\par}
\vspace{1mm}
\begingroup
\fontsize{8.5}{10}\selectfont
\renewcommand{\arraystretch}{1.0}
\setlength{\tabcolsep}{2pt}
\centering
\begin{tabularx}{\linewidth}{M{12mm} Y Y Y Y Y}
\toprule
TS ID & Reactant & Product & DFT $\Delta G^{\ddagger}$ & DFT $\Delta E^{\ddagger}$ & \multicolumn{1}{c}{$\Delta G_{\rm rxn}$} \\
\midrule
TS01 & KHP & IM1 & 36.472 & 36.059 & -0.957 \\
TS02 & KHP & IM2 & 40.931 & 40.634 & -79.756 \\
TS03 & KHP & IM3 & 41.540 & 42.663 & 8.102 \\
TS04 & KHP & IM4 & 44.184 & 49.153 & -48.899 \\
TS05 & KHP & IM5 & 46.198 & 50.860 & -62.768 \\
TS06 & IM1 & IM2 & 27.142 & 32.175 & -78.560 \\
TS07 & IM1 & IM1-1 & 29.246 & 35.103 & -71.694 \\
TS08 & IM1 & KHP & 37.674 & 41.138 & 2.107 \\
TS09 & IM1 & IM1-2 & 43.057 & 44.615 & 40.226 \\
TS10 & IM1 & IM1-3 & 44.224 & 49.721 & -44.681 \\
TS11 & IM2 & IM2-1 & 18.907 & 18.164 & 5.684 \\
TS12 & IM2 & IM2-2 & 24.556 & 25.480 & 13.069 \\
TS13 & IM2 & IM2 & 36.576 & 36.589 & 0.902 \\
TS14 & IM2 & IM2-3 & 36.603 & 38.249 & -13.036 \\
TS15 & IM2 & IM2-2 & 42.844 & 44.905 & 16.354 \\
TS16 & IM3 & IM5 & 24.021 & 29.003 & -69.141 \\
TS17 & IM3 & IM3-1 & 24.272 & 28.423 & -2.376 \\
TS18 & IM3 & KHP & 33.439 & 36.685 & -7.197 \\
TS19 & IM3 & IM3-2 & 33.561 & 37.767 & -42.883 \\
TS20 & IM3 & IM3-3 & 37.149 & 40.644 & 34.236 \\
TS21 & IM4 & IM4-1 & 23.154 & 15.914 & -28.439 \\
TS22 & IM4 & IM4-2 & 31.904 & 26.178 & -14.841 \\
TS23 & IM4 & IM4-3 & 32.446 & 28.788 & -2.647 \\
TS24 & IM4 & IM4-4 & 32.911 & 26.113 & 2.325 \\
TS25 & IM4 & IM4-5 & 33.356 & 26.123 & -25.425 \\
TS26 & IM5 & IM5-1 & 28.730 & 26.998 & -3.372 \\
TS27 & IM5 & IM5-2 & 35.963 & 37.291 & 2.739 \\
TS28 & IM5 & IM5-2 & 37.835 & 39.416 & 2.735 \\
TS29 & IM5 & IM2-2 & 38.607 & 35.536 & -3.722 \\
TS30 & IM5 & IM5-3 & 41.698 & 38.989 & 1.179 \\
\bottomrule
\end{tabularx}\par
\endgroup
\vspace{1.5mm}
{\raggedright\fontsize{8.5}{10}\selectfont
DFT calculations were performed using Gaussian~16 at the gas-phase B3LYP-D3/TZVP level with an UltraFine integration grid. For pathways from the same starting state, the optimized reactant conformer with the lowest Gibbs free energy was used as a reference. The free-energy barrier, electronic-energy barrier and reaction free energy are defined as $\Delta G^{\ddagger}=G_{\mathrm{TS}}-G_{\mathrm{R}}$, $\Delta E^{\ddagger}=E_{\mathrm{TS}}-E_{\mathrm{R}}$ and $\Delta G_{\mathrm{rxn}}=G_{\mathrm{P}}-G_{\mathrm{R}}$, respectively. All energies are reported in kcal~mol$^{-1}$. TS IDs correspond to the structures shown in Supplementary Fig.~19.\par}
\end{minipage}

\clearpage
{\centering\KHPtableheading{T9}{Reaction channels identified from KHP}\par}
{\SIcaptionfont 
The 64 reaction channels validated at the g-xTB level in the first-round KHP search are ranked by their lowest TS electronic energies. DFT results are reported where available.\par}
\begingroup
\fontsize{10}{12}\selectfont
\renewcommand{\arraystretch}{1.08}
\setlength{\tabcolsep}{2pt}
\setlength{\LTpre}{2mm}\setlength{\LTpost}{2mm}
\def\LTcaptype{}\stepcounter{LT@tables}
\begin{longtable}{@{} >{\raggedright\arraybackslash}m{9mm} >{\raggedright\arraybackslash}m{12mm} >{\raggedright\arraybackslash}m{58mm} M{18mm} M{18mm} M{18mm} M{31mm} @{}}
\toprule
Rank & TS ID & Product (SMILES/ID)  & g-xTB $\Delta E^{\ddagger}$ & DFT $\Delta E^{\ddagger}$ & DFT $\Delta G^{\ddagger}$ & DFT outcome \\
\midrule\endfirsthead
\multicolumn{7}{l}{\bfseries Supplementary Table~\thetable\ (continued)} \\
\toprule
Rank & TS ID & Product (SMILES/ID) & g-xTB $\Delta E^{\ddagger}$ & DFT $\Delta E^{\ddagger}$ & DFT $\Delta G^{\ddagger}$ & DFT outcome \\
\midrule\endhead
\midrule\multicolumn{7}{r}{\footnotesize Continued on next page} \\\endfoot
\bottomrule\endlastfoot
01$^{*}$ & TS01 & \texttt{O\allowbreak{}C\allowbreak{}1\allowbreak{}C\allowbreak{}C\allowbreak{}O\allowbreak{}O\allowbreak{}1} & 37.43 & 36.06 & 36.47 & Validated \\
02$^{*}$ & TS02 & \texttt{C\allowbreak{}C\allowbreak{}=\allowbreak{}O\allowbreak{}.\allowbreak{}O\allowbreak{}=\allowbreak{}C\allowbreak{}O} & 39.49 & 40.63 & 40.93 & Validated \\
03$^{*}$ & TS03 & \texttt{O\allowbreak{}C\allowbreak{}=\allowbreak{}C\allowbreak{}C\allowbreak{}O\allowbreak{}O} & 43.93 & 42.66 & 41.54 & Validated \\
04$^{*}$ & TS05 & \texttt{O\allowbreak{}.\allowbreak{}O\allowbreak{}=\allowbreak{}C\allowbreak{}C\allowbreak{}C\allowbreak{}=\allowbreak{}O} & 48.14 & 50.86 & 46.20 & Validated \\
05 & \textemdash{} & \texttt{O\allowbreak{}=\allowbreak{}C\allowbreak{}C\allowbreak{}C\allowbreak{}[\allowbreak{}O\allowbreak{}H\allowbreak{}+\allowbreak{}]\allowbreak{}[\allowbreak{}O\allowbreak{}-\allowbreak{}]} & 49.64 & 50.88 & 48.58 & Validated \\
06 & \textemdash{} & \texttt{C\allowbreak{}=\allowbreak{}C\allowbreak{}O\allowbreak{}.\allowbreak{}C\allowbreak{}=\allowbreak{}[\allowbreak{}O\allowbreak{}+\allowbreak{}]\allowbreak{}[\allowbreak{}O\allowbreak{}-\allowbreak{}]} & 54.31 & 48.05 & 47.35 & Validated \\
07 & \textemdash{} & \texttt{C\allowbreak{}=\allowbreak{}C\allowbreak{}C\allowbreak{}=\allowbreak{}O\allowbreak{}.\allowbreak{}[\allowbreak{}O\allowbreak{}-\allowbreak{}]\allowbreak{}[\allowbreak{}O\allowbreak{}H\allowbreak{}2\allowbreak{}+\allowbreak{}]} & 55.30 & \textemdash{} & \textemdash{} & Reactant mismatch \\
08 & \textemdash{} & \texttt{C\allowbreak{}=\allowbreak{}C\allowbreak{}C\allowbreak{}=\allowbreak{}O\allowbreak{}.\allowbreak{}O\allowbreak{}O} & 56.27 & 51.30 & 49.89 & Validated \\
09 & \textemdash{} & \texttt{C\allowbreak{}=\allowbreak{}C\allowbreak{}.\allowbreak{}O\allowbreak{}=\allowbreak{}C\allowbreak{}O\allowbreak{}O} & 57.38 & 56.97 & 56.58 & Validated \\
10 & \textemdash{} & \texttt{O\allowbreak{}.\allowbreak{}O\allowbreak{}=\allowbreak{}C\allowbreak{}C\allowbreak{}C\allowbreak{}=\allowbreak{}O} & 57.81 & 57.26 & 53.41 & Validated \\
11 & \textemdash{} & \texttt{[\allowbreak{}O\allowbreak{}-\allowbreak{}]\allowbreak{}[\allowbreak{}O\allowbreak{}+\allowbreak{}]\allowbreak{}=\allowbreak{}C\allowbreak{}C\allowbreak{}C\allowbreak{}O} & 59.62 & 63.37 & 62.38 & Validated \\
12 & \textemdash{} & \texttt{O\allowbreak{}C\allowbreak{}=\allowbreak{}C\allowbreak{}C\allowbreak{}O\allowbreak{}O} & 60.27 & \textemdash{} & \textemdash{} & Reactant mismatch \\
13 & \textemdash{} & \texttt{[\allowbreak{}O\allowbreak{}-\allowbreak{}]\allowbreak{}[\allowbreak{}O\allowbreak{}+\allowbreak{}]\allowbreak{}1\allowbreak{}C\allowbreak{}C\allowbreak{}C\allowbreak{}1\allowbreak{}O} & 60.77 & \textemdash{} & \textemdash{} & Endpoint check failed \\
14 & \textemdash{} & \texttt{O\allowbreak{}.\allowbreak{}O\allowbreak{}=\allowbreak{}C\allowbreak{}C\allowbreak{}=\allowbreak{}C\allowbreak{}O} & 61.56 & 58.84 & 55.06 & Validated \\
15$^{*}$ & TS04 & \texttt{C\allowbreak{}=\allowbreak{}C\allowbreak{}=\allowbreak{}O\allowbreak{}.\allowbreak{}C\allowbreak{}=\allowbreak{}O\allowbreak{}.\allowbreak{}O} & 62.54 & 49.15 & 44.18 & Validated \\
16 & \textemdash{} & \texttt{C\allowbreak{}=\allowbreak{}C\allowbreak{}O\allowbreak{}.\allowbreak{}O\allowbreak{}.\allowbreak{}[\allowbreak{}C\allowbreak{}-\allowbreak{}]\allowbreak{}\#\allowbreak{}[\allowbreak{}O\allowbreak{}+\allowbreak{}]} & 67.54 & \textemdash{} & \textemdash{} & Endpoint check failed \\
17 & \textemdash{} & \texttt{O\allowbreak{}=\allowbreak{}C\allowbreak{}C\allowbreak{}C\allowbreak{}O\allowbreak{}O} & 68.73 & 69.91 & 67.86 & Validated \\
18 & \textemdash{} & \texttt{O\allowbreak{}=\allowbreak{}C\allowbreak{}1\allowbreak{}C\allowbreak{}C\allowbreak{}O\allowbreak{}O\allowbreak{}1\allowbreak{}.\allowbreak{}[\allowbreak{}H\allowbreak{}]\allowbreak{}[\allowbreak{}H\allowbreak{}]} & 71.42 & \textemdash{} & \textemdash{} & Not calculated \\
19 & \textemdash{} & \texttt{C\allowbreak{}\#\allowbreak{}[\allowbreak{}O\allowbreak{}+\allowbreak{}]\allowbreak{}.\allowbreak{}C\allowbreak{}C\allowbreak{}=\allowbreak{}O\allowbreak{}.\allowbreak{}[\allowbreak{}O\allowbreak{}H\allowbreak{}-\allowbreak{}]} & 71.72 & \textemdash{} & \textemdash{} & Not calculated \\
20 & \textemdash{} & \texttt{O\allowbreak{}O\allowbreak{}C\allowbreak{}1\allowbreak{}C\allowbreak{}C\allowbreak{}O\allowbreak{}1} & 73.81 & \textemdash{} & \textemdash{} & Not calculated \\
21 & \textemdash{} & \texttt{C\allowbreak{}=\allowbreak{}C\allowbreak{}.\allowbreak{}O\allowbreak{}=\allowbreak{}C\allowbreak{}[\allowbreak{}O\allowbreak{}H\allowbreak{}+\allowbreak{}]\allowbreak{}[\allowbreak{}O\allowbreak{}-\allowbreak{}]} & 73.97 & \textemdash{} & \textemdash{} & Not calculated \\
22 & \textemdash{} & \texttt{C\allowbreak{}=\allowbreak{}[\allowbreak{}O\allowbreak{}+\allowbreak{}]\allowbreak{}[\allowbreak{}O\allowbreak{}-\allowbreak{}]\allowbreak{}.\allowbreak{}C\allowbreak{}C\allowbreak{}=\allowbreak{}O} & 73.99 & \textemdash{} & \textemdash{} & Not calculated \\
23 & \textemdash{} & \texttt{O\allowbreak{}.\allowbreak{}O\allowbreak{}=\allowbreak{}C\allowbreak{}C\allowbreak{}1\allowbreak{}C\allowbreak{}O\allowbreak{}1} & 74.71 & \textemdash{} & \textemdash{} & Not calculated \\
24 & \textemdash{} & \texttt{C\allowbreak{}=\allowbreak{}C\allowbreak{}.\allowbreak{}O\allowbreak{}O\allowbreak{}.\allowbreak{}[\allowbreak{}C\allowbreak{}-\allowbreak{}]\allowbreak{}\#\allowbreak{}[\allowbreak{}O\allowbreak{}+\allowbreak{}]} & 75.23 & \textemdash{} & \textemdash{} & Not calculated \\
25 & \textemdash{} & \texttt{C\allowbreak{}C\allowbreak{}(\allowbreak{}=\allowbreak{}O\allowbreak{})\allowbreak{}C\allowbreak{}O\allowbreak{}O} & 75.48 & \textemdash{} & \textemdash{} & Not calculated \\
26 & \textemdash{} & \texttt{[\allowbreak{}O\allowbreak{}-\allowbreak{}]\allowbreak{}[\allowbreak{}O\allowbreak{}+\allowbreak{}]\allowbreak{}=\allowbreak{}C\allowbreak{}C\allowbreak{}C\allowbreak{}O} & 75.61 & \textemdash{} & \textemdash{} & Not calculated \\
27 & \textemdash{} & \texttt{O\allowbreak{}=\allowbreak{}C\allowbreak{}C\allowbreak{}C\allowbreak{}=\allowbreak{}[\allowbreak{}O\allowbreak{}+\allowbreak{}]\allowbreak{}[\allowbreak{}O\allowbreak{}-\allowbreak{}]\allowbreak{}.\allowbreak{}[\allowbreak{}H\allowbreak{}]\allowbreak{}[\allowbreak{}H\allowbreak{}]} & 76.27 & \textemdash{} & \textemdash{} & Not calculated \\
28 & \textemdash{} & \texttt{O\allowbreak{}=\allowbreak{}C\allowbreak{}C\allowbreak{}C\allowbreak{}[\allowbreak{}O\allowbreak{}H\allowbreak{}+\allowbreak{}]\allowbreak{}[\allowbreak{}O\allowbreak{}-\allowbreak{}]} & 76.40 & \textemdash{} & \textemdash{} & Not calculated \\
29 & \textemdash{} & \texttt{O\allowbreak{}O\allowbreak{}C\allowbreak{}C\allowbreak{}1\allowbreak{}C\allowbreak{}O\allowbreak{}1} & 76.77 & \textemdash{} & \textemdash{} & Not calculated \\
30 & \textemdash{} & \texttt{C\allowbreak{}=\allowbreak{}C\allowbreak{}O\allowbreak{}O\allowbreak{}.\allowbreak{}C\allowbreak{}=\allowbreak{}O} & 77.13 & \textemdash{} & \textemdash{} & Not calculated \\
31 & \textemdash{} & \texttt{O\allowbreak{}=\allowbreak{}C\allowbreak{}=\allowbreak{}C\allowbreak{}C\allowbreak{}O\allowbreak{}O\allowbreak{}.\allowbreak{}[\allowbreak{}H\allowbreak{}]\allowbreak{}[\allowbreak{}H\allowbreak{}]} & 77.35 & \textemdash{} & \textemdash{} & Not calculated \\
32 & \textemdash{} & \texttt{O\allowbreak{}=\allowbreak{}O\allowbreak{}.\allowbreak{}O\allowbreak{}C\allowbreak{}1\allowbreak{}C\allowbreak{}C\allowbreak{}1} & 78.06 & \textemdash{} & \textemdash{} & Not calculated \\
\noalign{\penalty-10000}
33 & \textemdash{} & \texttt{C\allowbreak{}=\allowbreak{}O\allowbreak{}.\allowbreak{}O\allowbreak{}=\allowbreak{}C\allowbreak{}C\allowbreak{}O} & 78.13 & \textemdash{} & \textemdash{} & Not calculated \\
34$^{\dagger}$ & \textemdash{} & \texttt{O\allowbreak{}O\allowbreak{}.\allowbreak{}[\allowbreak{}C\allowbreak{}H\allowbreak{}]\allowbreak{}C\allowbreak{}[\allowbreak{}C\allowbreak{}H\allowbreak{}+\allowbreak{}]\allowbreak{}[\allowbreak{}O\allowbreak{}-\allowbreak{}]} & 78.20 & \textemdash{} & \textemdash{} & Not calculated \\
35 & \textemdash{} & \texttt{C\allowbreak{}=\allowbreak{}C\allowbreak{}O\allowbreak{}O\allowbreak{}.\allowbreak{}[\allowbreak{}C\allowbreak{}-\allowbreak{}]\allowbreak{}\#\allowbreak{}[\allowbreak{}O\allowbreak{}+\allowbreak{}]\allowbreak{}.\allowbreak{}[\allowbreak{}H\allowbreak{}]\allowbreak{}[\allowbreak{}H\allowbreak{}]} & 79.02 & \textemdash{} & \textemdash{} & Not calculated \\
36 & \textemdash{} & \texttt{O\allowbreak{}=\allowbreak{}C\allowbreak{}C\allowbreak{}C\allowbreak{}\#\allowbreak{}[\allowbreak{}O\allowbreak{}+\allowbreak{}]\allowbreak{}.\allowbreak{}[\allowbreak{}H\allowbreak{}]\allowbreak{}[\allowbreak{}H\allowbreak{}]\allowbreak{}.\allowbreak{}[\allowbreak{}O\allowbreak{}H\allowbreak{}-\allowbreak{}]} & 79.29 & \textemdash{} & \textemdash{} & Not calculated \\
37 & \textemdash{} & \texttt{C\allowbreak{}C\allowbreak{}(\allowbreak{}O\allowbreak{})\allowbreak{}C\allowbreak{}=\allowbreak{}[\allowbreak{}O\allowbreak{}+\allowbreak{}]\allowbreak{}[\allowbreak{}O\allowbreak{}-\allowbreak{}]} & 84.43 & \textemdash{} & \textemdash{} & Not calculated \\
38 & \textemdash{} & \texttt{C\allowbreak{}[\allowbreak{}O\allowbreak{}+\allowbreak{}]\allowbreak{}=\allowbreak{}[\allowbreak{}O\allowbreak{}H\allowbreak{}+\allowbreak{}]\allowbreak{}.\allowbreak{}[\allowbreak{}C\allowbreak{}H\allowbreak{}-\allowbreak{}]\allowbreak{}=\allowbreak{}C\allowbreak{}[\allowbreak{}O\allowbreak{}-\allowbreak{}]} & 84.55 & \textemdash{} & \textemdash{} & Not calculated \\
39 & \textemdash{} & \texttt{O\allowbreak{}=\allowbreak{}C\allowbreak{}C\allowbreak{}C\allowbreak{}[\allowbreak{}O\allowbreak{}H\allowbreak{}+\allowbreak{}]\allowbreak{}[\allowbreak{}O\allowbreak{}-\allowbreak{}]} & 85.50 & \textemdash{} & \textemdash{} & Not calculated \\
40 & \textemdash{} & \texttt{[\allowbreak{}O\allowbreak{}-\allowbreak{}]\allowbreak{}[\allowbreak{}O\allowbreak{}H\allowbreak{}+\allowbreak{}]\allowbreak{}C\allowbreak{}C\allowbreak{}[\allowbreak{}C\allowbreak{}-\allowbreak{}]\allowbreak{}=\allowbreak{}[\allowbreak{}O\allowbreak{}H\allowbreak{}+\allowbreak{}]} & 85.50 & \textemdash{} & \textemdash{} & Not calculated \\
41 & \textemdash{} & \texttt{C\allowbreak{}C\allowbreak{}O\allowbreak{}O\allowbreak{}C\allowbreak{}=\allowbreak{}O} & 86.38 & \textemdash{} & \textemdash{} & Not calculated \\
42 & \textemdash{} & \texttt{C\allowbreak{}C\allowbreak{}O\allowbreak{}O\allowbreak{}.\allowbreak{}[\allowbreak{}C\allowbreak{}-\allowbreak{}]\allowbreak{}\#\allowbreak{}[\allowbreak{}O\allowbreak{}+\allowbreak{}]} & 87.48 & \textemdash{} & \textemdash{} & Not calculated \\
43 & \textemdash{} & \texttt{O\allowbreak{}=\allowbreak{}C\allowbreak{}C\allowbreak{}[\allowbreak{}C\allowbreak{}-\allowbreak{}]\allowbreak{}=\allowbreak{}[\allowbreak{}O\allowbreak{}+\allowbreak{}]\allowbreak{}O\allowbreak{}.\allowbreak{}[\allowbreak{}H\allowbreak{}]\allowbreak{}[\allowbreak{}H\allowbreak{}]} & 88.20 & \textemdash{} & \textemdash{} & Not calculated \\
44 & \textemdash{} & \texttt{O\allowbreak{}=\allowbreak{}C\allowbreak{}=\allowbreak{}C\allowbreak{}C\allowbreak{}O\allowbreak{}O\allowbreak{}.\allowbreak{}[\allowbreak{}H\allowbreak{}]\allowbreak{}[\allowbreak{}H\allowbreak{}]} & 89.28 & \textemdash{} & \textemdash{} & Not calculated \\
45 & \textemdash{} & \texttt{O\allowbreak{}=\allowbreak{}C\allowbreak{}=\allowbreak{}C\allowbreak{}C\allowbreak{}O\allowbreak{}O\allowbreak{}.\allowbreak{}[\allowbreak{}H\allowbreak{}]\allowbreak{}[\allowbreak{}H\allowbreak{}]} & 89.55 & \textemdash{} & \textemdash{} & Not calculated \\
46 & \textemdash{} & \texttt{O\allowbreak{}O\allowbreak{}C\allowbreak{}C\allowbreak{}[\allowbreak{}C\allowbreak{}-\allowbreak{}]\allowbreak{}=\allowbreak{}[\allowbreak{}O\allowbreak{}H\allowbreak{}+\allowbreak{}]} & 90.36 & \textemdash{} & \textemdash{} & Not calculated \\
47$^{\dagger}$ & \textemdash{} & \texttt{[\allowbreak{}C\allowbreak{}H\allowbreak{}2\allowbreak{}+\allowbreak{}]\allowbreak{}[\allowbreak{}O\allowbreak{}-\allowbreak{}]\allowbreak{}.\allowbreak{}[\allowbreak{}C\allowbreak{}H\allowbreak{}]\allowbreak{}C\allowbreak{}O\allowbreak{}O} & 91.01 & \textemdash{} & \textemdash{} & Not calculated \\
48 & \textemdash{} & \texttt{C\allowbreak{}C\allowbreak{}(\allowbreak{}C\allowbreak{}=\allowbreak{}O\allowbreak{})\allowbreak{}[\allowbreak{}O\allowbreak{}H\allowbreak{}+\allowbreak{}]\allowbreak{}[\allowbreak{}O\allowbreak{}-\allowbreak{}]} & 91.68 & \textemdash{} & \textemdash{} & Not calculated \\
49 & \textemdash{} & \texttt{[\allowbreak{}H\allowbreak{}]\allowbreak{}[\allowbreak{}H\allowbreak{}]\allowbreak{}.\allowbreak{}[\allowbreak{}O\allowbreak{}-\allowbreak{}]\allowbreak{}C\allowbreak{}=\allowbreak{}C\allowbreak{}1\allowbreak{}C\allowbreak{}[\allowbreak{}O\allowbreak{}+\allowbreak{}]\allowbreak{}1\allowbreak{}O} & 93.18 & \textemdash{} & \textemdash{} & Not calculated \\
50 & \textemdash{} & \texttt{[\allowbreak{}H\allowbreak{}]\allowbreak{}[\allowbreak{}H\allowbreak{}]\allowbreak{}.\allowbreak{}[\allowbreak{}O\allowbreak{}-\allowbreak{}]\allowbreak{}C\allowbreak{}=\allowbreak{}[\allowbreak{}C\allowbreak{}-\allowbreak{}]\allowbreak{}C\allowbreak{}[\allowbreak{}O\allowbreak{}+\allowbreak{}]\allowbreak{}=\allowbreak{}[\allowbreak{}O\allowbreak{}H\allowbreak{}+\allowbreak{}]} & 93.50 & \textemdash{} & \textemdash{} & Not calculated \\
51 & \textemdash{} & \texttt{O\allowbreak{}O\allowbreak{}C\allowbreak{}1\allowbreak{}C\allowbreak{}C\allowbreak{}1\allowbreak{}O} & 93.51 & \textemdash{} & \textemdash{} & Not calculated \\
52 & \textemdash{} & \texttt{O\allowbreak{}=\allowbreak{}C\allowbreak{}C\allowbreak{}C\allowbreak{}[\allowbreak{}O\allowbreak{}H\allowbreak{}+\allowbreak{}]\allowbreak{}[\allowbreak{}O\allowbreak{}-\allowbreak{}]} & 97.01 & \textemdash{} & \textemdash{} & Not calculated \\
53 & \textemdash{} & \texttt{O\allowbreak{}O\allowbreak{}O\allowbreak{}C\allowbreak{}1\allowbreak{}C\allowbreak{}C\allowbreak{}1} & 97.04 & \textemdash{} & \textemdash{} & Not calculated \\
54 & \textemdash{} & \texttt{C\allowbreak{}=\allowbreak{}O\allowbreak{}.\allowbreak{}[\allowbreak{}O\allowbreak{}-\allowbreak{}]\allowbreak{}[\allowbreak{}O\allowbreak{}+\allowbreak{}]\allowbreak{}1\allowbreak{}C\allowbreak{}C\allowbreak{}1} & 99.28 & \textemdash{} & \textemdash{} & Not calculated \\
55 & \textemdash{} & \texttt{C\allowbreak{}=\allowbreak{}C\allowbreak{}C\allowbreak{}=\allowbreak{}O\allowbreak{}.\allowbreak{}O\allowbreak{}=\allowbreak{}O\allowbreak{}.\allowbreak{}[\allowbreak{}H\allowbreak{}]\allowbreak{}[\allowbreak{}H\allowbreak{}]} & 99.87 & \textemdash{} & \textemdash{} & Not calculated \\
56 & \textemdash{} & \texttt{O\allowbreak{}=\allowbreak{}C\allowbreak{}C\allowbreak{}1\allowbreak{}C\allowbreak{}[\allowbreak{}O\allowbreak{}+\allowbreak{}]\allowbreak{}1\allowbreak{}[\allowbreak{}O\allowbreak{}-\allowbreak{}]\allowbreak{}.\allowbreak{}[\allowbreak{}H\allowbreak{}]\allowbreak{}[\allowbreak{}H\allowbreak{}]} & 101.12 & \textemdash{} & \textemdash{} & Not calculated \\
57 & \textemdash{} & \texttt{C\allowbreak{}=\allowbreak{}C\allowbreak{}.\allowbreak{}C\allowbreak{}=\allowbreak{}O\allowbreak{}.\allowbreak{}O\allowbreak{}=\allowbreak{}O} & 101.54 & \textemdash{} & \textemdash{} & Not calculated \\
58 & \textemdash{} & \texttt{C\allowbreak{}C\allowbreak{}(\allowbreak{}C\allowbreak{}=\allowbreak{}O\allowbreak{})\allowbreak{}O\allowbreak{}O} & 102.46 & \textemdash{} & \textemdash{} & Not calculated \\
59 & \textemdash{} & \texttt{C\allowbreak{}C\allowbreak{}=\allowbreak{}C\allowbreak{}=\allowbreak{}O\allowbreak{}.\allowbreak{}O\allowbreak{}O} & 105.03 & \textemdash{} & \textemdash{} & Not calculated \\
60 & \textemdash{} & \texttt{O\allowbreak{}=\allowbreak{}C\allowbreak{}C\allowbreak{}=\allowbreak{}C\allowbreak{}O\allowbreak{}O\allowbreak{}.\allowbreak{}[\allowbreak{}H\allowbreak{}]\allowbreak{}[\allowbreak{}H\allowbreak{}]} & 105.74 & \textemdash{} & \textemdash{} & Not calculated \\
61 & \textemdash{} & \texttt{[\allowbreak{}O\allowbreak{}-\allowbreak{}]\allowbreak{}[\allowbreak{}O\allowbreak{}H\allowbreak{}+\allowbreak{}]\allowbreak{}C\allowbreak{}1\allowbreak{}C\allowbreak{}C\allowbreak{}1\allowbreak{}O} & 108.13 & \textemdash{} & \textemdash{} & Not calculated \\
62 & \textemdash{} & \texttt{[\allowbreak{}C\allowbreak{}H\allowbreak{}-\allowbreak{}]\allowbreak{}=\allowbreak{}[\allowbreak{}O\allowbreak{}+\allowbreak{}]\allowbreak{}C\allowbreak{}C\allowbreak{}O\allowbreak{}O} & 108.90 & \textemdash{} & \textemdash{} & Not calculated \\
63 & \textemdash{} & \texttt{[\allowbreak{}C\allowbreak{}H\allowbreak{}-\allowbreak{}]\allowbreak{}=\allowbreak{}[\allowbreak{}O\allowbreak{}+\allowbreak{}]\allowbreak{}C\allowbreak{}(\allowbreak{}C\allowbreak{})\allowbreak{}O\allowbreak{}O} & 118.89 & \textemdash{} & \textemdash{} & Not calculated \\
64 & \textemdash{} & \texttt{C\allowbreak{}C\allowbreak{}=\allowbreak{}[\allowbreak{}O\allowbreak{}+\allowbreak{}]\allowbreak{}[\allowbreak{}O\allowbreak{}-\allowbreak{}]\allowbreak{}.\allowbreak{}[\allowbreak{}C\allowbreak{}H\allowbreak{}-\allowbreak{}]\allowbreak{}=\allowbreak{}[\allowbreak{}O\allowbreak{}H\allowbreak{}+\allowbreak{}]} & 123.22 & \textemdash{} & \textemdash{} & Not calculated \\
\end{longtable}
\endgroup
{\raggedright\SIcaptionfont
Each row represents a distinct reactant--product reaction channel. The 64 channels were validated at the g-xTB level in the first-round KHP search (Supplementary Fig.~15) and ranked by their lowest optimized g-xTB TS electronic energies. These g-xTB TS electronic energies are referenced to the original KHP single-point energy. For DFT, the electronic-energy barrier and Gibbs free-energy barrier are defined as $\Delta E^{\ddagger}=E_{\mathrm{TS}}-E_{\mathrm{R}}$ and $\Delta G^{\ddagger}=G_{\mathrm{TS}}-G_{\mathrm{R}}$, respectively, using the optimized reactant reference described in Supplementary Table~2. DFT results are reported where available and do not necessarily correspond to the lowest-energy g-xTB TS within each channel. All energies are reported in kcal~mol$^{-1}$.

DFT outcome indicates the results of TS optimization and IRC validation. Dashes indicate unavailable energy values, while ``Not calculated'' denotes channels for which DFT calculations were not performed. Asterisks mark pathways included in main-text Fig.~4c, and TS01--TS30 correspond to Supplementary Fig.~19 and Supplementary Table~2. Products are represented by SMILES where bond-order assignments are available, with stereochemistry omitted. A dagger ($\dagger$) in the Rank column marks products requiring candidate bond-order assignments based on their atom-connectivity graphs. These products are illustrated in Supplementary Fig.~20, with bond orders inferred using valence-based rules where possible. The resulting representations may include formal charge separation and radical centres.

The same definitions and notation apply to Supplementary Tables~4--8, except that their g-xTB TS electronic energies are referenced to the optimized starting reactant of each search.
\par}

\clearpage
{\centering\KHPtableheading{T10}{Reaction channels identified from IM1}\par}
{\SIcaptionfont The 57 reaction channels are ranked by their lowest g-xTB TS electronic energies relative to the optimized IM1 reactant. DFT results are reported where available. Other definitions and notation follow Supplementary Table~3.\par}
\begingroup
\fontsize{10}{12}\selectfont
\renewcommand{\arraystretch}{1.08}
\setlength{\tabcolsep}{2pt}
\setlength{\LTpre}{2mm}\setlength{\LTpost}{2mm}
\def\LTcaptype{}\stepcounter{LT@tables}
\begin{longtable}{@{} >{\raggedright\arraybackslash}m{9mm} >{\raggedright\arraybackslash}m{12mm} >{\raggedright\arraybackslash}m{58mm} M{18mm} M{18mm} M{18mm} M{31mm} @{}}
\toprule
Rank & TS ID & Product (SMILES/ID) & g-xTB $\Delta E^{\ddagger}$ & DFT $\Delta E^{\ddagger}$ & DFT $\Delta G^{\ddagger}$ & DFT outcome \\
\midrule\endfirsthead
\multicolumn{7}{l}{\bfseries Supplementary Table~\thetable\ (continued)} \\
\toprule
Rank & TS ID & Product (SMILES/ID) & g-xTB $\Delta E^{\ddagger}$ & DFT $\Delta E^{\ddagger}$ & DFT $\Delta G^{\ddagger}$ & DFT outcome \\
\midrule\endhead
\midrule\multicolumn{7}{r}{\footnotesize Continued on next page} \\\endfoot
\bottomrule\endlastfoot
01$^{*}$ & TS06 & \texttt{C\allowbreak{}C\allowbreak{}=\allowbreak{}O\allowbreak{}.\allowbreak{}O\allowbreak{}=\allowbreak{}C\allowbreak{}O} & 32.86 & 32.17 & 27.14 & Validated \\
02$^{*}$ & TS07 & \texttt{C\allowbreak{}=\allowbreak{}O\allowbreak{}.\allowbreak{}C\allowbreak{}C\allowbreak{}(\allowbreak{}=\allowbreak{}O\allowbreak{})\allowbreak{}O} & 34.57 & 35.10 & 29.25 & Validated \\
03$^{*}$ & TS08 & \texttt{O\allowbreak{}=\allowbreak{}C\allowbreak{}C\allowbreak{}C\allowbreak{}O\allowbreak{}O} & 45.93 & 41.14 & 37.67 & Validated \\
04$^{*}$ & TS09 & \texttt{O\allowbreak{}=\allowbreak{}C\allowbreak{}C\allowbreak{}C\allowbreak{}[\allowbreak{}O\allowbreak{}H\allowbreak{}+\allowbreak{}]\allowbreak{}[\allowbreak{}O\allowbreak{}-\allowbreak{}]} & 49.61 & 44.62 & 43.06 & Validated \\
05$^{*}$ & TS10 & \texttt{C\allowbreak{}=\allowbreak{}C\allowbreak{}(\allowbreak{}O\allowbreak{})\allowbreak{}O\allowbreak{}.\allowbreak{}C\allowbreak{}=\allowbreak{}O} & 53.37 & 49.72 & 44.22 & Validated \\
06 & \textemdash{} & \texttt{O\allowbreak{}.\allowbreak{}O\allowbreak{}=\allowbreak{}C\allowbreak{}C\allowbreak{}C\allowbreak{}=\allowbreak{}O} & 54.85 & 55.93 & 49.90 & Validated \\
07 & \textemdash{} & \texttt{C\allowbreak{}=\allowbreak{}C\allowbreak{}.\allowbreak{}O\allowbreak{}=\allowbreak{}C\allowbreak{}O\allowbreak{}O} & 58.71 & 54.77 & 51.64 & Validated \\
08 & \textemdash{} & \texttt{O\allowbreak{}=\allowbreak{}C\allowbreak{}(\allowbreak{}O\allowbreak{})\allowbreak{}C\allowbreak{}C\allowbreak{}O} & 63.55 & 64.80 & 59.02 & Validated \\
\end{longtable}
\endgroup

\clearpage
{\centering\KHPtableheading{T13}{Reaction channels identified from IM2}\par}
{\SIcaptionfont The 166 reaction channels are ranked by their lowest g-xTB TS electronic energies relative to the optimized IM2 reactant. DFT results are reported where available. Other definitions and notation follow Supplementary Table~3.\par}
\begingroup
\fontsize{10}{12}\selectfont
\renewcommand{\arraystretch}{1.08}
\setlength{\tabcolsep}{2pt}
\setlength{\LTpre}{2mm}\setlength{\LTpost}{2mm}
\def\LTcaptype{}\stepcounter{LT@tables}
\begin{longtable}{@{} >{\raggedright\arraybackslash}m{9mm} >{\raggedright\arraybackslash}m{12mm} >{\raggedright\arraybackslash}m{58mm} M{18mm} M{18mm} M{18mm} M{31mm} @{}}
\toprule
Rank & TS ID & Product (SMILES/ID)& g-xTB $\Delta E^{\ddagger}$ & DFT $\Delta E^{\ddagger}$ & DFT $\Delta G^{\ddagger}$ & DFT outcome \\
\midrule\endfirsthead
\multicolumn{7}{l}{\bfseries Supplementary Table~\thetable\ (continued)} \\
\toprule
Rank & TS ID & Product (SMILES/ID)& g-xTB $\Delta E^{\ddagger}$ & DFT $\Delta E^{\ddagger}$ & DFT $\Delta G^{\ddagger}$ & DFT outcome \\
\midrule\endhead
\midrule\multicolumn{7}{r}{\footnotesize Continued on next page} \\\endfoot
\bottomrule\endlastfoot
01$^{*}$ & TS11 & \texttt{C\allowbreak{}C\allowbreak{}(\allowbreak{}O\allowbreak{})\allowbreak{}O\allowbreak{}C\allowbreak{}=\allowbreak{}O} & 19.20 & 18.16 & 18.91 & Validated \\
02$^{*}$ & TS12 & \texttt{C\allowbreak{}=\allowbreak{}C\allowbreak{}O\allowbreak{}.\allowbreak{}O\allowbreak{}=\allowbreak{}C\allowbreak{}O} & 27.10 & 25.48 & 24.56 & Validated \\
03$^{*}$ & TS13 & \texttt{C\allowbreak{}C\allowbreak{}=\allowbreak{}O\allowbreak{}.\allowbreak{}O\allowbreak{}=\allowbreak{}C\allowbreak{}O} & 31.91 & 36.59 & 36.58 & Validated \\
04$^{*}$ & TS14 & \texttt{C\allowbreak{}C\allowbreak{}O\allowbreak{}.\allowbreak{}O\allowbreak{}=\allowbreak{}C\allowbreak{}=\allowbreak{}O} & 32.36 & 38.25 & 36.60 & Validated \\
05 & \textemdash{} & \texttt{C\allowbreak{}[\allowbreak{}C\allowbreak{}-\allowbreak{}]\allowbreak{}=\allowbreak{}[\allowbreak{}O\allowbreak{}H\allowbreak{}+\allowbreak{}]\allowbreak{}.\allowbreak{}O\allowbreak{}=\allowbreak{}C\allowbreak{}O} & 38.51 & \textemdash{} & \textemdash{} & TS/frequency/IRC failed \\
06$^{*}$ & TS15 & \texttt{C\allowbreak{}=\allowbreak{}C\allowbreak{}O\allowbreak{}.\allowbreak{}O\allowbreak{}=\allowbreak{}C\allowbreak{}O} & 43.62 & 44.90 & 42.84 & Validated \\
07 & \textemdash{} & \texttt{C\allowbreak{}C\allowbreak{}(\allowbreak{}O\allowbreak{})\allowbreak{}O\allowbreak{}C\allowbreak{}=\allowbreak{}O} & 43.80 & 42.84 & 42.89 & Validated \\
08 & \textemdash{} & \texttt{C\allowbreak{}=\allowbreak{}C\allowbreak{}O\allowbreak{}C\allowbreak{}(\allowbreak{}O\allowbreak{})\allowbreak{}O} & 48.82 & 49.56 & 49.89 & Validated \\
\end{longtable}
\endgroup

\clearpage
{\centering\KHPtableheading{T11}{Reaction channels identified from IM3}\par}
{\SIcaptionfont The 84 reaction channels are ranked by their lowest g-xTB TS electronic energies relative to the optimized IM3 reactant. DFT results are reported where available. Other definitions and notation follow Supplementary Table~3.\par}
\begingroup
\fontsize{10}{12}\selectfont
\renewcommand{\arraystretch}{1.08}
\setlength{\tabcolsep}{2pt}
\setlength{\LTpre}{2mm}\setlength{\LTpost}{2mm}
\def\LTcaptype{}\stepcounter{LT@tables}
\begin{longtable}{@{} >{\raggedright\arraybackslash}m{9mm} >{\raggedright\arraybackslash}m{12mm} >{\raggedright\arraybackslash}m{58mm} M{18mm} M{18mm} M{18mm} M{31mm} @{}}
\toprule
Rank & TS ID & Product (SMILES/ID)& g-xTB $\Delta E^{\ddagger}$ & DFT $\Delta E^{\ddagger}$ & DFT $\Delta G^{\ddagger}$ & DFT outcome \\
\midrule\endfirsthead
\multicolumn{7}{l}{\bfseries Supplementary Table~\thetable\ (continued)} \\
\toprule
Rank & TS ID & Product (SMILES/ID)& g-xTB $\Delta E^{\ddagger}$ & DFT $\Delta E^{\ddagger}$ & DFT $\Delta G^{\ddagger}$ & DFT outcome \\
\midrule\endhead
\midrule\multicolumn{7}{r}{\footnotesize Continued on next page} \\\endfoot
\bottomrule\endlastfoot
01$^{*}$ & TS17 & \texttt{C\allowbreak{}=\allowbreak{}C\allowbreak{}C\allowbreak{}=\allowbreak{}O\allowbreak{}.\allowbreak{}O\allowbreak{}O} & 31.81 & 28.42 & 24.27 & Validated \\
02$^{*}$ & TS16 & \texttt{O\allowbreak{}.\allowbreak{}O\allowbreak{}=\allowbreak{}C\allowbreak{}C\allowbreak{}C\allowbreak{}=\allowbreak{}O} & 33.71 & 29.00 & 24.02 & Validated \\
03$^{*}$ & TS18 & \texttt{O\allowbreak{}=\allowbreak{}C\allowbreak{}C\allowbreak{}C\allowbreak{}O\allowbreak{}O} & 40.09 & 36.69 & 33.44 & Validated \\
04$^{*}$ & TS20 & \texttt{C\allowbreak{}=\allowbreak{}C\allowbreak{}C\allowbreak{}=\allowbreak{}O\allowbreak{}.\allowbreak{}[\allowbreak{}O\allowbreak{}-\allowbreak{}]\allowbreak{}[\allowbreak{}O\allowbreak{}H\allowbreak{}2\allowbreak{}+\allowbreak{}]} & 43.43 & 40.64 & 37.15 & Validated \\
05 & \textemdash{} & \texttt{O\allowbreak{}.\allowbreak{}O\allowbreak{}=\allowbreak{}C\allowbreak{}C\allowbreak{}=\allowbreak{}C\allowbreak{}O} & 43.82 & \textemdash{} & \textemdash{} & Endpoint check failed \\
06$^{*}$ & TS19 & \texttt{O\allowbreak{}.\allowbreak{}O\allowbreak{}=\allowbreak{}C\allowbreak{}C\allowbreak{}1\allowbreak{}C\allowbreak{}O\allowbreak{}1} & 44.03 & 37.77 & 33.56 & Validated \\
07 & \textemdash{} & \texttt{O\allowbreak{}=\allowbreak{}C\allowbreak{}C\allowbreak{}C\allowbreak{}[\allowbreak{}O\allowbreak{}H\allowbreak{}+\allowbreak{}]\allowbreak{}[\allowbreak{}O\allowbreak{}-\allowbreak{}]} & 45.57 & 42.63 & 39.17 & Validated \\
08 & \textemdash{} & \texttt{[\allowbreak{}O\allowbreak{}-\allowbreak{}]\allowbreak{}[\allowbreak{}O\allowbreak{}H\allowbreak{}+\allowbreak{}]\allowbreak{}C\allowbreak{}C\allowbreak{}=\allowbreak{}C\allowbreak{}O} & 48.26 & 46.99 & 44.05 & Validated \\
\end{longtable}
\endgroup

\clearpage
{\centering\KHPtableheading{T14}{Reaction channels identified from IM4}\par}
{\SIcaptionfont The 114 reaction channels are ranked by their lowest g-xTB TS electronic energies relative to the optimized IM4 reactant. DFT results are reported where available. GX0494 denotes the only product for which no bond-order-resolved SMILES could be assigned; it is therefore represented by its connectivity graph in Supplementary Fig.~20. Other definitions and notation follow Supplementary Table~3.\par}
\begingroup
\fontsize{10}{12}\selectfont
\renewcommand{\arraystretch}{1.08}
\setlength{\tabcolsep}{2pt}
\setlength{\LTpre}{2mm}\setlength{\LTpost}{2mm}
\def\LTcaptype{}\stepcounter{LT@tables}
\begin{longtable}{@{} >{\raggedright\arraybackslash}m{9mm} >{\raggedright\arraybackslash}m{12mm} >{\raggedright\arraybackslash}m{58mm} M{18mm} M{18mm} M{18mm} M{31mm} @{}}
\toprule
Rank & TS ID & Product (SMILES/ID)& g-xTB $\Delta E^{\ddagger}$ & DFT $\Delta E^{\ddagger}$ & DFT $\Delta G^{\ddagger}$ & DFT outcome \\
\midrule\endfirsthead
\multicolumn{7}{l}{\bfseries Supplementary Table~\thetable\ (continued)} \\
\toprule
Rank & TS ID & Product (SMILES/ID)& g-xTB $\Delta E^{\ddagger}$ & DFT $\Delta E^{\ddagger}$ & DFT $\Delta G^{\ddagger}$ & DFT outcome \\
\midrule\endhead
\midrule\multicolumn{7}{r}{\footnotesize Continued on next page} \\\endfoot
\bottomrule\endlastfoot
01$^{*}$ & TS21 & \texttt{O\allowbreak{}=\allowbreak{}C\allowbreak{}(\allowbreak{}O\allowbreak{})\allowbreak{}C\allowbreak{}C\allowbreak{}O} & 19.58 & 15.91 & 23.15 & Validated \\
02$^{*}$ & TS22 & \texttt{O\allowbreak{}.\allowbreak{}O\allowbreak{}=\allowbreak{}C\allowbreak{}1\allowbreak{}C\allowbreak{}C\allowbreak{}O\allowbreak{}1} & 25.23 & 26.18 & 31.90 & Validated \\
03 & \textemdash{} & \texttt{C\allowbreak{}=\allowbreak{}C\allowbreak{}(\allowbreak{}O\allowbreak{})\allowbreak{}O\allowbreak{}C\allowbreak{}O} & 25.87 & \textemdash{} & \textemdash{} & Endpoint check failed \\
04$^{*}$ & TS24 & \texttt{C\allowbreak{}=\allowbreak{}C\allowbreak{}(\allowbreak{}O\allowbreak{})\allowbreak{}O\allowbreak{}C\allowbreak{}O} & 26.85 & 26.11 & 32.91 & Validated \\
05$^{*}$ & TS23 & \texttt{O\allowbreak{}.\allowbreak{}O\allowbreak{}=\allowbreak{}C\allowbreak{}=\allowbreak{}C\allowbreak{}C\allowbreak{}O} & 31.63 & 28.79 & 32.45 & Validated \\
06$^{*}$ & TS25 & \texttt{C\allowbreak{}C\allowbreak{}(\allowbreak{}=\allowbreak{}O\allowbreak{})\allowbreak{}O\allowbreak{}C\allowbreak{}O} & 32.96 & 26.12 & 33.36 & Validated \\
07 & \textemdash{} & \texttt{C\allowbreak{}=\allowbreak{}C\allowbreak{}(\allowbreak{}O\allowbreak{})\allowbreak{}O\allowbreak{}.\allowbreak{}C\allowbreak{}=\allowbreak{}O} & 36.16 & \textemdash{} & \textemdash{} & TS/frequency/IRC failed \\
08 & \textemdash{} & \texttt{C\allowbreak{}\#\allowbreak{}C\allowbreak{}O\allowbreak{}.\allowbreak{}O\allowbreak{}C\allowbreak{}O} & 39.52 & 36.14 & 41.60 & Validated \\
\end{longtable}
\endgroup

\clearpage
{\centering\KHPtableheading{T12}{Reaction channels identified from IM5}\par}
{\SIcaptionfont The 52 reaction channels are ranked by their lowest g-xTB TS electronic energies relative to the optimized IM5 reactant. DFT results are reported where available. Other definitions and notation follow Supplementary Table~3.\par}
\begingroup
\fontsize{10}{12}\selectfont
\renewcommand{\arraystretch}{1.08}
\setlength{\tabcolsep}{2pt}
\setlength{\LTpre}{2mm}\setlength{\LTpost}{2mm}
\def\LTcaptype{}\stepcounter{LT@tables}
\begin{longtable}{@{} >{\raggedright\arraybackslash}m{9mm} >{\raggedright\arraybackslash}m{12mm} >{\raggedright\arraybackslash}m{58mm} M{18mm} M{18mm} M{18mm} M{31mm} @{}}
\toprule
Rank & TS ID & Product (SMILES/ID)& g-xTB $\Delta E^{\ddagger}$ & DFT $\Delta E^{\ddagger}$ & DFT $\Delta G^{\ddagger}$ & DFT outcome \\
\midrule\endfirsthead
\multicolumn{7}{l}{\bfseries Supplementary Table~\thetable\ (continued)} \\
\toprule
Rank & TS ID & Product (SMILES/ID)& g-xTB $\Delta E^{\ddagger}$ & DFT $\Delta E^{\ddagger}$ & DFT $\Delta G^{\ddagger}$ & DFT outcome \\
\midrule\endhead
\midrule\multicolumn{7}{r}{\footnotesize Continued on next page} \\\endfoot
\bottomrule\endlastfoot
01$^{*}$ & TS26 & \texttt{O\allowbreak{}.\allowbreak{}O\allowbreak{}=\allowbreak{}C\allowbreak{}C\allowbreak{}=\allowbreak{}C\allowbreak{}O} & 31.45 & 27.00 & 28.73 & Validated \\
02$^{*}$ & TS29 & \texttt{C\allowbreak{}=\allowbreak{}C\allowbreak{}O\allowbreak{}.\allowbreak{}O\allowbreak{}=\allowbreak{}C\allowbreak{}O} & 38.23 & 35.54 & 38.61 & Validated \\
03 & \textemdash{} & \texttt{O\allowbreak{}C\allowbreak{}1\allowbreak{}C\allowbreak{}C\allowbreak{}(\allowbreak{}O\allowbreak{})\allowbreak{}O\allowbreak{}1} & 42.19 & 40.41 & 45.20 & Validated \\
04$^{*}$ & TS30 & \texttt{O\allowbreak{}=\allowbreak{}C\allowbreak{}C\allowbreak{}C\allowbreak{}(\allowbreak{}O\allowbreak{})\allowbreak{}O} & 42.97 & 38.99 & 41.70 & Validated \\
05$^{*}$ & TS28 & \texttt{C\allowbreak{}=\allowbreak{}C\allowbreak{}O\allowbreak{}.\allowbreak{}O\allowbreak{}.\allowbreak{}[\allowbreak{}C\allowbreak{}-\allowbreak{}]\allowbreak{}\#\allowbreak{}[\allowbreak{}O\allowbreak{}+\allowbreak{}]} & 49.55 & 39.42 & 37.84 & Validated \\
06 & \textemdash{} & \texttt{C\allowbreak{}=\allowbreak{}C\allowbreak{}O\allowbreak{}C\allowbreak{}=\allowbreak{}O\allowbreak{}.\allowbreak{}O} & 50.07 & 46.35 & 47.42 & Validated \\
07$^{*}$ & TS27 & \texttt{C\allowbreak{}=\allowbreak{}C\allowbreak{}O\allowbreak{}.\allowbreak{}O\allowbreak{}.\allowbreak{}[\allowbreak{}C\allowbreak{}-\allowbreak{}]\allowbreak{}\#\allowbreak{}[\allowbreak{}O\allowbreak{}+\allowbreak{}]} & 51.22 & 37.29 & 35.96 & Validated \\
08 & \textemdash{} & \texttt{O\allowbreak{}.\allowbreak{}O\allowbreak{}=\allowbreak{}C\allowbreak{}C\allowbreak{}[\allowbreak{}C\allowbreak{}-\allowbreak{}]\allowbreak{}=\allowbreak{}[\allowbreak{}O\allowbreak{}H\allowbreak{}+\allowbreak{}]} & 62.49 & 55.32 & 54.30 & Validated \\
\end{longtable}
\endgroup

\end{document}